\documentclass[12pt]{article}
\usepackage{eqsection,subeqnarray,indent,amsfonts,epsf}

 \def\subdominantcolor{red}          

\begin{document}

\bibliographystyle{plain}

\date{April 19, 2000 \\[2mm] revised December 13, 2000}

\title{\vspace*{-1cm}
       Transfer Matrices and Partition-Function Zeros \\
       for Antiferromagnetic Potts Models \\[5mm]
       \large\bf I.~General Theory and Square-Lattice Chromatic Polynomial}

\author{
  \\
  {\small Jes\'us Salas}                                    \\[-0.2cm]
  {\small\it Departamento de F\'{\i}sica Te\'orica}         \\[-0.2cm]
  {\small\it Facultad de Ciencias, Universidad de Zaragoza} \\[-0.2cm]
  {\small\it Zaragoza 50009, SPAIN}                         \\[-0.2cm]
  {\small\tt JESUS@MELKWEG.UNIZAR.ES}                        \\[5mm]
  {\small Alan D.~Sokal
                   }                  \\[-0.2cm]
  {\small\it Department of Physics}       \\[-0.2cm]
  {\small\it New York University}         \\[-0.2cm]
  {\small\it 4 Washington Place}          \\[-0.2cm]
  {\small\it New York, NY 10003 USA}      \\[-0.2cm]
  {\small\tt SOKAL@NYU.EDU}               \\[-0.2cm]
  {\protect\makebox[5in]{\quad}}  
  \\
}

\maketitle
\thispagestyle{empty}   


\begin{abstract}
We study the chromatic polynomials (= zero-temperature antiferromagnetic
Potts-model partition functions) $P_G(q)$ for
$m \times n$ rectangular subsets of the square lattice,
with $m \le 8$ (free or periodic transverse boundary conditions)
and $n$ arbitrary (free longitudinal boundary conditions),
using a transfer matrix in the Fortuin-Kasteleyn representation.
In particular, we extract the limiting curves of partition-function zeros
when $n \to\infty$, which arise from the crossing in modulus
of dominant eigenvalues (Beraha--Kahane--Weiss theorem).
We also provide evidence that the Beraha numbers $B_2,B_3,B_4,B_5$
are limiting points of partition-function zeros as $n \to\infty$
whenever the strip width $m$ is
$\ge 7$ (periodic transverse b.c.)\ or $\ge 8$ (free transverse b.c.).
Along the way, we prove that a noninteger Beraha number
(except perhaps $B_{10}$) cannot be a chromatic root of any graph.
\end{abstract}

\bigskip
\noindent
{\bf Key Words:}  Chromatic polynomial; chromatic root;
antiferromagnetic Potts model; square lattice;
transfer matrix; Fortuin-Kasteleyn representation; Temperley-Lieb algebra;
Beraha--Kahane--Weiss theorem; Beraha numbers.

\clearpage

\newcommand{\be}{\begin{equation}}
\newcommand{\ee}{\end{equation}}
\newcommand{\<}{\langle}
\renewcommand{\>}{\rangle}
\newcommand{\widebar}{\overline}
\def\reff#1{(\protect\ref{#1})}
\def\spose#1{\hbox to 0pt{#1\hss}}
\def\ltapprox{\mathrel{\spose{\lower 3pt\hbox{$\mathchar"218$}}
 \raise 2.0pt\hbox{$\mathchar"13C$}}}
\def\gtapprox{\mathrel{\spose{\lower 3pt\hbox{$\mathchar"218$}}
 \raise 2.0pt\hbox{$\mathchar"13E$}}}
\def\textprime{${}^\prime$}
\def\proof{\par\medskip\noindent{\sc Proof.\ }}
\def\qed{\hbox{\hskip 6pt\vrule width6pt height7pt depth1pt \hskip1pt}\bigskip}
\def\proofof#1{\bigskip\noindent{\sc Proof of #1.\ }}
\def\half{ {1 \over 2} }
\def\third{ {1 \over 3} }
\def\twothird{ {2 \over 3} }
\def\smfrac#1#2{\textstyle{#1\over #2}}
\def\smhalf{ \smfrac{1}{2} }
\newcommand{\real}{\mathop{\rm Re}\nolimits}
\renewcommand{\Re}{\mathop{\rm Re}\nolimits}
\newcommand{\imag}{\mathop{\rm Im}\nolimits}
\renewcommand{\Im}{\mathop{\rm Im}\nolimits}
\newcommand{\sgn}{\mathop{\rm sgn}\nolimits}
\newcommand{\tr}{\mathop{\rm tr}\nolimits}
\newcommand{\diag}{\mathop{\rm diag}\nolimits}
\newcommand{\Gal}{\mathop{\rm Gal}\nolimits}
\newcommand{\mycup}{\mathop{\cup}}
\def\hboxscript#1{ {\hbox{\scriptsize\em #1}} }
\def\zhat{ {\widehat{Z}} }
\def\phat{ {\widehat{P}} }
\def\qtilde{ {\widetilde{q}} }

\def\scra{\mathcal{A}}
\def\scrb{\mathcal{B}}
\def\scrc{\mathcal{C}}
\def\scrd{\mathcal{D}}
\def\scrf{\mathcal{F}}
\def\scrg{\mathcal{G}}
\def\scrl{\mathcal{L}}
\def\scro{\mathcal{O}}
\def\scrp{\mathcal{P}}
\def\scrq{\mathcal{Q}}
\def\scrr{\mathcal{R}}
\def\scrs{\mathcal{S}}
\def\scrt{\mathcal{T}}
\def\scrv{\mathcal{V}}
\def\scrz{\mathcal{Z}}

\def\q{{\sf q}}

\def\Z{{\mathbb Z}}
\def\R{{\mathbb R}}
\def\C{{\mathbb C}}
\def\Q{{\mathbb Q}}

\def\T{{\mathsf T}}
\def\H{{\mathsf H}}
\def\V{{\mathsf V}}
\def\D{{\mathsf D}}
\def\J{{\mathsf J}}
\def\P{{\mathsf P}}
\def\QQ{{\mathsf Q}}
\def\RR{{\mathsf R}}

\def\bsigma{\mbox{\protect\boldmath $\sigma$}}
\def\bone{{\mathbf 1}}
\def\vv{{\bf v}}
\def\w{{\bf w}}

\newtheorem{theorem}{Theorem}[section]
\newtheorem{proposition}[theorem]{Proposition}
\newtheorem{lemma}[theorem]{Lemma}
\newtheorem{corollary}[theorem]{Corollary}
\newtheorem{conjecture}[theorem]{Conjecture}


\newenvironment{sarray}{
          \textfont0=\scriptfont0
          \scriptfont0=\scriptscriptfont0
          \textfont1=\scriptfont1
          \scriptfont1=\scriptscriptfont1
          \textfont2=\scriptfont2
          \scriptfont2=\scriptscriptfont2
          \textfont3=\scriptfont3
          \scriptfont3=\scriptscriptfont3
        \renewcommand{\arraystretch}{0.7}
        \begin{array}{l}}{\end{array}}

\newenvironment{scarray}{
          \textfont0=\scriptfont0
          \scriptfont0=\scriptscriptfont0
          \textfont1=\scriptfont1
          \scriptfont1=\scriptscriptfont1
          \textfont2=\scriptfont2
          \scriptfont2=\scriptscriptfont2
          \textfont3=\scriptfont3
          \scriptfont3=\scriptscriptfont3
        \renewcommand{\arraystretch}{0.7}
        \begin{array}{c}}{\end{array}}

\section{Introduction}   \label{sec1}

The Potts model \cite{Potts_52,Wu_82,Wu_84}
plays an important role in the general theory of critical phenomena,
especially in two dimensions
\cite{Baxter_82,Itzykson_collection,DiFrancesco_97},
and has applications to various condensed-matter systems \cite{Wu_82}.
Ferromagnetic Potts models have been extensively studied over the
last two decades, and much is known about their
phase diagrams \cite{Wu_82,Wu_84}
and critical exponents \cite{Itzykson_collection,DiFrancesco_97,Nienhuis_84}.
But for antiferromagnetic Potts models, many basic questions remain open:
Is there a phase transition at finite temperature, and if so, of what order?
What is the nature of the low-temperature phase(s)?
If there is a critical point, what are the critical exponents and the
universality classes?
The answers to these questions are expected to be highly lattice-dependent,
in sharp contrast to the universality typically enjoyed by ferromagnets.

According to the Yang-Lee picture of phase transitions \cite{Yang-Lee_52},
information about the possible loci of phase transitions can be obtained
by investigating the zeros of the partition function
when one or more physical parameters (e.g.\ temperature or magnetic field)
are allowed to take {\em complex}\/ values.
For the Potts model on a finite graph $G$,
the partition function $Z_G(q,v)$
depends on the number $q$ of Potts states
and on the temperature-like variable $v = e^{\beta J} - 1$.
The Fortuin-Kasteleyn representation \cite{Kasteleyn_69,Fortuin_72}
shows that $Z_G(q,v)$ is a polynomial in $q$ and $v$
(see Section~\ref{sec2.1}),
so it makes sense for either or both of these variables to be made complex.
In particular, the chromatic polynomial $P_G(q) = Z_G(q,-1)$
corresponds to the zero-temperature limit of the antiferromagnetic
Potts model ($J=-\infty$, $v=-1$).

Many investigations of the zeros of Potts partition functions
in the complex $q$- and/or $v$-plane
have been performed in the last few years,
notably by Shrock and collaborators
\cite{Shrock_97a,Shrock_97b,Shrock_97c,Shrock_97d,Feldmann_98a,Feldmann_98b,%
Shrock_98a,Shrock_98b,Shrock_98c,Tsai_98,Shrock_98e,%
Shrock_99a,Shrock_99b,Shrock_99c,Shrock_99d,Shrock_99e,Shrock_99f,Shrock_99g,%
Shrock_00a,Shrock_00b,Shrock_00c,Shrock_00d,Shrock_00e,Biggs_99a,Biggs_99b}.
The best results concern families $G_n$ of graphs for which
the partition function can be expressed via a transfer matrix $T$
of fixed size $M \times M$:
\begin{subeqnarray}
   Z_{G_n}(q,v)  & = &   \tr[ A(q,v) \, T(q,v)^n ]  \\[2mm]
                 & = &   \sum\limits_{k=1}^{M}
                             \alpha_k(q,v) \, \lambda_k(q,v)^n
   \;,
\end{subeqnarray}
where the transfer matrix $T(q,v)$ and the boundary-condition matrix $A(q,v)$
are polynomials in $q$ and $v$,
so that the eigenvalues $\{ \lambda_k \}$ of $T$
and the amplitudes $\{ \alpha_k \}$
are algebraic functions of $q$ and $v$.
It then follows, using a theorem of Beraha--Kahane--Weiss
\cite{BKW_75,BKW_78,Beraha_79,Beraha_80,Sokal_hierarchical},
that the zeros of $Z_{G_n}(q,v)$ accumulate along the curves $\scrb$
where $T$ has two or more dominant eigenvalues
(i.e.\ eigenvalues of maximum modulus),
as well as at the isolated points where
$T$ has a single dominant eigenvalue $\lambda_k$
whose corresponding amplitude $\alpha_k$ vanishes.
See Section~\ref{sec2.2} for more details.

For ferromagnetic Potts models on the
square, triangular and hexagonal lattices,
the exact critical curves $v_c(q)$ in the {\em real}\/ $(q,v)$-plane
have long been known \cite{Baxter_82}.
For antiferromagnetic Potts models, by contrast,
there are some tantalizing conjectures concerning the critical loci,
but many aspects remain obscure.\footnote{
   For a more detailed review, see \cite[Section 1]{Salas_97}.
}
The two best-understood cases appear to be the square and triangular
lattices:

{\em Square lattice.}\/
Baxter \cite{Baxter_82,Baxter_82b} has determined the exact free energy
(among other quantities) for the square-lattice Potts model
on two special curves in the $(q,v)$-plane:
\begin{eqnarray}
   v   & = &     \pm \sqrt{q}             \label{eqDOB1.1}  \\[1mm]
   v   & = &  -2 \pm \sqrt{4-q}           \label{eqDOB1.2}
\end{eqnarray}
Curve (\ref{eqDOB1.1}${}_+$) is known to correspond to the ferromagnetic
critical point,
and Baxter \cite{Baxter_82b} conjectured that
curve (\ref{eqDOB1.2}${}_+$) corresponds to the
antiferromagnetic critical point.
For $q=2$ this gives the known exact value \cite{Onsager_44};
for $q=3$ it predicts a zero-temperature critical point ($v_c = -1$),
in accordance with strong analytical and numerical evidence
\cite{Lenard_67,Baxter_70,Nijs_82,Burton_Henley_97,%
Salas_98,deQueiroz_99,Ferreira_99};
and for $q>3$ it predicts that the putative critical point lies in the
unphysical region ($v < -1$ or $v$ complex),
so that the entire physical region $-1 \le v \le 0$
lies in the disordered phase,
in agreement with numerical evidence for $q=4$ \cite{Ferreira_99}.
For some interesting further speculations,
see Saleur \cite{Saleur_90,Saleur_91}.

{\em Triangular lattice.}\/
Baxter and collaborators \cite{Baxter_78,Baxter_86,Baxter_87}
have determined the exact free energy
(among other quantities) for the triangular-lattice Potts model
on two special curves in the $(q,v)$-plane:
\begin{eqnarray}
   v^3 + 3v^2 - q   & = &  0             \label{eqDOB1.3}  \\[1mm]
   v                & = &  -1            \label{eqDOB1.4}
\end{eqnarray}
The uppermost branch ($v \ge 0$) of curve \reff{eqDOB1.3}
is known to correspond to the ferromagnetic critical point
\cite{Baxter_78,Baxter_82};
and Baxter \cite{Baxter_86} initially conjectured
(following a hint of Nienhuis \cite{Nienhuis_82})
that \reff{eqDOB1.4}
--- which is the zero-temperature antiferromagnetic model,
hence the chromatic polynomial ---
corresponds in the interval $0 \le q \le 4$
to the antiferromagnetic critical point.
This prediction of a zero-temperature critical point
is known to be correct for $q=2$ \cite{Stephenson_64,Blote_82b,Nienhuis_84b}
and is believed to be correct also for $q=4$
\cite{Baxter_70_TRI,Henley_unpublished,Salas_TRI4state}.
On the other hand, for $q=3$ this prediction contradicts the rigorous result
\cite{vEFS_unpub}, based on Pirogov-Sinai theory,
that there is a low-temperature phase with long-range order
and small correlation length.\footnote{
   A Monte Carlo study of the $q=3$ model
   found strong evidence for a first-order transition
   to an ordered phase at $\beta J \approx -1.594$ \cite{Adler_95}.
}
%
%
%
For the model \reff{eqDOB1.4},
Baxter \cite{Baxter_86} computed three different expressions
$\lambda_i(q)$ [$i=1,2,3$] that he argued correspond to the
dominant eigenvalues of the transfer matrix in different regions
$\scrd_i$ of the complex $q$-plane;
in a second paper \cite{Baxter_87} he provided corrected estimates
for the precise locations of $\scrd_1,\scrd_2,\scrd_3$.
Unfortunately, no analogous analytic prediction is available
for the chromatic polynomials of other two-dimensional lattices.

One way to test the conjecture that (\ref{eqDOB1.2}${}_+$)
is a critical curve for the square-lattice Potts model
is to compute the partition function $Z_{m \times n}(q,v)$
for $m \times n$ strips of the square lattice,
investigate its zero variety in the complex $(q,v)$-space,
and test whether the zeros of $Z_{m \times n}(q,v)$
appear to be converging to (\ref{eqDOB1.2}${}_+$) as $m,n \to\infty$.
Here we shall carry out this program for the zero-temperature
antiferromagnetic model ($v=-1$).\footnote{
   In future work \cite{transfer5} we plan to extend this analysis
   to the antiferromagnetic model at nonzero temperature
   ($-1 < v < 0$).
   See also \cite{Shrock_00b}.
}
Using a transfer matrix in the Fortuin-Kasteleyn representation
\cite{Blote_82}, we shall compute the chromatic polynomials
$P_{m \times n}(q)$ for $m \times n$ square-lattice strips
of width $m \le 8$ (free or periodic transverse boundary conditions)
and arbitrary length $n$ (free longitudinal boundary conditions).
In particular, we shall extract the limiting curves $\scrb$
of partition-function zeros when $n \to\infty$,
which arise from the crossing in modulus of dominant eigenvalues
in accordance with the Beraha--Kahane--Weiss theorem.\footnote{
   Here we follow in the footsteps of
   Shrock and collaborators \cite{Shrock_98a,Shrock_98c,Shrock_00d},
   who have been carrying out this program
   using a generating-function approach that is equivalent to
   transfer matrices;
   they determine the recurrence relations by repeated use of
   the deletion-contraction identity.
   In particular, Shrock {\em et al.}\/ have computed the transfer matrices
   for square-lattice strips of width $m \le 5_{\rm F}$ and $m \le 6_{\rm P}$
   (leading to matrices of size up to $7 \times 7$),
   and have computed the limiting curves $\scrb$ of partition-function zeros
   for $m \le 4_{\rm F}$ and $m \le 5_{\rm P}$.
   [Here the subscript F (resp.\ P) denotes free (resp.\ periodic)
   boundary conditions.]
   By explicit use of transfer matrices,
   we are able to automate the former calculation
   and handle much larger transfer matrices
   (here up to $127 \times 127$);
   and using the resultant method (Section~\ref{sec4.1.1})
   we are able to detect small gaps and other fine details
   in the limiting curves.
}
Finally, we shall attempt to understand the behavior
of these limiting curves as $m \to\infty$.
Of course, there is little doubt in this case that they will converge
to the critical point of the zero-temperature model, $q_c = 3$;
but it is illuminating to see this convergence explicitly
and to view the critical {\em point}\/ $q_c = 3$
as simply one (real) point on a complex critical {\em curve}\/.
Not surprisingly, we find for this critical curve
a shape that is qualitatively similar
to that found by Baxter \cite{Baxter_87} for the triangular lattice,
with the zero-temperature critical point lying now at $q_c = 3$
rather than $q_c = 4$.

A special role in the theory of chromatic polynomials
appears to be played by the {\em Beraha numbers}\/ $B_n = 4 \cos^2 (\pi/n)$
[see Table~\ref{table_monomials} for the first few $B_n$].
As we shall show in Section~\ref{sec2.3},
a noninteger Beraha number (except possibly $B_{10}$)
cannot be a chromatic root of any graph.
Nevertheless, Beraha \cite{Beraha_unpub}
observed that planar graphs frequently have chromatic roots
very {\em near}\/ one or more of the $B_n$.\footnote{
   For $B_5$ this was observed earlier by Berman and Tutte \cite{Berman_69}.
}
Indeed, Beraha, Kahane and Weiss \cite{Beraha_79,Beraha_80}
found families of planar graphs that have chromatic roots
{\em converging}\/ to $B_5$, $B_7$ or $B_{10}$.\footnote{
   The graphs in question are $4_{\rm P} \times n_{\rm F}$,
   $5_{\rm P} \times n_{\rm F}$ and $2_{\rm F} \times n_{\rm P}$
   strips of the triangular lattice (with an extra vertex
   adjoined at top and bottom in the first two cases).
   In the case of $B_5$ and $B_7$,
   Beraha, Kahane and Weiss \cite{Beraha_80} proved that there are even
   {\em real}\/ chromatic roots converging to them.
}

Here we shall provide additional curious evidence
in favor of the idea that chromatic roots
tend to accumulate at the Beraha numbers.
We find empirically (at least for $m \le 8$)
that on a square-lattice strip of width $m$
with either free or periodic transverse b.c.,
there is at least one vanishing amplitude $\alpha_i(q)$
at each of the first $m$ Beraha numbers $B_2,\ldots,B_{m+1}$
(but not higher ones).
Assuming that this behavior persists for all $m$,
in the limit $m \to\infty$ all the Beraha numbers will be
zeros of some amplitude.
Moreover, in all the cases except $m=7,8$ with free transverse b.c.\ (where
our computer power gave out) and $m=8$ with periodic transverse
b.c.\ (see Section~\ref{sec7.2} for a discussion on this point),
we verified that the vanishing amplitude corresponds to
the eigenvalue obtained by analytic continuation in $q$
from the one that is dominant at small real $q$ (e.g.\ at $q=1$),
in agreement with a conjecture of Baxter \cite[p.~5255]{Baxter_87}.
Thus, the first few Beraha numbers ---
namely, those (up to at most $B_{m+1}$) that lie below the point $q_0(m)$
where the dominant-eigenvalue-crossing locus $\scrb$
intersects the real axis ---
correspond to the vanishing of a dominant amplitude
and hence (via the Beraha--Kahane--Weiss theorem)
to a limit point of chromatic roots, while the remaining Beraha numbers do not.
As the strip width $m$ grows, this crossing point $q_0(m)$
increases and presumably tends to a limiting value $q_0(\infty)$;
for the square lattice, we expect $q_0(\infty)$ to lie somewhere around 2.9,
i.e.\ strictly between $B_5$ and $B_6$.
Therefore, for all sufficiently large strip widths,
we expect the Beraha numbers $B_2,B_3,B_4,B_5$
--- but not higher ones --- to be limiting points of chromatic roots.
Our data confirm (at least up to $m=8$)
that $B_2,B_3,B_4$ are limiting points of zeros for all widths $m \ge 4$,
and that $B_5$ is a limiting point of zeros for all widths $m \ge 7$
(cylindrical b.c.)\ or $m \ge 8$ (free b.c.).
This scenario for the accumulation of chromatic roots
at {\em some}\/ of the Beraha numbers
was set forth by Baxter \cite{Baxter_87}
and elaborated by Saleur \cite{Saleur_90}.
For further speculations on the special role of the Beraha numbers
in the Potts model, and especially for the chromatic polynomials
of planar graphs, see
\cite{Berman_69,Tutte_70a,Tutte_70b,Tutte_74a,Tutte_75a,Tutte_82a,Tutte_82b,%
Baxter_87,Martin_87,Martin_89,Martin_91,%
Saleur_90,Saleur_91,Jaeger_91,Kauffmann_93,Temperley_93,
Tutte_93,Jackson_94,Maillard_94a,Maillard_94b,Maillard_97}.

The plan of this paper is as follows:
In Section~\ref{sec2} we review the Fortuin-Kasteleyn representation,
the Beraha--Kahane--Weiss theorem, and some algebraic number theory
related to the Beraha numbers.
In Section~\ref{sec3} we explain how to construct transfer matrices
for the Potts model in the spin representation and in the
Fortuin-Kasteleyn representation,
and we compute the dimensions of these transfer matrices.
In Section~\ref{sec4} we discuss the general properties
of the dominant-eigenvalue-crossing curves $\scrb$
and the isolated limiting points of zeros.
In Sections~\ref{sec5} and \ref{sec6} we present our numerical results
for square-lattice strips with free and cylindrical boundary conditions.
In Section~\ref{sec7} we analyze the theoretical import of our
calculations, and discuss prospects for future work
\cite{transfer2,transfer3,transfer4,transfer5}.

\section{Preliminaries}   \label{sec2}

In Sections~\ref{sec2.1} and \ref{sec2.2} we review some well-known facts
about the Fortuin-Kasteleyn representation and
the Beraha--Kahane--Weiss theorem,
which will play a fundamental role in the remainder of the paper.
We also use these sections to set the notation.
In Section~\ref{sec2.3} we discuss some algebraic number theory
related to the Beraha numbers;
this section contains a few new results, notably Corollary~\ref{cor2.4}.

\subsection{Fortuin-Kasteleyn representation of the Potts model}
    \label{sec2.1}

Let $G = (V,E)$ be a finite undirected graph
with vertex set $V$ and edge set $E$,
let $\{ J_e \} _{e \in E}$ be a set of couplings,
and let $q$ be a positive integer.
Then the {\em $q$-state Potts model}\/ on $G$ with couplings $\{ J_e \}$
is, by definition, the canonical ensemble at inverse temperature $\beta$
for a model of spins $\{ \sigma_x \} _{x \in V}$
taking values in the set $\{ 1,2,\ldots,q \}$,
interacting via a Hamiltonian
\be
   H(\{\sigma\})   \;=\;
   - \! \sum_{e = \< xy \> \in E} J_e \delta(\sigma_x, \sigma_y)
\ee
where $\delta$ is the Kronecker delta.
The partition function is thus
\be
   Z_G(q, \{v_e\})   \;=\;
   \sum_{ \{\sigma\} }  \,  \prod_{e = \< xy \> \in E}  \,
      [ 1 + v_e \delta(\sigma_x, \sigma_y) ]
   \;,
 \label{def_Z_G}
\ee
where we have written
\be
   v_e  \;=\;  e^{\beta J_e} - 1  \;.
\ee
A coupling $J_e$ (or $v_e$)
is called ferromagnetic if $J_e \ge 0$ ($v_e \ge 0$)
and antiferromagnetic if $-\infty \le J_e \le 0$ ($-1 \le v_e \le 0$).
The $q$-coloring problem, in which adjacent spins are required
to take different values, corresponds to the
zero-temperature limit of the antiferromagnetic Potts model
(namely $J_e = -\infty$, $v_e = -1$).

In fact, $Z_G(q, \{v_e\})$ is the restriction to
positive integers $q$ of a {\em polynomial}\/ in $q$ and $\{v_e\}$
(with coefficients that are in fact 0 or 1).
To see this, expand out the product over $e \in E$ in \reff{def_Z_G},
and let $E' \subseteq E$ be the set of edges for which the term
$v_e \delta(\sigma_x, \sigma_y)$ is taken.
Now perform the sum over configurations $\{ \sigma \}$:
in each connected component of the subgraph $(V,E')$
the spin value $\sigma_x$ must be constant,
and there are no other constraints.
Therefore,
\be
   Z_G(q, \{v_e\})   \;=\;
   \sum_{ E' \subseteq E }  q^{k(E')}  \prod_{e \in E'}  v_e
   \;,
 \label{FK_rep}
\ee
where $k(E')$ is the number of connected components
(including isolated vertices) in the subgraph $(V,E')$.
The expansion \reff{FK_rep} was discovered by
Birkhoff \cite{Birkhoff_12} and Whitney \cite{Whitney_32a}
for the special case $v_e = -1$ (see also Tutte \cite{Tutte_47,Tutte_54});
in its general form it is due to
Fortuin and Kasteleyn \cite{Kasteleyn_69,Fortuin_72}
(see also \cite{Edwards-Sokal}).
We shall henceforth take \reff{FK_rep} as the {\em definition}\/
of $Z_G(q, \{v_e\})$ for arbitrary complex numbers $q$ and $\{v_e\}$.
When $v_e$ takes the same value $v$ for all edges $e$, we write $Z_G(q,v)$.
When $v_e = -1$ for all edges $e$, this defines the
chromatic polynomial $P_G(q)$.
Note that the chromatic polynomial $P_G(q)$ of any loopless graph $G$
is a monic polynomial in $q$ with integer coefficients.\footnote{
   A {\em loop}\/, in graph-theoretic terminology, is an edge connecting
   a vertex to itself.  Obviously, if $G$ has a loop,
   then its chromatic polynomial is identically zero.
}
See \cite{Read_68,Read_88} for excellent reviews on chromatic polynomials,
and \cite{Chia_97} for an extensive bibliography.

\subsection{Beraha--Kahane--Weiss theorem}   \label{sec2.2}

A central role in our work is played by a theorem
on analytic functions due to
Beraha, Kahane and Weiss \cite{BKW_75,BKW_78,Beraha_79,Beraha_80}
and generalized slightly by one of us \cite{Sokal_hierarchical}.
The situation is as follows:
Let $D$ be a domain (connected open set) in the complex plane,
and let $\alpha_1,\ldots,\alpha_M,\lambda_1,\ldots,\lambda_M$ ($M \ge 2$)
be analytic functions on $D$, none of which is identically zero.
For each integer $n \ge 0$, define
\be
   f_n(z)   \;=\;   \sum\limits_{k=1}^M \alpha_k(z) \, \lambda_k(z)^n
   \;.
   \label{def_fn}
\ee
We are interested in the zero sets
\be
   \scrz(f_n)   \;=\;   \{z \in D \colon\;  f_n(z) = 0 \}
\ee
and in particular in their limit sets as $n\to\infty$:
\begin{eqnarray}
   \liminf \scrz(f_n)   & = &  \{z \in D \colon\;
   \hbox{every neighborhood $U \ni z$ has a nonempty intersection} \nonumber\\
      & & \qquad \hbox{with all but finitely many of the sets } \scrz(f_n) \}
   \\[4mm]
   \limsup \scrz(f_n)   & = &  \{z \in D \colon\;
   \hbox{every neighborhood $U \ni z$ has a nonempty intersection} \nonumber\\
      & & \qquad \hbox{with infinitely many of the sets } \scrz(f_n) \}
\end{eqnarray}
Let us call an index $k$ {\em dominant at $z$}\/ if
$|\lambda_k(z)| \ge |\lambda_l(z)|$ for all $l$ ($1 \le l \le M$);
and let us write
\be
   D_k  \;=\;  \{ z \in D \colon\;  k \hbox{ is dominant at } z  \}
   \;.
\ee
Then the limiting zero sets can be completely characterized as follows:

\begin{theorem}
  {\bf \protect\cite{BKW_75,BKW_78,Beraha_79,Beraha_80,Sokal_hierarchical}}
   \label{BKW_thm}
Let $D$ be a domain in $\C$,
and let $\alpha_1,\ldots,\alpha_M,\lambda_1,\ldots,\lambda_M$ ($M \ge 2$)
be analytic functions on $D$, none of which is identically zero.
Let us further assume a ``no-degenerate-dominance'' condition:
there do not exist indices $k \neq k'$
such that $\lambda_k \equiv \omega \lambda_{k'}$ for some constant $\omega$
with $|\omega| = 1$ and such that $D_k$ ($= D_{k'}$)
has nonempty interior.
For each integer $n \ge 0$, define $f_n$ by
$$
   f_n(z)   \;=\;   \sum\limits_{k=1}^M \alpha_k(z) \, \lambda_k(z)^n
   \;.
$$
Then $\liminf \scrz(f_n) = \limsup \scrz(f_n)$,
and a point $z$ lies in this set if and only if either
\begin{itemize}
   \item[(a)]  There is a unique dominant index $k$ at $z$,
       and $\alpha_k(z) =0$;  or
   \item[(b)]  There are two or more dominant indices at $z$.
\end{itemize}
\end{theorem}
Note that case (a) consists of isolated points in $D$,
while case (b) consists of curves
(plus possibly isolated points where all the $\lambda_k$ vanish simultaneously).
Beraha--Kahane--Weiss considered the special case of Theorem~\ref{BKW_thm}
in which the $f_n$ are polynomials satisfying a linear finite-order
recurrence relation, and they assumed a slightly stronger nondegeneracy
condition.  (This is all we really need in this paper.)
Henceforth we shall denote by $\scrb$ the locus of points
satisfying condition (b).

In the applications considered in this paper,
the functions $f_n$ will be of the form
$f_n(z) = \tr [A(z) T(z)^n]$ where the $M \times M$ matrices
$T(z)$ and $A(z)$ are polynomials in $z$;
therefore, the functions $\lambda_k(z)$
[which are the eigenvalues of $T(z)$] and $\alpha_k(z)$
will be algebraic functions of $z$,
i.e.\ locally analytic except at isolated branch points.
So one can cover the complex plane minus branch points
by a family of simply connected domains $D$,
and then apply Theorem~\ref{BKW_thm} separately to each such domain $D$.
The branch points are not covered by this analysis,
but they will always be endpoints of curves of type (b),
hence also limit points of zeros.

It is interesting to ask about the rate at which the zeros of $f_n$
converge to the limit set.  In case (a), it is easy to see that the
convergence is exponentially fast.  In case (b),
simple expansions suggest that the rate is $1/n$ near regular points
of the curve $\scrb$, and $1/n^2$ at branch points
(see Sections \ref{sec4.2.1} and \ref{sec4.2.2}).

In checking for isolated limiting points of zeros [case (a)],
the following result will be useful (see Section \ref{sec4.3}):

\begin{lemma} {\bf \protect\cite{Beraha_80}}
   \label{lemma2.2}
Suppose that $f_n = \sum\limits_{k=1}^M \alpha_k \, \lambda_k^n$,
and define
\be
   D   \;=\;
        \left( \begin{array}{cccc}
                    f_0 & f_1 & \cdots & f_{M-1} \\
                    f_1 & f_2 & \cdots & f_{M}   \\
                    \vdots & \vdots &        & \vdots   \\
                    f_{M-1} & f_M & \cdots & f_{2M-2}
               \end{array}
        \right)
   \;.
\ee
Then
\be
   \det D   \;=\;
   \prod_{k=1}^M \alpha_k \,
   \prod_{1 \le i < j \le M}  (\lambda_j - \lambda_i)^2
   \;.
\ee
\end{lemma}

\proof
It is not difficult to see that
$D = \Lambda^{\rm T} \diag(\alpha_1,\ldots,\alpha_M) \Lambda$
where $\Lambda$ is the $M \times M$ Vandermonde matrix
$\Lambda_{ij} = \lambda_i^{j-1}$.
Lemma~\ref{lemma2.2} then follows from the well-known formula
\be
   \det \Lambda   \;=\;  \prod_{1 \le i < j \le M}  (\lambda_j - \lambda_i)
\ee
for the Vandermonde determinant.
\qed

\noindent
We remark that the product $\prod_{i < j}  (\lambda_j - \lambda_i)^2$
is the discriminant of the characteristic polynomial of the transfer matrix:
see \reff{discriminant} below.

\subsection{Beraha numbers}   \label{sec2.3}

We recall \cite{Stewart_87} that a complex number $\zeta$
is called an {\em algebraic number}\/ (resp.\ an {\em algebraic integer}\/)
if it is a root of some monic polynomial with rational (resp.\ integer)
coefficients.  Corresponding to any algebraic number $\zeta$,
there is a unique monic polynomial $p$ with rational coefficients,
called the {\em minimal polynomial of $\zeta$}\/ (over the rationals),
with the property that $p$ divides every polynomial with rational coefficients
having $\zeta$ as a root.  (The minimal polynomial of $\zeta$
has integer coefficients if and only if $\zeta$ is an algebraic integer.)
Two algebraic numbers $\zeta$ and $\zeta'$ are called {\em conjugate}\/
if they have the same minimal polynomial.
Conjugacy is obviously an equivalence relation,
and so divides the set of algebraic numbers into equivalence classes.

\bigskip

{\bf Examples.}
1.  The numbers $B_5 = (3+\sqrt{5})/2$ and $B_5^* = (3-\sqrt{5})/2$
are conjugates, as they have the same minimal polynomial
$p(x) = x^2 -3x+1$.

2.  The numbers $2^{1/3}$, $2^{1/3} e^{2 \pi i/3}$ and $2^{1/3} e^{4 \pi i/3}$
are all conjugates, as they have the same minimal polynomial $p(x) = x^3 -2$.

3.  If $\zeta$ is a primitive $n^{th}$ root of unity,
the minimal polynomial of $\zeta$ is the {\em cyclotomic polynomial}\/
\be
   \Phi_n(x)  \;=\;  \prod_{\begin{scarray}
                               1 \le k \le n \\[1mm]
                               \gcd(k,n) = 1
                            \end{scarray} }
                     (x \,-\, e^{2\pi ki/n})
   \;,
\ee
where the product runs over all positive integers $k \le n$
that are relatively prime to $n$
(see e.g.\ \cite[Section 3.7]{Cohn_89} for a proof).
In particular, $\Phi_n$ is a monic polynomial with integer coefficients,
irreducible over the field of rational numbers, of degree
\be
   \deg \Phi_n  \;=\;  \varphi(n)
\ee
where $\varphi(n)$ is the Euler totient function (i.e.\ the number of
positive integers $k \le n$ that are relatively prime to $n$).
Thus, all the primitive $n^{th}$ roots of unity are mutually conjugate.

4.  For $n \ge 2$, let us define the generalized Beraha numbers
\be
   B_n^{(k)}   \;=\;  4 \cos^2 {k\pi \over n}
               \;=\;  2 \,+\, 2 \cos {2\pi k \over n}
   \;,
\ee
and let us call $B_n^{(k)}$ a primitive $n^{th}$ generalized Beraha number
in case $k$ is relatively prime to $n$.
Then the minimal polynomial of $B_n$
(or of any primitive $B_n^{(k)}$) is\footnote{
   The proof of \reff{def_pn} requires some elementary Galois theory
   \cite[Chapter 3]{Cohn_89} \cite{Stewart_89}:
   Fix $n \ge 3$ and $\zeta = e^{2\pi i/n}$,
   and let $F$ be the extension field $\Q(\zeta)$.
   The irreducibility of the cyclotomic polynomial $\Phi_n$
   implies that $F$ has dimension $\varphi(n)$ over $\Q$,
   and that its Galois group $G = \Gal(F/\Q)$
   is the abelian group $\{ \sigma_k \}_{1 \le k \le n, \, \gcd(k,n)=1}$
   where $\sigma_k(r)=r$ for $r \in \Q$ and $\sigma_k(\zeta) = \zeta^k$.
   Now define $c \equiv 2 \cos(2\pi/n) = \zeta + \bar{\zeta} \in \R$.
   By repeated application of the equation $\zeta^2 - c\zeta + 1 = 0$
   it follows that $\Q(\zeta) = \Q(c) \oplus \zeta \Q(c)$,
   so that $F$ has dimension 2 over $\Q(c)$.
   %
   %
   Therefore, $\Q(c)$ has dimension $\varphi(n)/2$ over $\Q$,
   so that the minimal polynomial $P(x)$ of $c$ over $\Q$
   has degree $\varphi(n)/2$.
   Now, for every $k$ that is relatively prime to $n$,
   let us define $c_k = \sigma_k(c) = \zeta^k + \bar{\zeta}^k$;
   there are clearly $\varphi(n)/2$ distinct such numbers $c_k$
   (note that $c_{n-k} = c_k$).
   And we have $P(c_k) = P(\sigma_k(c)) = \sigma_k(P(c)) = 0$
   since $\sigma_k$ is a field automorphism and $P$ has rational coefficients.
   It follows that $P(x) = \prod_{1 \le k \le n/2, \, \gcd(k,n)=1} (x - c_k)$.
   A trivial shift $x \to x-2$ gives the same result for the
   $B_n^{(k)} = 2 + c_k$.
   We thank Dan Segal for explaining this proof to us.
}
\be
   p_n(x)   \;=\;  \prod_{\begin{scarray}
                               1 \le k \le n/2 \\[1mm]
                               \gcd(k,n) = 1
                            \end{scarray} }
                     (x \,-\, B_n^{(k)})
 \label{def_pn}
   \;.
\ee
In Table~\ref{table_monomials} we show the $p_n$ for $2 \le n \le 16$.
$p_n$ is a monic polynomial with integer coefficients,
irreducible over the field of rational numbers, of degree
\be
   \deg p_n  \;=\;  \cases{ 1             & for $n=2$  \cr
                            \varphi(n)/2  & for $n \ge 3$ \cr
                         }
\ee
Thus, all the primitive $n^{th}$ generalized Beraha numbers are
mutually conjugate.

\bigskip

The point of this digression into algebraic number theory
is, of course, that the chromatic polynomial $P_G(q)$
of any loopless graph $G$ is a monic polynomial with integer coefficients.
Therefore, all the chromatic roots of $G$ are algebraic integers;
and if $\zeta$ is a chromatic root of $G$, then so are all its conjugates.
We therefore have:

\begin{proposition}
   \label{prop2.3}
\begin{itemize}
   \item[(a)]  Suppose that $\zeta$ has a conjugate lying in one of the
intervals $(-\infty,0)$, $(0,1)$ or $(1, 32/27]$.  Then $\zeta$ is not
a chromatic root of any loopless graph.
   \item[(b)]  Suppose that $\zeta$ has a conjugate lying in the
interval $[5,\infty)$.  Then $\zeta$ is not a chromatic root
of any loopless {\em planar} graph.
   \item[(c)]  Suppose that $\zeta$ has a conjugate lying in the
interval $(1,2)$.  Then $\zeta$ is not a chromatic root
of any {\em plane near-triangulation}.
   \item[(d)]  Suppose that $\zeta$ has a conjugate lying in the
interval $(2,2.546602\ldots)$, where $2.546602\ldots$ is shorthand for the
unique real solution of $q^3 - 9q^2 + 29q - 32 = 0$
(which is the nontrivial chromatic root of the octahedron).
Then $\zeta$ is not a chromatic root of any {\em plane triangulation}.
\end{itemize}
\end{proposition}

\proof
This follows immediately from the facts that there are no chromatic roots
in the relevant intervals:
see ref.~\cite{Read_88} for $(-\infty,0)$ and $(0,1)$,
ref.~\cite{Jackson_93} for $(1, 32/27]$,
refs.~\cite{Birkhoff_46,Woodall_97,Thomassen_97} for $[5,\infty)$,
refs.~\cite{Birkhoff_46,Woodall_77,Woodall_92a} for $(1,2)$,
and refs.~\cite{Woodall_77,Woodall_92b} for $(2,2.546602\ldots)$.\footnote{
   See also \cite{Woodall_97,Thomassen_97}
   for related results on zero-free intervals.
}
\qed

{\bf Examples.}
1.  A special case of Proposition \ref{prop2.3}(a) is the well-known fact
\cite[p.~23]{Read_88} that $B_5 = (3+\sqrt{5})/2$ is not a
chromatic root of any (loopless) graph, since its conjugate
$B_5^* = (3-\sqrt{5})/2$ lies in $(0,1)$
[their minimal polynomial is $q^2 -3q+1$].
We shall prove a vastly stronger result in Corollary~\ref{cor2.4} below.

2.  The number $1 + \sqrt{2}$ is not a chromatic root of any (loopless) graph,
since its conjugate $1 - \sqrt{2}$ lies in $(-\infty,0)$
[their minimal polynomial is $q^2 -2q-1$].

3.  For $n \ge 5$, none of the numbers $2^{1/n} e^{2\pi k i/n}$
($k$ integer) is a chromatic root of any (loopless) graph,
since their conjugate $2^{1/n}$ lies in $(1, 32/27]$
(their minimal polynomial is $q^n -2$).\footnote{
   This barely fails for $n=4$.  It is amusing to note that
   $32/27 \approx 2^{1/4}$ is equivalent to $(3/2)^{12} \approx 2^7$,
   which underlies the construction of the well-tempered scale.
}

4.  The numbers $2 + \sqrt{2\sqrt{2} -2}$ and $2 \pm i\sqrt{2\sqrt{2} +2}$
are not chromatic roots of any (loopless) graph,
since their conjugate $2 - \sqrt{2\sqrt{2} -2}$ lies in $(1, 32/27]$
(their minimal polynomial is $q^4 - 8q^3 + 28q^2 - 48q + 28$).

\bigskip

We now state and prove the promised generalization of the fact that
$B_5$ is not a chromatic root:

\begin{corollary}
   \label{cor2.4}
A noninteger primitive $n^{th}$ generalized Beraha number $B_n^{(k)}$
is not a chromatic root of any graph, except possibly when $n=10$.
\end{corollary}

\noindent
This is an immediate consequence of
the case $(0,1)$ of Proposition \ref{prop2.3}(a)
together with the following number-theoretic lemma:

\begin{lemma}
   \label{lemma2.5}
Let $n=5,7,8,9$ or $n \ge 11$.
Then there exists an integer $k$, relatively prime to $n$,
in the interval $n/3 < k < n/2$.
\end{lemma}

\proof
If $n$ is a prime $\ge 5$, the result is trivial
(take $k = \lfloor n/2 \rfloor$).
The cases $n=8,12$ can be verified by hand (take $k=3,5$, respectively).
Otherwise, note first that it suffices to find $k$ in the interval
$n/3 < k < 2n/3$ (if $k > n/2$, then replace $k$ by $n-k$).
So let $p$ be the least prime divisor of $n$;
define $r = \lceil p/3 \rceil \ge 1$, so that $rn/p \ge n/3 > (r-1)n/p$;
and let $k$ be either $rn/p + 1$ or $rn/p + 2$.
Now the only possible common prime factor of $k$ and $n$ is $p$:
for if $p' \neq p$ were to divide $n$,
then it would also divide $n/p$;
but if $p'$ also divided $k$, then it would divide
$k - rn/p = 1$ or 2, which is impossible since $p' \ge 3$.
But $p$ cannot divide {\em both}\/ choices of $k$.
Therefore, this construction allows $k$ to be chosen relatively prime to $n$;
and $k$ lies in the interval $(n/3, 2n/3)$ provided that
$n > 12$ (when $p=2$), $n > 6$ (when $p=3$) or $n \ge 15$ (when $p \ge 5$).
\qed

{\bf Remarks.}
1.  As mentioned earlier, the case $n=5$ of Corollary~\ref{cor2.4}
is ancient folklore;
the case $n=7$ was stated by Tutte \cite[p.~372]{Tutte_75a}.
Beraha, Kahane and Weiss \cite[p.~53]{Beraha_80}
asserted that the argument for $B_5$
``can be extended without much difficulty to the case of arbitrary
 nonintegral $B_n$'', apparently overlooking the problem with $B_{10}$.

2.  The exceptional case $n=10$ is very curious.
We do not know whether $B_{10} = (5 + \sqrt{5})/2 \approx 3.6180339887$
and $B_{10}^* = (5 - \sqrt{5})/2 \approx 1.3819660113$
can be chromatic roots.
But Proposition~\ref{prop2.3}(c) shows that $B_{10}$ is not a chromatic
root of any {\em plane near-triangulation}\/.
Note also that when $G$ is a {\em plane triangulation}\/ having $n$ vertices,
Tutte's ``golden identity'' \cite{Tutte_70b} \cite[pp.~26--27]{Read_88}
\be
   P_G(B_{10})   \;=\;  (\tau+2) \tau^{3n-10} P_G(B_5)^2
\ee
where $\tau = B_5 - 1$ is the golden ratio,
together with the fact that $P_G(B_5) \neq 0$,
yields the slightly stronger result $P_G(B_{10}) > 0$.

\section{Transfer-Matrix Theory}   \label{sec3}

In this section we briefly review the transfer-matrix formalism
for the Potts model, both in the spin representation (Section~\ref{sec3.1})
and in the Fortuin-Kasteleyn representation (Section~\ref{sec3.2}).
Thereafter we use only the Fortuin-Kasteleyn representation,
which allows us to perform computations for noninteger $q$.
We conclude by computing the size of the FK-representation transfer matrix
for the square and triangular lattices
with free longitudinal boundary conditions and either
free or periodic transverse boundary conditions (Section~\ref{sec3.3}).

\subsection{Transfer matrix in the spin representation}   \label{sec3.1}

Consider a graph $G_n=(V_n,E_n)$ consisting of $n$ identical ``layers'',
with connections between adjacent layers repeated in a regular fashion.
Then the partition function of the Potts model on $G$ can be written
in terms of a transfer matrix.

To make this precise, let $V^0$ be the set of vertices in a single layer,
let $E^0$ be the set of edges within a single layer
 (we call these {\em horizontal}\/ edges),
and let $E^*$ be the set of edges connecting each layer to the next one
 (we call these {\em vertical}\/ edges).
Note that $E^0$ is a set of unordered pairs of elements of $V^0$,
while $E^*$ is a set of ordered pairs of elements of $V^0$.
The vertex set of the graph $G_n$ is then
\be
   V_n   \;=\;  V^0 \,\times\, \{1,\ldots,n\}
         \;=\;  \{ (x,i) \colon\; x \in V^0 \hbox{ and } 1 \le i \le n \}
   \;,
 \label{eq3.1}
\ee
while the edge set is either
\be
   E_n^{\rm free}   \;=\;  \mycup\limits_{i=1}^n E^{\rm horiz}_i  \,\bigcup\,
                           \mycup\limits_{i=1}^{n-1} E^{\rm vert}_i
 \label{eq3.2}
\ee
for free longitudinal boundary conditions or
\be
   E_n^{\rm per}   \;=\;  \mycup\limits_{i=1}^n E^{\rm horiz}_i  \,\bigcup\,
                          \mycup\limits_{i=1}^n E^{\rm vert}_i
\ee
for periodic longitudinal boundary conditions,
where
\be
   E^{\rm horiz}_i   \;=\;
   \{ \< (x,i), (x',i) \>  \colon\;  \<xx'\> \in E^0 \}
\ee
and
\be
   E^{\rm vert}_i   \;=\;
   \{ \< (x,i), (x',i+1) \>  \colon\;  (x,x') \in E^* \}
\ee
and of course layer $n+1$ is identified with layer 1.
We also assume that the couplings are identical from layer to layer:
that is, we are given weights $\{ v_e \} _{e \in E^0 \cup E^*}$,
and we define the edge weights for $G_n$ by
\begin{subeqnarray}
   v_{\< (x,i), (x',i) \>}   & = &
      v_{\<xx'\>}  \qquad \hbox{for } \<xx'\> \in E^0   \\[2mm]
   v_{\< (x,i), (x',i+1) \>}   & = &
      v_{(x,x)}  \qquad \hbox{for } (x,x') \in E^*
 \label{eq3.6}
\end{subeqnarray}

We now fix an integer $q \ge 1$,
and let $\Sigma = \{1,\ldots,q\}^{V^0}$ be the space of
spin configurations on a single layer;
we denote a generic such configuration by $\bsigma = \{\sigma_x\}_{x \in V^0}$.
We then define the matrices $\H$ and $\V$,
which encode the Boltzmann factors corresponding to
horizontal and vertical edges, respectively:
\begin{eqnarray}
   \H(\bsigma',\bsigma)   & = &
   \delta(\bsigma,\bsigma')
     \prod_{\<xx'\> \in E^0} [1 + v_{\<xx'\>} \delta(\sigma_x, \sigma_{x'}) ]
   \\[4mm]
   \V(\bsigma',\bsigma)   & = &
     \prod_{(x,x') \in E^*} [1 + v_{(x,x')} \delta(\sigma_x, \sigma'_{x'}) ]
\end{eqnarray}
The partition function of the Potts model on $G_n$ is then
\begin{eqnarray}
   Z_{G_n^{\rm free}}(q, \{v_e\})   & = &
     \bone^{\rm T} \H (\V \H)^{n-1} \bone   \label{Z_free} \\[2mm]
   Z_{G_n^{\rm per}}(q, \{v_e\})   & = &   \tr[ (\V \H)^n ]  \label{Z_per}
\end{eqnarray}
where $\bone$ is the vector whose entries all equal 1,
and ${}^{\rm T}$ denotes transpose.
The transfer matrix is thus
\be
   \T  \;=\;  \V \H   \;.
\ee
It has size $q^m \times q^m$,
where $m = |V^0|$ is the number of sites in each layer.

The horizontal matrix $\H$ is diagonal in the spin basis, hence sparse;
its computation takes a time of order $|E^0| q^m \ltapprox m^2 q^m$.
But since the vertical matrix $\V$ is dense,
its computation takes in general a time of order $|E^*| q^{2m}$.
This is a severe constraint on the practical applicability of the
transfer-matrix method.
However, when the graphs $G_n$ are {\em planar}\/,
$\V$ can be written as a product of sparse matrices
that correspond to the replacement of {\em one}\/ site
on layer $i$ by the corresponding site on layer $i+1$,
so that its computation takes only a time of order $m q^m$.
(Indeed, this can be done for an arbitrary planar graph,
 whether or not it consists of repeated layers \cite{Kauffmann_93}.)
This factorization of $\V$ can also be performed for some non-planar graphs,
including all those in which the single-layer vertex set $V^0$
can be ordered in such a way that $(x,x') \in E^*$ implies
$x \succeq x'$.\footnote{
   This is equivalent to requiring that the directed graph $(V^0, E^{**})$
   be {\em acyclic}\/,
   where $E^{**} = \{ (x,x') \in E^* \colon\; x \neq x' \}$.
}
(This latter situation includes, in particular, all the lattices $\Z^d$,
 since the only vertical edges are of the form $(x,x)$.
 On the other hand, it excludes the triangular lattice with
 {\em periodic}\/ transverse boundary conditions,
 which needs a separate treatment.)
For simplicity we describe this construction
only for the cases of greatest practical interest,
namely when $G_n$ is a square or triangular lattice
with free or periodic transverse boundary conditions.
The single-layer vertex set is thus $V^0 = \{1,\ldots,m\}$;
the horizontal edge set is either
\be
   E^0_{\rm free}   \;=\;  \{ \<x, x+1 \>  \colon\;  1 \le x \le m-1 \}
\ee
for free transverse boundary conditions or
\be
   E^0_{\rm per}   \;=\;  \{ \<x, x+1 \>  \colon\;  1 \le x \le m \}
\ee
for periodic transverse boundary conditions
(where site $m+1$ is of course identified with site 1);
and the vertical edge set is either
\be
   E^0_{\rm SQ}   \;=\;  \{ (x,x)  \colon\;  1 \le x \le m \}
\ee
for the square lattice,
\be
   E^0_{\rm TRI,free}   \;=\;  \{ (x,x)  \colon\;  1 \le x \le m \}
            \,\bigcup\, \{ (x+1,x)  \colon\;  1 \le x \le m-1 \}
 \label{E0_TRI_free}
\ee
for the triangular lattice with free transverse boundary conditions, or
\be
   E^0_{\rm TRI,per}   \;=\;  \{ (x,x)  \colon\;  1 \le x \le m \}
            \,\bigcup\, \{ (x+1,x)  \colon\;  1 \le x \le m \}
 \label{E0_TRI_per}
\ee
for the triangular lattice with periodic transverse boundary conditions.
Note that the diagonal edges in \reff{E0_TRI_free}--\reff{E0_TRI_per}
point ``southeast--northwest''.

We now define $q^m \times q^m$ matrices $\D_x$ and $\J_{xx'}$ by
\begin{subeqnarray}
   \D_x(\bsigma',\bsigma)   & = &
      \prod_{y \neq x}  \delta(\sigma_y,\sigma'_y)   \\[2mm]
   \J_{xx'}(\bsigma',\bsigma)   & = &
      \delta(\sigma_x, \sigma_{x'}) \, \delta(\bsigma,\bsigma')
\end{subeqnarray}
Thus, $\D_x$ (for ``detach'' or ``disconnect'') is of the form
$I \otimes I \otimes \cdots \otimes E \otimes \cdots \otimes I \otimes I$,
where $I$ is the $q \times q$ identity matrix,
$E$ is the $q \times q$ matrix whose entries are all 1,
and the $E$ factor occurs at position $x$;
informally, $\D_x$ disconnects the two rows at site $x$.
Since $I$ and $E$ commute, all the operators $\D_x$ commute among themselves.
Likewise, $\J_{xx'}$ identifies (``joins'') spins $x$ and $x'$
in a single layer.
Since the operators $\J_{xx'}$ are diagonal in the spin basis,
they also commute among themselves.
Finally, $\D_x$ commutes with $\J_{x'x''}$ whenever
$x$ is different from both $x'$ and $x''$.

We now define matrices
\begin{eqnarray}
   \P_j   & = & v_{(j,j)} I  \,+\,  \D_j   \\[1mm]
   \QQ_j  & = & I  \,+\,  v_{\<j,j+1\>} \J_{j,j+1}   \\[1mm]
   \RR_j  & = & I  \,+\,  v_{(j+1,j)} \J_{j,j+1}
\end{eqnarray}
corresponding to the Boltzmann factors for
vertical, horizontal and diagonal edges, respectively.\footnote{
   In this sentence we are of course using the term ``vertical edges''
   in its ordinary geometrical sense, not in the technical sense
   defined previously.
   We follow the notation of Baxter \cite[Section 2]{Baxter_87}
   with some modifications.
}
%
%
The horizontal and vertical parts of the transfer matrix are then
\begin{subeqnarray}
   \H^{\rm free}_{\rm SQ}  \;=\; \H^{\rm free}_{\rm TRI}
           & = &        \QQ_{m-1} \cdots \QQ_2 \QQ_1       \\[1mm]
   \H^{\rm per}_{\rm SQ}   & = &
                        \QQ_m \QQ_{m-1} \cdots \QQ_2 \QQ_1
 \label{def_horiz}
\end{subeqnarray}
and
\begin{subeqnarray}
  \V^{\rm free}_{\rm SQ}  \;=\;  \V^{\rm per}_{\rm SQ}
           & = &        \P_m \P_{m-1} \cdots \P_2 \P_1
  \slabel{def_V_SQ}       \\[1mm]
  \V^{\rm free}_{\rm TRI}   & = &
           \P_m \RR_{m-1} \P_{m-1} \cdots \RR_2 \P_2 \RR_1 \P_1
  \slabel{def_V_TRI_free}
 \label{def_vert}
\end{subeqnarray}
where the action of these matrices should always be read from right to left.
Note that $\RR_j$ corresponds to a diagonal edge whenever it is applied
after $\P_j$ and before $\P_{j+1}$.

For the {\em triangular}\/ lattice with {\em periodic}\/ transverse
boundary conditions, a slight trick is needed in order to treat correctly
the last diagonal bond joining columns $m$ and 1 \cite{Baxter_87}.
The idea is to work with $q^{m+1} \times q^{m+1}$ matrices indexed by
spins $\bsigma = (\sigma_1,\ldots,\sigma_{m+1})$, and include in the
horizontal matrix an operator that identifies $\sigma_{m+1}$ with
$\sigma_1$ on each row.  Thus,
\begin{eqnarray}
   \H^{\rm per}_{\rm TRI}   & = &
      \J_{m+1,1} \QQ_m \QQ_{m-1} \cdots \QQ_2 \QQ_1
   \label{def_H_TRI_per}   \\[1mm]
   \V^{\rm per}_{\rm TRI}   & = &
         \P_{m+1} \RR_m \P_m \RR_{m-1} \P_{m-1} \cdots \RR_2 \P_2 \RR_1 \P_1
   \label{def_V_TRI_per}
\end{eqnarray}
In particular, the vertical matrix $\V^{\rm per}_{\rm TRI}$
is just $\V^{\rm free}_{\rm TRI}$ with $m$ replaced by $m+1$.

\bigskip

{\bf Remarks.}
1.  The rewriting \reff{def_horiz} of the horizontal matrix $\H$
is of course unnecessary, as $\H$ was already diagonal (hence sparse)
in the spin basis.  But \reff{def_horiz} will be useful when we use
the partition basis (see below).

2. The order of operators in \reff{def_horiz},
\reff{def_V_SQ} and \reff{def_H_TRI_per} is irrelevant,
as all the operators in question commute.
Only in \reff{def_V_TRI_free} and \reff{def_V_TRI_per}
is the operator ordering crucial.

3.  The operators $e_1,\ldots,e_{2m-1}$ defined by
\begin{subeqnarray}
   e_{2j-1}  & = &   q^{-1/2} \D_j  \qquad \hbox{for $1 \le j \le m$} \\[2mm]
   e_{2j}    & = &   q^{1/2} \J_{j,j+1}  \quad \hbox{for $1 \le j \le m-1$}
\end{subeqnarray}
satisfy the Temperley-Lieb algebra \cite{Temperley_71,Baxter_82,Kauffmann_93}
\begin{subeqnarray}
   e_i^2                  & = &  q^{1/2} e_i   \\[1mm]
   e_i e_{i \pm 1} e_i    & = &  e_i           \\[1mm]
   e_i e_j                & = &  e_j e_i  \qquad\hbox{for } |i-j| > 1
\end{subeqnarray}

\bigskip

All these operators act, of course, in the $q^m$-dimensional vector space
consisting of all functions of $\sigma_1,\ldots,\sigma_m$.\footnote{
   Or, in the case of the triangular lattice with periodic transverse b.c.,
   in the $q^{m+1}$-dimensional space of all functions of
   $\sigma_1,\ldots,\sigma_{m+1}$.
}
But for the case of {\em free}\/ longitudinal boundary conditions
\reff{Z_free}, we need only consider the subspace spanned by those vectors
that can be obtained from the constant vector $\bone$ by action of the
operators $\H$ and $\V$.
All such functions are sums of products of
delta functions $\delta(\sigma_x,\sigma_y)$, and we can take as a basis
the functions $\vv_{\scrp}$ associated to a partition
$\scrp = \{ P_1, \ldots, P_k \}$ of $\{1,\ldots,m\}$ by
\be
  \vv_{\scrp}(\sigma_1,\ldots,\sigma_m)   \;=\;
  \prod\limits_{P \in \scrp} \prod\limits_{x,y \in P} \delta(\sigma_x,\sigma_y)
  \;.
 \label{def_vP_spin}
\ee
For example, for $m=3$ we have the five basis vectors
\begin{subeqnarray}
   \vv_{ \{\, \{1\}, \{2\}, \{3\} \,\} }   & = &   1  \\[1mm]
   \vv_{ \{\, \{1,2\}, \{3\} \,\} }        & = &
                                \delta(\sigma_1,\sigma_2)  \\[1mm]
   \vv_{ \{\, \{1,3\}, \{2\} \,\} }        & = &
                                \delta(\sigma_1,\sigma_3)  \\[1mm]
   \vv_{ \{\, \{1\}, \{2,3\} \,\} }        & = &
                                \delta(\sigma_2,\sigma_3)  \\[1mm]
   \vv_{ \{\, \{1,2,3\} \,\} }        & = &
                                \delta(\sigma_1,\sigma_2,\sigma_3)
\end{subeqnarray}
Indeed, when the graph $G$ is {\em planar}\/,
it is not hard to see on topological grounds
that only {\em non-crossing}\/ partitions can arise.
(A partition is said to be {\em non-crossing}\/ if
 $a < b < c < d$ with $a,c$ in the same block and $b,d$ in the same block
 imply that $a,b,c,d$ are all in the same block.)
Moreover, spatial symmetries can further restrict the subspace.
Finally, when the horizontal couplings $v_{\<xx'\>}$ are all equal to $-1$
--- which is the case for the chromatic polynomial ---
then the horizontal operator $\H$ is a {\em projection}\/,
and we can work in its image subspace by using the modified transfer matrix
$\T' = \H \V \H$ in place of $\T = \V \H$,
and using the basis vectors
\be
   \w_{\scrp}   \;=\;   \H \vv_{\scrp}
 \label{def_wP}
\ee
in place of $\vv_{\scrp}$.
Note that $\w_{\scrp} = 0$ if $\scrp$ has any pair of nearest neighbors
in the same block.
In Section~\ref{sec3.3} we shall compute the dimensions of all these subspaces.

The action of the operators $\D_x$ and $\J_{xx'}$
on the basis vectors $\vv_{\scrp}$ is quite simple.
As one might expect, $\J_{xx'}$ joins sites $x$ and $x'$, i.e.\
\be
   \J_{xx'} \vv_{\scrp} \;=\; \vv_{\scrp \bullet xx'}
\ee
where $\scrp \bullet xx'$ is the partition obtained from $\scrp$
by amalgamating the blocks containing $x$ and $x'$
(if they were not already in the same block).
$\D_x$ detaches site $x$ from the block it currently belongs to,
multiplying by a factor $q$ if $x$ is currently a singleton:
\be
   \D_x \vv_{\scrp} \;=\;
        \cases{ \vv_{\scrp \setminus x}   & if $\{x\} \notin \scrp$ \cr
                \noalign{\vskip 2mm}
                q \vv_{\scrp}             & if $\{x\} \in \scrp$    \cr
              }
 \label{D_vP}
\ee
where $\scrp \setminus x$ is the partition obtained from $\scrp$
by detaching $x$ from its block (and thus making it a singleton).

At the final stage we need to compute the inner products
$\bone^{\rm T} \vv_{\scrp}$, which are easy:
\be
   \bone^{\rm T} \vv_{\scrp}   \;=\;   q^{|\scrp|}
 \label{leftvector_vP}
\ee
where $|\scrp|$ is the number of blocks in $\scrp$.

\bigskip

To summarize:
The Potts-model partition function \reff{Z_free} or \reff{Z_per}
can always be computed by the transfer-matrix method working in the
spin basis $f(\sigma_1,\ldots,\sigma_m)$, which has dimension $q^m$.
Of course, each value of $q$ has to be treated separately;
and it goes without saying that $q$ must be a positive integer.
On the other hand, for {\em free}\/ longitudinal boundary conditions,
there is an alternative method for computing the partition function
\reff{Z_free}, using the partition basis $\vv_{\scrp}$.
In this formalism $q$ appears only as a parameter
[in \reff{D_vP} and \reff{leftvector_vP}];
and since the final result will obviously be a polynomial in $q$,
it must perforce be equal to the Fortuin-Kasteleyn partition function
\reff{FK_rep}.
This latter formalism is thus essentially equivalent to
constructing the transfer matrix directly in the Fortuin-Kasteleyn
representation, as we shall see explicitly in the next subsection.

\subsection{Transfer matrix in the Fortuin-Kasteleyn representation}
      \label{sec3.2}

Let us now temporarily forget the Potts spin model
--- and in particular forget the transfer-matrix formalism
constructed in the preceding subsection ---
and try instead to devise a transfer-matrix method for computing
the Fortuin-Kasteleyn partition function \reff{FK_rep}
when the graph $G$ has a layered structure \reff{eq3.1}--\reff{eq3.6}.
What makes this a bit tricky is, of course, the nonlocality of the
factor $q^{k(E')}$ in \reff{FK_rep}.
The technique for handling this nonlocality was foreshadowed
in the work of Lieb and Beyer \cite{Lieb_69} on percolation
and was made explicit (for the case of the chromatic polynomial)
in the work of Biggs and collaborators \cite{Biggs_72,Sands_72,Biggs_76}.
In the physics literature, this approach was first used
(to our knowledge) by Derrida and Vannimenus \cite{Derrida_80}
in their study of percolation,
and was applied to the $q$-state Potts model
(and explained very clearly) by Bl\"ote and Nightingale \cite{Blote_82};
it was subsequently employed by several groups
\cite{Blote_84,Blote_89,Jacobsen_98,Dotsenko_99}.
Here we limit ourselves to giving a brief summary.

The basic idea is to build up the subgraph $E' \subseteq E$ layer by layer.
At the end we will need to know the number of connected components
in this subgraph;  in order to be able to compute this,
we shall keep track, as we go along, of which sites in the
current ``top'' layer are connected to which other sites in that layer
by a path of occupied bonds (i.e.\ bonds of $E'$) in lower layers.
Thus, we shall work in the basis of connectivities of the top layer,
whose basis elements $\vv_{\scrp}$ are indexed by
partitions $\scrp$ of the single-layer vertex set $V^0$.
[The reader is reminded to forget \reff{def_vP_spin} for the time being.
 The $\vv_{\scrp}$ should here be thought of simply as abstract basis elements
 indexed by partitions $\scrp$.]
The elementary operators we shall need are the {\em join operators}\/
\be
   \J_{xx'} \vv_{\scrp} \;=\; \vv_{\scrp \bullet xx'}
\ee
(note that all these operators commute)
and the {\em detach operators}\/
\be
   \D_x \vv_{\scrp} \;=\;
        \cases{ \vv_{\scrp \setminus x}   & if $\{x\} \notin \scrp$ \cr
                \noalign{\vskip 2mm}
                q \vv_{\scrp}             & if $\{x\} \in \scrp$    \cr
              }
\ee
where $\scrp \bullet xx'$ and $\scrp \setminus x$ were defined previously
(don't forget {\em those}\/ definitions!).
The horizontal transfer matrix is then
\be
   \H   \;=\;
     \prod_{\<xx'\> \in E^0} [1 + v_{\<xx'\>} \J_{xx'}]
   \;.
\ee
The vertical transfer matrix is slightly more complicated:
\be
   \V  \vv_{\scrp}  \;=\;
   \sum_{\widetilde{E} \subseteq E^*}  q^{A(\scrp,\widetilde{E})}
       \left( \prod_{(x,x') \in \widetilde{E}}  v_{(x,x')} \right)
       \vv_{\scrp | \widetilde{E}}
\ee
where $A(\scrp,\widetilde{E})$ is the number of ``abandoned clusters'',
i.e.\ the number of blocks $P \in \scrp$ such that no vertex in $P$
is an endpoint of an edge in $\widetilde{E}$;
and $\scrp | \widetilde{E}$ is the partition of $V^0$
in which vertices $x',y'$ lie in the same block
if and only if there exist vertices $x,y$ in the same block of $\scrp$
such that both $(x,x')$ and $(y,y')$ lie in $\widetilde{E}$.

In many cases (including all planar graphs),
$\V$ can be written as a product of sparse matrices
that correspond to the replacement of {\em one}\/ site
on layer $i$ by the corresponding site on layer $i+1$;
and these sparse matrices have a simple expression
in terms of join and detach operators.
Suppose, for concreteness, that the single-layer vertex set $V^0$
can be ordered in such a way that $(x,x') \in E^*$ implies $x \succeq x'$;
using this ordering, let us number the sites of $V^0$ as $1,\ldots,m$.
Then
\be
   \V   \;=\;
   \P_m \RR_{m-1,m} \P_{m-1}
   \left( \prod\limits_{y > m-2} \RR_{m-2,y} \right)  \P_{m-2}
   \,\cdots\,
   \left( \prod\limits_{y > 2} \RR_{2,y} \right)  \P_2
   \left( \prod\limits_{y > 1} \RR_{1,y} \right)  \P_1
   \;,
\ee
where
\begin{eqnarray}
   \P_x       & = & v_{(x,x)} I  \,+\,  \D_x   \\[1mm]
   \RR_{x,y}  & = & I  \,+\,  v_{(y,x)} \J_{xy}
\end{eqnarray}
(The triangular lattice with periodic transverse b.c.\ is handled
 using the trick discussed in the preceding subsection.)

Finally, the partition function for {\em free}\/ longitudinal boundary
conditions can be obtained by sandwiching the transfer matrix between
suitable vectors on right and left:
\be
   Z_{G_n^{\rm free}}(q, \{v_e\})   \;=\;
     {\bf u}^{\rm T} \H (\V \H)^{n-1} \vv_{\rm id}
   \;,
   \label{Z_free_FK} 
\ee
where ``id'' denotes the partition in which each site $x \in V^0$
is a singleton, and ${\bf u}^{\rm T}$ is defined by
\be
   {\bf u}^{\rm T} \vv_{\scrp}   \;=\;  q^{|\scrp|}
   \;.
 \label{def_uT}
\ee

Of course, it will not have escaped the reader's notice
that all the formulae in this subsection are {\em identical}\/
to those developed in the preceding subsection;
only the interpretation is different.
In particular, the operators $\D_x$ and $\J_{xx'}$ in the FK representation
act in precisely the same way as the operators of the same name
act in the spin representation {\em with respect to the partition basis}\/.

Let us observe, in conclusion, that
the Potts model with {\em free}\/ longitudinal boundary conditions
is much easier to handle than that with {\em periodic}\/ longitudinal
boundary conditions.  We can understand the difficulty posed by
periodic longitudinal b.c.\ from two complementary points of view.
In the spin representation, the trouble is that, because of the trace
in \reff{Z_per}, we must compute the action of $\T$ in the entire
$q^m$-dimensional space of functions of $\sigma_1,\ldots,\sigma_m$,
not merely in the subspace spanned by sums of products of delta functions.
In some cases this can be done by careful counting of subspaces and dimensions
\cite{Biggs_99a,Biggs_99b}, 
but it is not entirely trivial.
In the FK representation, by contrast, everything is fully automated,
but it is not sufficient to keep track of the connectivities of the sites
in the top layer alone, as this layer will eventually need to be joined up
to the bottom layer.  Rather, it is necessary to keep track of the
combined connectivities of the sites in the {\em top and bottom}\/ layers:
this method for handling periodic longitudinal b.c.\ will be sketched
in Section~\ref{sec7.4}
and explained in detail in a subsequent paper \cite{transfer4}.

\subsection{Dimension of the transfer matrix}   \label{sec3.3}

Let us consider a strip of width $m$ with free longitudinal
boundary conditions.
We want to know the dimension of the corresponding
FK-representation transfer matrix
for different lattices (square and triangular)
and different transverse boundary conditions (free and periodic).
Each of these cases corresponds to a different class of allowed partitions
$\scrp$.
The more stringent the restrictions on the allowed partitions,
the smaller the dimension of the transfer matrix.

We shall in general follow the notation of Stanley's 2-volume work
{\em Enumerative Combinatorics}\/ \cite{Stanley_86,Stanley_99},
where more detail on various integer sequences can be found.
Another useful reference is the On-Line Encyclopedia of Integer Sequences
\cite{Sloane_on-line}.

\bigskip

1)  Let ${\bf P}_{m}$ denote the set of all partitions of $\{ 1,\ldots,m \}$.
Its cardinality is given by the {\em Bell numbers}\/
(or exponential numbers) $B(m)$,
which can be computed \cite[Sections 6.1 and 6.3]{deBruijn_61}
from the exponential generating function
\be
   E_B(x)  \;\equiv\; \sum\limits_{m=0}^\infty B(m) \, {x^m \over m!}
           \;=\;  \exp(e^x - 1)
\ee
or from the remarkable formula
\be
   B(m)   \;=\;  e^{-1} \sum\limits_{k=0}^\infty {k^m \over k!}
\ee
(it is far from obvious at first sight that the right-hand side
 defines an integer!).
For example, for $m=4$ we have $B(4) = 15$,
corresponding to the partitions\footnote{
   Henceforth we shall abbreviate delta functions
   $\delta(\sigma_1,\sigma_3)$ by $\delta_{13}$, etc.
   Moreover, we shall usually abbreviate partitions $\scrp$
   by writing instead the corresponding product $\vv_{\scrp}$
   of delta functions [cf.\ \reff{def_vP_spin}]:
   e.g.\ in place of $\scrp = \{ \, \{1,3\}, \{2,4\}, \{5\} \, \}$
   we shall write simply $\scrp = \delta_{13} \delta_{24}$.
}
\begin{eqnarray}
{\bf P}_{4} &=& \{ 1, \delta_{12}, \delta_{13}, \delta_{14},
                 \delta_{23}, \delta_{24}, \delta_{34},
                 \delta_{123}, \delta_{124}, \delta_{134},
                 \delta_{234}, \delta_{1234}, \nonumber \\
                   & & \qquad
                 \delta_{12}\delta_{34}, \delta_{13}\delta_{24},
                 \delta_{14}\delta_{23} \}
\;.
\end{eqnarray}
The Bell numbers have the asymptotic behavior
\cite[Sections 6.1--6.3]{deBruijn_61}
\cite[Section 5.8]{Sachkov_96}
\cite[Examples 5.4, 5.10, 12.5 and 12.6]{Odlyzko_95}
\begin{subeqnarray}
   \log B(m)   & = &  m W(m+1) \,-\, m \,+\, {m+1 \over W(m+1)}
       \,-\, \smhalf \log [W(m+1) +1] \,-\, 2  \,+\,
       O\!\left( {\log m \over m} \right)
   \nonumber \\ \\[2mm]
      & = &  m W(m) \,-\, m \,+\, {m \over W(m)}
       \,-\, \smhalf \log [W(m) +1] \,-\, 1  \,+\,  o(1)
   \\[2mm]
               & = &  m \log m \,-\, m \log\log m \,-\, m \,+\,
                       {m \log\log m \over \log m} \,+\,
                       {m \over \log m} \,+\,
                       O\!\left( {m (\log\log m)^2 \over (\log m)^2}\right)
     \nonumber \\
  \label{bell_asymp}
\end{subeqnarray}
as $m \to\infty$,
where $W(z)$ denotes the unique real solution of $w e^w = z$
(this is the Lambert $W$ function \cite{Corless_96})
and has the (very slowly) convergent expansion in powers
of $1/\log z$ and $\log\log z / \log z$:
\be
   W(z)  \;=\;  \log z \,-\, \log\log z \,+\,
                \sum_{n=1}^\infty \sum_{k=1}^n
       (-1)^{n+1} \, {s(n, n-k+1) \over k!} \,
       {(\log\log z)^k \over (\log z)^n}
%
%
\ee
where the $s(n,\ell)$ are the Stirling numbers of the first kind.
[The expansions (\ref{bell_asymp}a,b) thus yield much better numerical
 approximations than (\ref{bell_asymp}c).]
The Bell numbers for $1 \le m \le 14$ are shown in
Table~\ref{table_dimensions}.

The Bell numbers can also be written as
\be
   B(m)  \;=\;  \sum_{k=1}^m S(m,k)   \;,
\ee
where the {\em Stirling number of the second kind}\/ $S(m,k)$
is the number of ways of partitioning the set $\{ 1,\ldots,m \}$
into $k$ nonempty subsets
(i.e., the number of ways of placing $m$ labeled balls into
   $k$ indistinguishable boxes, with each box containing at least one ball).

\bigskip

2) If the graph $G$ is planar, then only non-crossing partitions can occur.
The number of non-crossing partitions of $\{ 1,\ldots,m \}$
is given by the {\em Catalan number}\/\footnote{
   Stanley \cite[Exercises 6.19 and 6.25]{Stanley_99}
   gives 66 combinatorial interpretations and 9 algebraic interpretations
   of the Catalan numbers $C_n$.
   See also \cite[Exercise 6.24]{Stanley_99}.
}
\be
   C_m  \;=\; {(2m)! \over m! \, (m+1)!}  \;=\; {1 \over m+1} \, {2m \choose m}
   \;.
\ee
Thus, for $m=4$ we have $C_m = 14$, corresponding to the partitions
\begin{eqnarray}
{\bf P}_{\rm nc,4} &=& \{ 1, \delta_{12}, \delta_{13}, \delta_{14},
                 \delta_{23}, \delta_{24}, \delta_{34},
                 \delta_{123}, \delta_{124}, \delta_{134},
                 \delta_{234}, \delta_{1234}, \nonumber \\
           & & \quad
                 \delta_{12}\delta_{34},
                 \delta_{14}\delta_{23} \}
               \;,
\end{eqnarray}
since the only crossing partition of $\{1,2,3,4\}$ is $\delta_{13}\delta_{24}$.
The Catalan numbers have the asymptotic behavior
\be
   C_m   \;=\;  4^m m^{-3/2} \pi^{-1/2} [1 + O(1/m)]
\ee
as $m \to\infty$.
For $1 \le m \le 14$ they are shown in Table~\ref{table_dimensions}.

\bigskip

3)  For the zero-temperature antiferromagnetic model ($v=-1$),
we can work in the basis $\w_{\scrp}$ defined by \reff{def_wP}, so that
partitions with nearest neighbors in the same block are also forbidden.
Let us consider first the case of free transverse boundary conditions.
The number of non-crossing non-nearest-neighbor partitions
of $\{ 1,\ldots,m \}$ is given by the {\em Motzkin number}\/ $M_{m-1}$
\cite[Proposition 3.6]{Simion_91} \cite{Klazar_98}.\footnote{
   {\em Warning:}\/  Several references use the notation $m_n$
   to denote what we call $M_n$;
   and one reference \cite{Donaghey_77} writes $M_n$
   to denote a {\em different}\/ sequence.
}
Here $M_n$ is the number of ways of drawing any number of
nonintersecting chords among $n$ points on a circle\footnote{
   This interpretation of $M_n$ arises in enumerating the
   ``non-magnetic connectivities'' of a loop model on the square lattice
   \cite[pp.~1436--1437]{Blote_89}.
   See Stanley \cite[Exercises 6.38 and 6.46(b)]{Stanley_99}
   for 14 combinatorial interpretations of the Motzkin numbers $M_n$.
};
it is given by the explicit formula
\be
   M_n  \;=\;  \sum\limits_{j=0}^{\lfloor n/2 \rfloor}
                  {n \choose 2j} \, C_j
\ee
and satisfies both the linear recursion relation
\be
   M_n  \;=\;  {2n+1 \over n+2} M_{n-1}  \,+\, {3n-3 \over n+2} M_{n-2}
               \,+\, \delta_{n0}
\ee
and the nonlinear recursion relation
\be
   M_n  \;=\;  M_{n-1} \,+\, \sum_{k=0}^{n-2} M_k M_{n-2-k} \,+\, \delta_{n0}
\ee
with initial condition $M_n = 0$ for $n < 0$
(see e.g.\ \cite{Riordan_75,Donaghey_77,Gouyou_88,Bernhart_99}).
The generating function $M(x) = \sum_{n=0}^\infty M_n x^n$ is
\be
   M(x)  \;=\;  {1 - x - \sqrt{1-2x-3x^2}  \over  2x^2}
   \;.
\ee
For example, for $m=4$ we have $M_3 = 4$, corresponding to the partitions
\be
{\bf P}_{\rm nc+nnn,4}  \;=\;
    \{ 1, \delta_{13}, \delta_{14}, \delta_{24} \}
\;.
\ee
The Motzkin numbers have the asymptotic behavior
\be
   M_n   \;=\;  3^n n^{-3/2} {3 \sqrt{3} \over 2 \sqrt{\pi}} [1 + O(1/n)]
\ee
as $n \to\infty$.
The transfer matrix for a triangular-lattice strip of width $m$
with free boundary conditions at zero temperature has dimension
\be
\mbox{\rm TriFree}(m)  \;=\; M_{m-1}   \;.
\ee
The numbers $\mbox{\rm TriFree}(m) = M_{m-1}$ for $1 \le m \le 14$
are shown in Table~\ref{table_dimensions}.

\bigskip

4)  Let us now consider the zero-temperature antiferromagnetic model
with periodic transverse boundary conditions.
The number of non-crossing non-nearest-neighbor partitions
of $\{ 1,\ldots,m \}$ when it is considered periodically
(i.e.\ when 1 and $m$ also are considered to be nearest neighbors)
is given by \cite[Section 3.2, family R2]{Bernhart_99}\footnote{
   Let $d_m$ be the number of non-crossing non-nearest-neighbor partitions
   of $\{ 1,\ldots,m \}$ when it is considered periodically.
   We have $d_1 = d_2 = 1$ and $M_{m-1} = d_{m-1} + d_m$ for $m \ge 3$.
   [{\sc Proof:}  For $m \ge 3$, a non-crossing non-nearest-neighbor partition
    of $\{ 1,\ldots,m \}$ (considered linearly) either has $1$ and $m$
    in the same block or it doesn't.  There are $d_m$ partitions of the
    latter type.  And in any partition of the former type, we can consider
    $1$ and $m$ to be amalgamated into a single site $1'$ that is a
    neighbor of both $2$ and $m-1$;  so there are $d_{m-1}$ partitions
    of this type.]
   The formula \reff{Motzkin_sum} for $d_m$ then follows by induction.
}
\be
d_m   \;=\; \cases{ 1     & for $m=1$ \cr
                    R_m   & for $m \ge 2$ \cr
                  }
\ee
where the {\em Riordan numbers}\/ (or Motzkin alternating sums) $R_m$
\cite{Riordan_75,Donaghey_77,Bernhart_99}\footnote{
   In most of the literature (e.g.\ \cite{Riordan_75,Donaghey_77})
   these numbers are called $\gamma_m$.
   We have adopted the recent proposal of Bernhart \cite{Bernhart_99}
   to name them after Riordan \cite{Riordan_75} and denote them $R_m$.
}
are defined by $R_0 = 1$, $R_1 = 0$ and
\be
   R_m  \;=\;  \sum_{k=0}^{m-1}  (-1)^{m-k-1} M_k
    \qquad \hbox{for $m \ge 2$}
  \label{Motzkin_sum}
\ee
and satisfy the linear recursion relations
\begin{eqnarray}
   R_m  & = & -R_{m-1} \,+\, M_{m-1} \,+\, \delta_{m0}   \\[2mm]
%
%
   R_m  & = &
   {2m-2 \over m+1} R_{m-1}  \,+\, {3m-3 \over m+1} R_{m-2}
   \,+\, \delta_{m0}
\end{eqnarray}
and the nonlinear recursion relation
\be
   R_m  \;=\;
   \sum_{k=0}^{m-1} R_k R_{m-1-k}  \,+\, (-1)^m
\ee
with initial condition $R_m = 0$ for $m < 0$
(see e.g.\ \cite{Riordan_75,Donaghey_77,Bernhart_99}).
The generating function $R(x) = \sum_{m=0}^\infty R_m x^m$ is
\be
   R(x)  \;=\;  {1 + x - \sqrt{1-2x-3x^2}  \over  2x(1+x)}
   \;.
\ee
For example, for $m=4$ we have $d_4 = R_4 = 3$,
corresponding to the partitions
\be
{\bf P}_{\rm nc+nnnCyl,4}  \;=\;  \{ 1, \delta_{13}, \delta_{24} \}
\ee
The Riordan numbers have the asymptotic behavior
\be
   d_m \;=\; R_m   \;=\;  3^m m^{-3/2}   {3 \sqrt{3} \over 8 \sqrt{\pi}}
                               [1 + O(1/m)]
\ee
as $m \to\infty$.
The numbers $d_m$ for $1 \le m \le 14$
are shown in Table~\ref{table_dimensions}.

\bigskip

5)  Further restrictions arise from spatial symmetries.
For instance, if we consider the square-lattice strip with
free transverse boundary conditions, then reflection with respect to
the center of the strip is a symmetry of the system.
We can therefore define equivalence classes
of non-crossing non-nearest-neighbor partitions modulo reflection.\footnote{
   One could also drop the non-nearest-neighbor condition;
   this would be relevant at nonzero temperature.
}
The dimension $\mbox{\rm SqFree}(m)$ of the transfer matrix
is then given by the number of these equivalence classes.
Since each equivalence class contains either one or two partitions
(and some contain only one), we clearly have
\be
   {\mbox{\rm TriFree}(m) \over 2}  \;<\;  \mbox{\rm SqFree}(m)
      \;\le\;  \mbox{\rm TriFree}(m)
   \;.
\ee
For $m=4$, for example, we have three equivalence classes:
\be
   {\bf P}_{\rm SqFree,4}  \;=\;
       \{ 1, \delta_{13} + \delta_{24}, \delta_{14}\}
   \;.
\ee
We have been unable to compute an explicit formula for $\mbox{\rm SqFree}(m)$,
or to find any known integer sequence that corresponds
to $\mbox{\rm SqFree}(m)$.
The asymptotic behavior is clearly the same as that of
$\mbox{\rm TriFree}(m)$ within a factor 2,
hence of order $3^m m^{-3/2}$.
(We conjecture that $\mbox{\rm SqFree}(m) \approx \mbox{\rm TriFree}(m)/2$,
 since ``most'' partitions are asymmetric.
 The data in Table~\ref{table_dimensions} strongly suggest that
 this conjecture is true:
 the ratios $\mbox{\rm SqFree}(m)/\mbox{\rm TriFree}(m)$
 are roughly decreasing in $m$, albeit with some even-odd oscillation,
 and seem clearly to be approaching $1/2$.
 Indeed, at $m=14$ the ratio has already reached $\approx 0.5032$.)

\bigskip

6)  For the square lattice with periodic transverse boundary conditions,
both reflections and translations are symmetries.
We therefore define equivalence classes of non-crossing non-nearest-neighbor
partitions modulo reflections and translations
and the corresponding number $\mbox{\rm SqCyl}(m)$
of equivalence classes.\footnote{
   One could also drop the non-nearest-neighbor condition;
   this would be relevant at nonzero temperature.
}
Since each equivalence class contains at most $2m$ partitions
(and some contain less), we clearly have
\be
   {d_m \over 2m}  \;<\;  \mbox{\rm SqCyl}(m)  \;\le\;  d_m
   \;.
\ee
For $m=4$, there are only two such classes:
\be
   {\bf P}_{\rm SqCyl,4}  \;=\;
       \{ 1, \delta_{13} + \delta_{24} \}
   \;.
\ee
We have been unable to compute an explicit formula for $\mbox{\rm SqCyl}(m)$,
or to find any known integer sequence that corresponds
to $\mbox{\rm SqCyl}(m)$.
The asymptotic behavior is clearly the same as that of $d_m$
within a factor $2m$, hence the leading behavior is $\sim 3^m$.
(We conjecture that $\mbox{\rm SqCyl}(m) \approx d_m/(2m)$,
 since ``most'' partitions are asymmetric.
 The data in Table~\ref{table_dimensions} strongly suggest that
 this conjecture is true:
 the ratios $2m \mbox{\rm SqCyl}(m) / d_m$ are roughly decreasing in $m$,
 albeit with some even-odd oscillation, and seem to be approaching
 a value near 1.  At $m=14$ the ratio has already reached $\approx 1.14$.)

\bigskip

7)  For the triangular lattice, reflection is not a symmetry;
but if we have periodic transverse boundary conditions,
then translations are symmetries.
We therefore define equivalence classes of non-crossing non-nearest-neighbor
partitions modulo translations,
and the corresponding number $\mbox{\rm TriCyl}(m)$
of equivalence classes.\footnote{
   One could also drop the non-nearest-neighbor condition;
   this would be relevant at nonzero temperature.
}
Since each equivalence class contains at most $m$ partitions
(and some contain less), we clearly have
\be
   {d_m \over m}  \;<\;  \mbox{\rm TriCyl}(m)  \;\le\;  d_m
   \;.
\ee
For $m=4$, there are again only two such classes:
\be
   {\bf P}_{\rm TriCyl,4}  \;=\;
       \{ 1, \delta_{13} + \delta_{24} \}
   \;.
\ee
Indeed, we have found that $\mbox{\rm TriCyl}(m) = \mbox{\rm SqCyl}(m)$
for $m \le 7$;  only for $m \ge 8$ does reflection symmetry impose
additional constraints beyond those imposed by translation symmetry,
so that $\mbox{\rm TriCyl}(m) > \mbox{\rm SqCyl}(m)$.
We have been unable to compute an explicit formula for $\mbox{\rm TriCyl}(m)$,
or to find any known integer sequence that corresponds
to $\mbox{\rm TriCyl}(m)$.
The asymptotic behavior is clearly the same as that of $d_m$
within a factor $m$, hence the leading behavior is $\sim 3^m$.
(We conjecture that $\mbox{\rm TriCyl}(m) \approx d_m/m$,
 since ``most'' partitions are asymmetric.
 The data in Table~\ref{table_dimensions} strongly suggest that
 this conjecture is true:
 the ratios $m \mbox{\rm TriCyl}(m) / d_m$ are roughly decreasing in $m$,
 albeit with strong even-odd oscillation, and seem clearly to be
 approaching 1.  At $m=14$ the ratio has already reached $\approx 1.01$.)

\bigskip

Although the dimension of the transfer matrix is given by
$\mbox{\rm SqCyl}(m)$, $\mbox{\rm SqFree}(m)$,
$\mbox{\rm TriCyl}(m)$ or $\mbox{\rm TriFree}(m)$,
the basis with respect to which the vectors are expressed
is considerably larger,
namely the set ${\bf P}_{{\rm nc},m}$ of all non-crossing partitions
(or ${\bf P}_{m}$ in case $G$ is non-planar).
Thus, the vectors produced at intermediate stages of the computation
can be rather long;  this is the main limiting factor in our numerical work.
In order to save memory, we use the following tricks:

\begin{itemize}
\item
Label the partitions by an integer variable $k = 1,\ldots, B(m)$.
(For large widths $m$, however, it is vastly more efficient to include
only the smaller set of the {\em non-crossing}\/ partitions $k=1,\ldots, C_m$
\cite{Blote_82,Jacobsen_98};  this approach will be taken in
future work in collaboration with Jesper-Lykke Jacobsen
\cite{transfer2,transfer3}.)

\item Represent the vectors in sparse-vector format (i.e.\ representing
explicitly only nonzero coefficients).  This is automatic in {\sc Mathematica}.
(If, however, we consider only the set of non-crossing partitions,
then typical intermediate vectors are dense, so that sparse-vector format
is actually a disadvantage.)

\item
The coefficients of the chromatic polynomial are extremely large
integers, which grow rapidly with the width and length of the strip.
We can reduce the magnitude of these coefficients (and therefore save memory)
by performing a change of variables $u = q-q_0$, where $q_0$ lies at or near
the barycenter of the roots.  This barycenter lies at $|E|/|V|$,
where $|E|$ (resp.\ $|V|$) is the number of edges (resp.\ vertices)
in the graph $G$;  that is, it lies at half the average coordination number.
It is therefore convenient to take $q_0=2$ (resp.\ $q_0=3$)
for the square (resp.\ triangular) lattice;
this vastly reduces the size of the coefficients of the chromatic polynomial,
as was noted already by Baxter \cite{Baxter_87}.
\end{itemize}

\bigskip
\noindent
{\bf Remark.} As explained in Section~\ref{sec3.1},
when computing the transfer matrix for a triangular-lattice
strip of width $m$ and cylindrical boundary conditions, we use the
following technical trick \cite{Baxter_87}:
To take account of the diagonal bond joining the sites $m$ and 1,
it is convenient to consider a triangular strip of width $m+1$
and, at the end of the computation, identify the spins at sites $m+1$ and 1.
This means that the number of partitions arising in
intermediate steps of the computation is not $C_m$, but $C_{m+1}$.
However, the dimension of the transfer matrix is still $\mbox{\rm TriCyl}(m)$.

\section{Eigenvalue Crossing and Isolated Limiting Points}   \label{sec4}

\subsection{Computation of eigenvalue-crossing curves}   \label{sec4.1}

The limiting curve $\scrb$ is the locus of points in the $q$-plane
where there are two or more dominant eigenvalues.
Our approach is to compute first the locus of $q$ values where
there are two or more equimodular eigenvalues, {\em dominant or not}\/;
we then check the corresponding eigenvalues one-by-one for dominance.
We have used two methods to compute the locus of equimodularity:
the resultant method, and the direct-search method.

\subsubsection{The resultant method}  \label{sec4.1.1}

The resultant of two polynomials
$P(x) = A \prod_{i=1}^M (x - x_i)$ and
$Q(y) = B \prod_{i=1}^N (y - y_i)$ is defined to be
the product of all the differences of roots,
scaled suitably by the two leading coefficients:
\be
   {\rm Res}(P,Q)  \;=\;
      A^N B^M \prod\limits_{i=1}^M \prod\limits_{j=1}^N  (x_i-y_j)
   \;.
\ee
Thus, the resultant vanishes if and only if $P$ and $Q$
have at least one root in common
(or one or both of the leading coefficients vanishes).
It is a nontrivial fact that the resultant ${\rm Res}(P,Q)$
can be expressed as an $(M+N) \times (M+N)$ determinant
involving the coefficients of $P$ and $Q$;
it is {\em not}\/ necessary to know explicitly the roots
$\{x_i\}$ and $\{y_j\}$.

Likewise, the discriminant of a polynomial $P(x) = A \prod_{i=1}^M (x - x_i)$
is defined as
\be
   {\rm Disc}(P)   \;=\; A^{2M-2} \prod_{i < j} (x_i - x_j)^2
     \;=\; (-1)^{M(M-1)/2} A^{2M-2} \prod_{i \neq j} (x_i - x_j)
   \;.
 \label{discriminant}
\ee
(For a quadratic $P(x) = a x^2 + b x + c$, we have ${\rm Disc}(P) = b^2 - 4ac$,
 which agrees with the definition used in high-school algebra.)
It is not difficult to show that
\be
   {\rm Disc}(P)   \;=\;   (-1)^{M(M-1)/2} A^{-1} {\rm Res}(P,P')
   \;,
\ee
so that the discriminant vanishes if and only if $P$ has
at least one multiple root.\footnote{
   Lang \cite[p.~204]{Lang_93} warns that many books
   (including his own first edition!)\ contain sign errors
   (i.e.\ fail to include the factor $(-1)^{M(M-1)/2}$)
   either in the definition of the discriminant
   or in the formula relating it to the resultant.
}

For further information on resultants and discriminants
and algorithms for computing them, see e.g.\ \cite[Chapter 3]{Cox_98}.

Consider now the characteristic polynomial
of the transfer matrix $T(q)$:
\be
   P(\lambda,q)  \;=\;  \det[\lambda I - T(q)]
                 \;=\;  \prod_{i=1}^{\dim T} [\lambda - \lambda_i(q)]
   \;,
\ee
where $\{ \lambda_i(q) \}$ are the eigenvalues of $T(q)$.
Consider next the polynomial
\be
   P_\theta(\lambda,q)   \;=\;  P(e^{i \theta} \lambda, q)
   \;.
\ee
$P$ and $P_\theta$ are polynomials in $\lambda$ whose coefficients are
polynomials in $q$ (and in $e^{i \theta}$),
and they have a root in common if and only if
$T(q)$ has eigenvalues $\lambda_1$ and $\lambda_2$
satisfying $\lambda_1 = e^{i \theta} \lambda_2$.
(Note that, in addition to the desired case of an equimodular pair
of eigenvalues, this also occurs whenever $T(q)$ has a zero eigenvalue.)
We can therefore compute the locus of equimodularity by sweeping over
a closely spaced set of points $\theta \in (0,\pi]$,
and computing for each $\theta$ the roots of
\be
\label{def_resultant}
   R_\theta(q)  \;=\; {\rm Res}_\lambda (P, P_\theta)
   \;,
\ee
which is a polynomial in $q$ and $e^{i \theta}$.\footnote{
   This method was presumably known already to
   Beraha, Kahane and Weiss:  see \cite[footnote 2]{Beraha_79}.
}

Of course, $\theta=0$ is a special case:  here we are looking for
the multiple roots of $P$, which can be located by computing the zeros
of the discriminant
\be
\label{def_resultantzero}
   \widetilde{R}_0(q)  \;=\; {\rm Res}_\lambda (P, P')
   \;=\; \lim\limits_{\theta \to 0}
           {R_\theta(q) \over (-i\theta)^{\dim T} P(0,q)}
\ee
where $P'(\lambda,q) = \partial P(\lambda,q)/\partial \lambda$.
As we shall see, the zeros corresponding to $\theta=0$ are very important
because they correspond to the endpoints of the curves of equimodularity.

After finding a set of $q$ values for which $T(q)$ has a pair of
equimodular eigenvalues, we check them one-by-one by solving the
characteristic equation $P(\lambda,q) = 0$ and testing whether
the pair of equimodular eigenvalues is dominant or subdominant.
The limiting curve $\scrb$ of partition-function zeros
corresponds only to the crossing in modulus of {\em dominant}\/ eigenvalues;
nevertheless, it is sometimes of interest to depict the loci
of subdominant eigenvalue-crossing as well.
Whenever we do so in this paper,
we shall draw the dominant eigenvalue-crossing curves using solid black lines
and the subdominant curves with dashed \subdominantcolor\ lines.

In practice, we first use {\sc Mathematica} to compute
$R_\theta(q)$ symbolically as a polynomial in $q$ and $e^{i \theta}$
with integer coefficients.
To minimize the effect of round-off errors in the subsequent computation,
we use instead of $\theta$ the real variable
$t = \tan(\theta/2) \in (0,\infty]$
defined by
\be
   e^{i \theta} \;=\; {1 + it \over 1 - it}   \;,
 \label{def_t}
\ee
and we always choose $t$ to be a rational number.
In this way, $e^{i \theta}$ is a complex rational number
whose modulus is always exactly equal to 1,
and the polynomial $R_\theta(q)$ has complex rational coefficients.
This allows us to take advantage of arbitrary-precision polynomial
root-finders such as {\sc Mathematica}'s {\tt NSolve}
or (better yet) the package MPSolve 2.0 developed by Bini and Fiorentino
\cite{Bini_package,Bini-Fiorentino}.\footnote{
   MPSolve 2.0 is much faster than {\sc Mathematica}'s {\tt NSolve}
   for high-degree polynomials
   (this is reported in \cite{Bini-Fiorentino}, and we confirm it);
   it gives guaranteed error bounds for the roots,
   based on rigorous theorems \cite{Bini-Fiorentino};
   its algorithms are publicly documented \cite{Bini-Fiorentino};
   and its source code is freely available \cite{Bini_package}.
}

The drawback of the resultant method is that the degree of the resultant
polynomial $R_\theta(q)$ grows very rapidly with the width of the
lattice strip;
moreover, the coefficients in $R_\theta(q)$ also grow very rapidly
(even if we use the variable $u = q - q_0$).
This limits in practice the widths we can study:
$L_x \le 6$ for free transverse boundary conditions
and $L_x \le 8$ for periodic transverse boundary conditions.

\bigskip

\noindent
{\bf Remark}.
In some cases (notably with periodic and twisted-periodic
boundary conditions in the longitudinal direction
\cite{Shrock_99e,Shrock_99f,Shrock_00a,Shrock_00c,Shrock_00d,Shrock_00e}),
the amplitude corresponding to one or more eigenvalues can vanish identically.
If this occurs for one of the dominant eigenvalues,
blind application of the foregoing procedure can lead to erroneous results.
As a safeguard against this phenomenon,
we numerically computed the corresponding amplitudes
for a selected value of $q$, by numerically diagonalizing the transfer matrix
and rotating the corresponding vectors on left and right.
It suffices to find at least one value of $q$
for which none of the amplitudes vanishes.
In all the cases considered in this paper,
an identically-vanishing amplitude never arises.

\bigskip

\noindent
{\bf Example.}
In the simple special case of a $2 \times 2$ transfer matrix
\be
\label{transfermatrix2x2}
   T(q)  \;=\;  \left( \begin{array}{cc}
                         a(q) & b(q) \\
                         c(q) & d(q)
                       \end{array}  \right)
\ee
with left and right vectors
\begin{subeqnarray}
  \label{vectors2x2}
\vec{u}  & = & \left( \begin{array}{l}
                 f(q) \\
                 g(q)
                 \end{array}\right) \quad
   \\[2mm]
\vec{v} & = & \left( \begin{array}{l}
                 1  \\
                 0
                 \end{array}\right)
\end{subeqnarray}
and partition function
\be
   Z_n  \;=\;  \vec{u}^{\rm T} T^{n-1} \vec{v}
\ee
for a strip of length $n$ (where ${}^{\rm T}$ denotes transpose),
we can obtain explicit expressions.
We have
\be
\label{partitionfunction2x2}
Z_n(q) \;=\; \alpha_{-}(q) \lambda_{-}(q)^{n-1} \,+\,
             \alpha_{+}(q) \lambda_{+}(q)^{n-1}
   \;,
\ee
where the eigenvalues $\lambda_{\pm}$ and their corresponding amplitudes
$\alpha_{\pm}$ are
\begin{subeqnarray}
  \label{eigenvalues2x2}
\lambda_{\pm}(q) &=& {1 \over 2} \left( \tr T(q) \pm
                  \sqrt{ \tr^2 T(q) - 4 \det T(q)} \right)   \\[2mm]
    &\equiv&  {1 \over 2} \left( P_1(q) \pm \sqrt{P_2(q)} \right)
\end{subeqnarray}
and
\begin{subeqnarray}
  \label{amplitudes2x2}
\alpha_{\pm}(q)       &=& {1 \over 2} \left( f(q) \pm
                  { [a(q)-d(q)] f(q) + 2c(q) g(q) \over
                    \sqrt{ \tr^2 T(q) - 4 \det T(q)} } \right) \\[2mm]
              &\equiv&
                  {1 \over 2} \left( f(q) \pm { P_3(q) \over
                                     \sqrt{P_2(q)} } \right)
   \;.
\end{subeqnarray}
Note that even though the eigenvalues and the amplitudes are
non-polynomial algebraic functions of $q$, the partition function
$Z_n(q)$ is always a polynomial in $q$.
{}From these equations it is not difficult to prove the recurrence relation
\be
\label{recursion2x2}
Z_{n+2}(q) \,-\, [\tr T(q)] Z_{n+1}(q) \,+\, [\det T(q)] Z_n(q)  \;=\; 0
   \;.
\ee

According to the Beraha--Kahane--Weiss theorem (Theorem~\ref{BKW_thm}),
the limit points as $n \to\infty$ of the partition-function zeros
are of two types:
the isolated limiting points, which occur where one eigenvalue is dominant
 and its amplitude vanishes;
and curves of non-isolated limiting points,
where there is a crossing in modulus of two or more dominant eigenvalues.
In our simple case, the limiting curves are given by the condition
$|\lambda_{-}(q)| = |\lambda_{+}(q)|$,
which means that $P_2(q)/P_1(q)^2$ should be a negative real number
(call it $-t^2$), or in other words
\be
  \tr^2 T(q)  \;=\;  4 \widetilde{t} \det T(q)
  \quad\hbox{with } 0 \,\le\, \widetilde{t} \equiv {1 \over 1+t^2} \,\le\, 1
  \;.
 \label{limitcurve2x2}
\ee
An identical result is obtained using the resultant method
\reff{def_resultant}/\reff{def_resultantzero}:  we have
\be
   P(\lambda,q)   \;=\;  \lambda^2 \,-\, [\tr T(q)] \lambda \,+\, \det T(q)
\ee
and hence
\begin{subeqnarray}
\slabel{resultant2x2}
R_\theta(q) &=& (1-e^{i\theta})^2 \, [\det T(q)] \,
   \left[ (1+e^{i\theta})^2 \det T(q) - e^{i\theta} \tr^2 T(q) \right] \\[2mm]
\slabel{resultantzero2x2}
\widetilde{R}_0(q) &=& 4 \det T(q) - \tr^2 T(q)  \;=\; -P_2(q)
\end{subeqnarray}
We see that \reff{resultant2x2} vanishes precisely on the
curve \reff{limitcurve2x2}:  it suffices to insert \reff{def_t}
and note that
\be
   { (1+e^{i\theta})^2 \over e^{i\theta} }
   \;=\;  {4 \over 1+t^2}  \;=\;  4 \widetilde{t}
\ee
[equivalently, $\widetilde{t} = \cos^2 (\theta/2)$].

The isolated limiting points of zeros are given by the condition
that the amplitude of the leading eigenvalue vanishes.
We first compute the product of the two amplitudes:
\be
\label{amplitudeproduct2x2}
\alpha_{-}(q) \alpha_{+}(q)  \;=\;  {P_4(q) \over P_2(q)}
\ee
where
\be
\label{isolatedzeros2x2}
P_4(q)  \;\equiv\; {f(q)^2 P_2(q) - P_3(q)^2  \over  4}
        \;=\;  Z_1(q) Z_3(q) - Z_2(q)^2
   \;.
\ee
[Note that \reff{amplitudeproduct2x2}/\reff{isolatedzeros2x2}
 is just Lemma~\ref{lemma2.2} specialized to the case $M=2$.]
We now observe that a root of $P_4$ corresponds to the
vanishing of $\alpha_-$ (resp.\ $\alpha_+$) in case
$f \sqrt{P_2} = P_3$ (resp.\ $f \sqrt{P_2} = -P_3$);
and $\lambda_-$ (resp.\ $\lambda_+$) is dominant in case
$\Re (P_1/\sqrt{P_2}) < 0$ (resp.\ $> 0$).
It follows that a root of $P_4$ corresponds to the
vanishing of the amplitude corresponding to the {\em dominant}\/
eigenvalue in case
\be
   \Re {f(q) P_1(q) \over P_3(q)}   \;<\;  0
 \label{dominance_M=2}
\ee
there.
If $P_3(q) = 0$ at a root of $P_4$, then {\em both}\/ amplitudes vanish there.

\subsubsection{Direct-search method}  \label{sec4.1.2}

When we were unable to apply the resultant method
(i.e.\ for large lattice widths),
or when we wanted to study in detail a small region in the $q$-plane
(for any lattice width),
we used a direct-search method to obtain the curves of equimodularity.
The idea is to define a function that measures the difference
between the moduli of the two dominant eigenvalues of the transfer matrix.
Since an explicit expression of the eigenvalues as a function of $q$
is usually not available, the eigenvalues are obtained numerically
for each value of $q$.
Then we extract the two eigenvalues of largest modulus and compute
\be
   F(q)  \;\equiv\;  |\lambda_1(q)| - |\lambda_2(q)|
\ee
where $\omega \lambda_1(q) \preceq \omega \lambda_2(q)$ in lexicographic order
(here $\omega$ is some arbitrarily chosen nonzero complex number).
Clearly $F(q)$ vanishes if and only if the two dominant eigenvalues
are equimodular;  moreover, because generically the two dominant eigenvalues
will {\em not}\/ satisfy
$\real[\omega \lambda_1(q)] = \real[\omega \lambda_2(q)]$
precisely where they also satisfy $|\lambda_1(q)| = |\lambda_2(q)|$,
they will generically {\em not}\/ interchange roles at the
eigenvalue crossing, and hence $F(q)$ will be a {\em smooth}\/
function of $q$ there.\footnote{
   There is nothing special about lexicographic order;
   virtually any order will do, {\em except}\/ ordering by moduli.
   In practice, we used {\sc Mathematica}'s default ordering,
   which is slightly different from lexicographic.
   Note also that any nonzero $\omega \in \C$ will do;
   but when studying an eigenvalue crossing on the real $q$ axis
   (for which $\lambda_1$ and $\lambda_2$ will be complex conjugates),
   it is advantageous to choose $\omega$ non-real.
}
We then search for the zeros of $F(q)$ using
a Newton method in the complex $q$-plane.

Once a good approximation for a zero is found,
we also compute the phase $e^{i\theta}=\lambda_1/\lambda_2$
[or equivalently, the number $t$ defined by \reff{def_t}].
Knowledge of $\theta$ is very useful in trying to understand
the topology of the limiting curve.

\subsection{Qualitative structure of eigenvalue-crossing curves}
   \label{sec4.2}

In this section we want to discuss the types of qualitative behaviors
that can arise when studying the eigenvalue-crossing curves.
Recall first \cite[Chapter 2]{Kato_80}
that the eigenvalues $\lambda_i(q)$ of the transfer matrix $T(q)$
are analytic functions of $q$ except possibly where two or more
eigenvalues collide.
Indeed, this can be seen by expanding the characteristic polynomial
$P(\lambda,q) = \det[\lambda I - T(q)]$ around a root $(\lambda_0,q_0)$:
\be
   P(\lambda,q)  \;=\;  a(\lambda-\lambda_0) \,+\, b(q-q_0)
       \,+\, c(\lambda-\lambda_0)^2 \,+\, d(q-q_0)^2
       \,+\, e(\lambda-\lambda_0)(q-q_0) \,+\, \ldots
\ee
Provided that $\lambda_0$ is not a multiple eigenvalue of $T(q_0)$,
we have $a \equiv (\partial P/\partial\lambda)(\lambda_0,q_0) \neq 0$;
the implicit function theorem then guarantees that in a neighborhood of
$q=q_0$ there exists an analytic function $\lambda(q)$ solving
$P(\lambda(q),q) = 0$ with $\lambda(q_0) = \lambda_0$,
and it has the convergent expansion
\be
   \lambda(q)  \;=\;  \lambda_0  \,-\, {b \over a} (q-q_0)
     \,+\, {abe - a^2 d - b^3 c  \over a^3} (q-q_0)^2
     \,+\, \scro\bigl( (q-q_0)^3 \bigr)
   \;.
  \label{regular}
\ee

If, on the other hand, $\lambda_0$ is a $k$-fold eigenvalue of $T(q_0)$
with $k \ge 2$,
then $\lambda(q)$ can have an $l^{th}$-root branch point at $q_0$
for any $l \le k$.
More precisely, near $q=q_0$ the eigenvalues divide into groups
of cardinalities $l_1,\ldots,l_M$ (with $l_1 + \cdots + l_M = \dim T$)
such that the eigenvalues of the $i^{th}$ group
are the $l_i$ distinct determinations of an analytic function of
$(q-q_0)^{1/l_i}$, i.e.\ they have an $l_i^{th}$-root branch point at $q_0$
\cite[pp.~65--66]{Kato_80}.
For example, for $k=2$ we have $a=0$ and $c \neq 0$:
if $b \neq 0$, then a pair of eigenvalues $\lambda_\pm(q)$
bifurcate off $\lambda_0$ with a square-root branch point,
\be
   \lambda_\pm(q)  \;=\;  \lambda_0  \,\pm\,
     \left(\! - {b \over c}\right)^{\! 1/2} (q-q_0)^{1/2}
     \,-\, {e \over 2c} (q-q_0)  \,+\, \scro\bigl( (q-q_0)^{3/2} \bigr)
   \;;
  \label{square_root}
\ee
while if $b=0$, then $\lambda_\pm(q)$ are analytic functions
in a neighborhood of $q_0$,
\be
   \lambda_\pm(q)  \;=\;  \lambda_0  \,+\,
      {-e \pm \sqrt{e^2 -4cd} \over 2c} (q-q_0)
      \,+\, \scro\bigl( (q-q_0)^2 \bigr)
   \;.
  \label{lambda+-_b=0}
\ee

\subsubsection{Crossing of two simple eigenvalues}   \label{sec4.2.1}

Generically, the equimodularity of two eigenvalues defines an analytic
curve in the complex $q$-plane, along which the parameter
$\theta$ (or $t$) varies smoothly.
To see this, suppose that at $q=q_0$ we have a pair of
{\em simple}\/ (i.e.\ non-multiple) eigenvalues
$\lambda_{1,0}$ and $\lambda_{2,0}$
that have equal modulus ($|\lambda_{1,0}| = |\lambda_{2,0}| \neq 0$);
they satisfy $\lambda_{1,0} = e^{i\theta} \lambda_{2,0}$
with $\theta \neq 0$ (mod $2\pi$).
Each of these eigenvalues then extends to a single-valued analytic function
of $q$ in a neighborhood of $q=q_0$, as in \reff{regular}:
\be
   \lambda_i(q)  \;=\;  \lambda_{i,0}  \,-\, {b_i \over a_i} (q-q_0)
     \,+\, {a_i b_i e_i - a_i^2 d_i - b_i^3 c_i  \over a_i^3} (q-q_0)^2
     \,+\, \scro\bigl( (q-q_0)^3 \bigr)
 \label{lambda_regular1}
\ee
for $i=1,2$.
Their ratio is then
\begin{subeqnarray}
{\lambda_1(q) \over \lambda_2(q)} &=&
   e^{i\theta} \left[ 1 -
    \left( {b_1 \over a_1 \lambda_{1,0}} - {b_2 \over a_2 \lambda_{2,0}} \right)
    (q-q_0) + {\cal O}((q-q_0)^2) \right]   \slabel{eq_lambda12_a}  \\[2mm]
 & \equiv &   e^{i\theta}
      \left[ 1 + \rho e^{i \phi} (q-q_0)  + {\cal O}((q-q_0)^2) \right]
 \label{lambda_regular2}
\end{subeqnarray}
Provided that $\rho \neq 0$, the equimodularity locus
$|\lambda_1(q) / \lambda_2(q)| = 1$ defines near $q=q_0$
an analytic curve that passes through $q=q_0$ at angle $-\phi \pm \pi/2$.

If, however, $\rho = 0$ [i.e.\ the term in \reff{eq_lambda12_a}
that is linear in $q-q_0$ vanishes],
then the equimodularity locus can have multiple points.
(A {\em sufficient}\/ though not necessary condition for this to occur
 is for the linear terms to vanish in each eigenvalue separately,
 i.e.\ $b_1 = b_2 = 0$.)
If the first nonvanishing term in \reff{eq_lambda12_a} is the one
of order $(q-q_0)^k$, then we have a $k$-fold multiple point
in the sense of algebraic geometry \cite[Section 6.2]{Gibson_98}:
that is, the equimodularity locus $|\lambda_1(q) / \lambda_2(q)| = 1$
defines near $q=q_0$ an analytic image of the set $\Re z^k = 0$ near $z=0$.
In particular, this set can be interpreted locally as $k$ analytic curves
crossing at angles $\pi/k$.

\bigskip
{\bf Remark.}
Let us consider the partition function $Z_n(q)$
near a point $q_0$ where there are exactly two dominant simple eigenvalues:
\be
   Z_n(q)   \;=\;  \alpha_1(q) \, \lambda_1(q)^n \;+\;
                   \alpha_2(q) \, \lambda_2(q)^n \;+\; \ldots
 \label{partition_regular}
\ee
where the dots indicate the contributions of subdominant eigenvalues.
Let us assume for simplicity that $\alpha_1(q_0), \alpha_2(q_0) \neq 0$
and insert the expansions
\begin{eqnarray}
   \log {\lambda_1(q) \over \lambda_2(q)}   & = &
      2\pi i \psi   \,+\,  A (q-q_0)  \,+\,  {\cal O}( (q-q_0)^2 )
   \\[2mm]
   \log {\alpha_1(q) \over -\alpha_2(q)}   & = &
      2\pi i B   \,+\,  C (q-q_0)  \,+\,  {\cal O}( (q-q_0)^2 )
\end{eqnarray}
where $\psi \equiv \theta/2\pi \in \R$ and $A,B,C \in \C$.
[From \reff{lambda_regular2} we have
$A = \rho e^{i\phi} \equiv (b_2/a_2 \lambda_{2,0}) - (b_1/a_1 \lambda_{1,0})$.]
Then \reff{partition_regular} can be written as
\begin{eqnarray}
   Z_n(q)   &=&
   2i \left[ \alpha_1(q) \, \alpha_2(q) \, \lambda_1(q)^n \, \lambda_2(q)^n
      \right] ^{1/2}
   \,\times\,
   \nonumber \\[2mm]
   & & \quad
   \sinh\!\left[ \pi i (n\psi+B) \,+\, {An+C \over 2} (q-q_0)
                        \,+\,  {\cal O}( (q-q_0)^2 )  \right]
   \;,
 \label{partition_regular_Baxter}
\end{eqnarray}
which agrees with \cite[eqn.~(3.3)]{Baxter_87}
by trivial renaming of variables.
When $n$ is large, \reff{partition_regular_Baxter} has zeros at
\be
   q   \;=\;  q_0  \,-\, {2\pi i \over An} \left[ k+B+R(n\psi) \right]
                   \,+\, {\cal O}(1/n^2)
 \label{partition_regular_zeros}
\ee
where $k \in \Z$ is an arbitrary integer
and $R(x) \equiv x - \lfloor x \rfloor$ is $x$ modulo 1.
Therefore, we see that near a regular point $q_0$ of the
limiting curve $\scrb$,
the finite-volume partition function
$Z_n(q)$ has a sequence of evenly spaced zeros
[spacing = $2\pi/(|A|n)$] lying parallel to $\scrb$.
These zeros lie a distance $2\pi |\!\imag B| /(|A|n)$
away from $\scrb$, hence converge to it generically at rate $1/n$.

\bigskip

In addition to this generic behavior, there are other features exhibited by
the limiting curves that are worth studying in detail:

\subsubsection{Endpoints (collision of two eigenvalues)}   \label{sec4.2.2}

Suppose that $\lambda_0 \neq 0$ is a two-fold eigenvalue of $T(q_0)$.
Then generically (i.e.\ if $b \neq 0$) a pair of eigenvalues $\lambda_\pm(q)$
bifurcates off $\lambda_0$ with a square-root branch point,
as in \reff{square_root}.  The ratio of the two eigenvalues is
\be
{\lambda_+(q) \over \lambda_-(q)}  \;=\;
     1 \,+\, A (q-q_0)^{1/2} \,+\, {A^2 \over 2} (q-q_0)
       \,+\, {\cal O}((q-q_0)^{3/2})
\ee
where $A = (2/\lambda_0) (-b/c)^{1/2} \neq 0$.
Setting
\be
   {\lambda_+(q) \over \lambda_-(q)}   \;=\;
   e^{i \theta} \;=\; {1 + it \over 1 - it}
\ee
with $t$ real, we see that the equimodularity locus is given by
\be
   q   \;=\;  q_0  \,-\,  {4 \over A^2} t^2  \,+\, {\cal O}(t^4)
   \;.
 \label{lambda_endpoint_equimod}
\ee
Only even powers of $t$ appear in this expansion,
since $\lambda_+$ and $\lambda_-$ interchange roles as we
go around the branch point, and
$\lambda_+/\lambda_- = e^{i \theta}$ if and only if
$\lambda_-/\lambda_+ = e^{-i \theta}$.
Writing $A = \rho e^{i\phi}$, we see that the equimodularity locus
is an analytic curve ending at $q=q_0$ and tangent there to the ray
$\arg(q-q_0) = \pi - 2\phi$.

Because the two eigenvalues are equal at $q=q_0$,
the parameter $\theta$ (or $t$) takes the value 0 at endpoints.
Each such endpoint corresponds to a simple root of
the $t=0$ resultant $\widetilde{R}_0(q)$:
indeed, \reff{square_root} implies that for $q$ near $q_0$ we have
$\lambda_{+} - \lambda_{-} \sim (q-q_0)^{1/2}$ and hence
$\widetilde{R}_0(q) \sim (\lambda_{+} - \lambda_{-})^2  \sim  q - q_0$.
For this reason, the simple zeros of the $t=0$ resultant
play an essential role
when we try to determine with high accuracy the topology of the limiting curve.

By contrast, in the non-generic case $b=0$, where the
eigenvalues $\lambda_\pm(q)$ are analytic functions in a
neighborhood of $q_0$ [cf.\ \reff{lambda+-_b=0}],
the formulae \reff{lambda_regular1}/\reff{lambda_regular2}
of the preceding subsection apply with $\theta = 0$.
In this case, the root $q=q_0$ is a {\em double}\/ zero
of the $t=0$ resultant $\widetilde{R}_0(q)$,
as $\lambda_{+} - \lambda_{-} \sim q-q_0$ by \reff{lambda+-_b=0},
so that $\widetilde{R}_0(q) \sim (\lambda_{+} - \lambda_{-})^2 \sim (q-q_0)^2$.

\bigskip

{\bf Example.}
Consider a two-dimensional transfer matrix as in \reff{transfermatrix2x2}.
Collision of the two eigenvalues \reff{eigenvalues2x2}
occurs whenever $P_2(q) = 0$, i.e.\ whenever $\tr^2 T(q) = 4 \det T(q)$.
Suppose this occurs at $q=q_0$,
with $\lambda_0 \equiv \lambda_\pm(q_0) = {1 \over 2} \tr T(q_0)$,
and let us expand the eigenvalues \reff{eigenvalues2x2} around $q=q_0$:
\be
\lambda_\pm = {1\over 2} \tr T(q_0)  \,\pm\, C_1 (q-q_0)^{1/2}
              \,+\, C_2 (q-q_0) \,+\,
              {\cal O}( (q-q_0)^{3/2} )
\ee
where
\begin{subeqnarray}
C_1 &=& \left( {1 \over 2} \tr T(q_0) \left.{d \tr T(q) \over dq}\right|_{q=q_0}
          \,-\, \left.{d \det T(q) \over dq}\right|_{q=q_0} \right) ^{\! 1/2}
 \\[2mm]
C_2 &=& {1 \over 2} \left.{d \tr T(q) \over dq}\right|_{q=q_0}
\end{subeqnarray}
The ratio of the two eigenvalues is therefore
\be
{\lambda_+(q) \over \lambda_-(q)}  \;=\;
     1 \,+\, A (q-q_0)^{1/2} \,+\, {A^2 \over 2} (q-q_0)
       \,+\, {\cal O}((q-q_0)^{3/2})
\ee
with $A = 4 C_1 / [\tr T(q_0)]$.

\bigskip
{\bf Remark.}
Let us return now to the general case, and consider the partition function
$Z_n(q)$ close to an endpoint $q=q_0$ where two dominant eigenvalues collide:
\be
   Z_n(q)   \;=\;  \alpha_+(q) \, \lambda_+(q)^n \;+\;
                   \alpha_-(q) \, \lambda_-(q)^n \;+\; \ldots
 \label{partition_endpoint}
\ee
where the dots indicate the contributions of subdominant eigenvalues.
{}From \reff{square_root}
[cf.\ also \reff{eigenvalues2x2}/\reff{amplitudes2x2}]
we have
\begin{eqnarray}
   \log {\lambda_+(q) \over \lambda_-(q)}   & = &
      A (q-q_0)^{1/2}  \,+\,  {\cal O}( q-q_0 )
   \\[2mm]
   \log {\alpha_+(q) \over -\alpha_-(q)}   & = &
      2\pi i B   \,+\,  C (q-q_0)^{1/2}  \,+\,  {\cal O}( q-q_0 )
\end{eqnarray}
where $A = 2(-b/c)^{1/2}/\lambda_0$.
Then \reff{partition_endpoint} can be written as
\begin{eqnarray}
   Z_n(q)   &=&
   2i \left[ \alpha_+(q) \, \alpha_-(q) \, \lambda_+(q)^n \, \lambda_-(q)^n
      \right] ^{1/2}
   \,\times\,
   \nonumber \\[2mm]
   & & \qquad \sinh\!\left[ i\pi B \,+\, {An+C \over 2} (q-q_0)^{1/2}
                        \,+\,  {\cal O}( q-q_0 )  \right]
   \;.
 \label{partition_endpoint_bis}
\end{eqnarray}
When $n$ is large, the zeros of \reff{partition_endpoint_bis} are given by
\be
   q   \;=\;  q_0  \,-\, {4\pi^2 (k+B)^2 \over A^2 n^2}
                   \,+\, {\cal O}(1/n^3)
 \label{partition_endpoint_zeros}
\ee
where $k \in \Z$, in agreement with \reff{lambda_endpoint_equimod}.
Therefore, the convergence close to an endpoint is generically of order
$1/n^2$, in contrast with the $1/n$ convergence near a regular point
of the limiting curve [cf.\ \reff{partition_regular_zeros}].

\subsubsection{Crossing of three simple eigenvalues (T points)}

\begin{figure}
  \centering
  \epsfxsize=400pt\epsffile{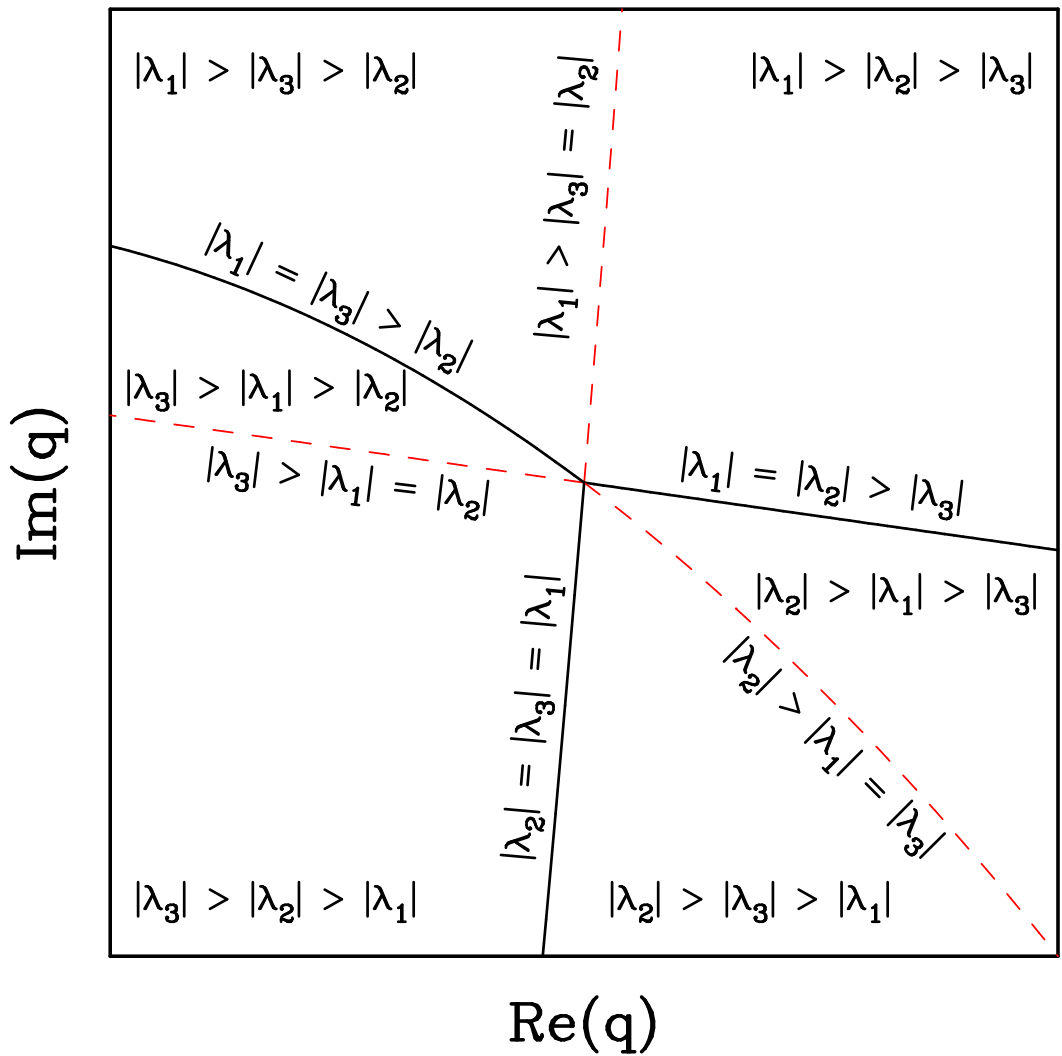}
  \caption{
  Schematic representation of a generic T point where three
  eigenvalues $\lambda_1, \lambda_2, \lambda_3$ are simultaneously dominant.
  The solid black lines represent the loci of dominant eigenvalue crossings,
  while the dashed \subdominantcolor\ lines represent the loci of
  subdominant eigenvalue crossings.
  Each line has been labeled by the inequalities and equalities it satisfies,
  and each region has been labeled by the inequalities it satisfies.
  }
\label{plot_T_point}
\end{figure}

Suppose that at $q=q_0$ we have three dominant simple eigenvalues
(call them $\lambda_1, \lambda_2, \lambda_3$):
then three smooth curves of equimodularity, corresponding to the three pairs
of eigenvalues $\lambda_{1/2}, \lambda_{1/3}, \lambda_{2/3}$,
pass through $q=q_0$.
One half of each curve of equimodularity corresponds to a
dominant pair of eigenvalues, while one half corresponds to
a subdominant pair;
therefore the locus of dominant equimodularity looks like a T,
so we call these crossings {\em T points}\/
(see Figure~\ref{plot_T_point}).
They are complex analogues of ``triple points'' in the thermodynamic sense
(but they are {\em not}\/ multiple points in the sense of
 algebraic geometry: see Remark 2 below).

More precisely, the eigenvalues $\lambda_i(q)$ for $i=1,2,3$
define single-valued analytic functions \reff{lambda_regular1}
in a neighborhood of $q_0$, so that their ratios are
\begin{subeqnarray}
{\lambda_1(q) \over \lambda_2(q)} &=& {\lambda_{1,0} \over \lambda_{2,0}} \left[
      1 - \left( {a_1 \over b_1 \lambda_{1,0}} - {a_2 \over b_2 \lambda_{2,0}}
          \right) (q-q_0) + {\cal O}((q-q_0)^2) \right] \\[2mm]
{\lambda_1(q) \over \lambda_3(q)} &=& {\lambda_{1,0} \over \lambda_{3,0}} \left[
      1 - \left( {a_1 \over b_1 \lambda_{1,0}} - {a_3 \over b_3 \lambda_{3,0}}
          \right) (q-q_0) + {\cal O}((q-q_0)^2) \right] \\[2mm]
{\lambda_2(q) \over \lambda_3(q)} &=& {\lambda_{2,0} \over \lambda_{3,0}} \left[
      1 - \left( {a_2 \over b_2 \lambda_{2,0}} - {a_3 \over b_3 \lambda_{3,0}}
          \right) (q-q_0) + {\cal O}((q-q_0)^2) \right]
 \label{lambda_ratios_T_point}
\end{subeqnarray}
Provided that the coefficients of $q-q_0$ in (\ref{lambda_ratios_T_point}a--c)
are not collinear, there are six possible phases for $q-q_0$ that render
one of the ratios unimodular to leading order in $q-q_0$;
three of these correspond to a dominant crossing,
while the three complementary phases correspond to a subdominant crossing.
If, on the curve where $|\lambda_i / \lambda_j| = 1$, we define
\be
   {\lambda_i(q) \over \lambda_j(q)}  \;=\;  e^{i \theta_{ij}(q)}
   \;,
\ee
then the phase $\theta_{ij}(q)$ varies smoothly along this curve.
No {\em pair}\/ of these three phases obeys in general any relation,
even as $q \to q_0$;
but at the T point we obviously have the relation
\be
   \theta_{12} + \theta_{23} + \theta_{31} \;=\; 2\pi n
 \label{theta_123_relation}
\ee
for some integer $n$.
If we take $-\pi < \theta_{ij} \le \pi$,
then $n$ must be $-1$, $0$ or $+1$.
In particular, the absolute values of the $\theta_{ij}$
(which are what we actually calculate in practice)
must satisfy either
$|\theta_{12}| + |\theta_{23}| + |\theta_{31}| = 2 \pi$
or some permutation of $|\theta_{12}| + |\theta_{23}| - |\theta_{31}| = 0$.
In practice, one way of detecting a T point is by noticing the discontinuity
in $\theta$ as we pass from one piece of the limiting curve to another.

\bigskip

\noindent
{\bf Remarks.}
1.  We frequently find that the limiting curve $\scrb$ contains
an arc starting at an endpoint ($t=0$)
and ending at a T point ($t=t_0$).\footnote{
   Similar features were found in the complex-temperature zeros of the
   two-dimensional spin-$s$ Ising model \cite{Matveev_95}, in the
   complex-temperature zeros of the Potts model on the hexagonal, Kagom\'e and
   triangular lattices \cite{Feldmann_98a,Feldmann_98b}, and in the
   complex-$q$ zeros of the Potts antiferromagnet on some families of strip
   graphs \cite{Tsai_98}.
}
If we are able to compute the $t=0$ resultant,
then all endpoints can be detected,
so that the corresponding arcs can be detected as well.
But using the direct-search method, a short arc can easily be overlooked:
for since the parameter $t$ grows smoothly
as we move along the arc towards the T point,
a short arc corresponds to a small value of $t_0$;
and if $t_0$ is smaller than our step size,
we may fail to detect the discontinuity in $t$ along the other two arcs
merging at the T point (unless we are very lucky!).
This means that in the cases where we were unable to obtain the zeros
of the $t=0$ resultant (namely, widths $7_{\rm F}$ and $8_{\rm F}$),
we may have missed some such ``close pairs'' of endpoints and T points;
so the lists of endpoints and T points reported
in Sections~\ref{sec7F} and \ref{sec8F} may be incomplete,
and the counts reported in Table~\ref{table_summary}
must be understood as {\em lower bounds}\/ on the true value.

2.  Ro\v{c}ek, Shrock and Tsai \cite[p.~528 top, item (ii)]{Shrock_98a}
state erroneously that a T point is a ``multiple point''
in the technical sense of algebraic geometry,
i.e.\ a point where two or more branches
of the {\em same}\/ irreducible algebraic curve meet.
In fact, a T point corresponds to the crossing {\em in modulus}\/
of three usually {\em unrelated}\/ analytic functions.
Our aim here is not to quibble about definitions,
but to emphasize a radical difference in behavior:
at a $k$-fold multiple point the branches always intersect at angles $\pi/k$,
while at a T point they can (and in general do)
intersect at {\em arbitrary}\/ angles.
(By contrast, the crossing \reff{lambda_regular2} with $\rho = 0$
is a true multiple point, as correctly observed by Ro\v{c}ek {\em et al.}\/
\cite[p.~528 top, item (i)]{Shrock_98a}.)

\subsection{Computation of isolated limiting points of zeros}   \label{sec4.3}

According to case (a) of the Beraha--Kahane--Weiss theorem,
the isolated limiting points of partition-function zeros
correspond to points where there is a unique dominant eigenvalue
whose amplitude vanishes.  We locate such points by a two-step process:
First we determine, using Lemma~\ref{lemma2.2},
the (finite) set of $q$ values where {\em at least one}\/ amplitude vanishes.
Then we check these $q$ values one by one, by diagonalizing $T(q)$,
rotating the left and right vectors $\vec{u}(q)$ and $\vec{v}(q)$,
and checking whether the vanishing amplitude(s) corresponds
to the dominant eigenvalue.
In the special case $\dim T = 2$,
we can test dominance using the analytic criterion \reff{dominance_M=2}.

\section{Numerical Results for the Square-Lattice Chromatic Polynomial:
   \hfill\break Free Transverse Boundary Conditions}
   \label{sec5}

We have computed the transfer matrix $T(q)$ and the limiting curves $\scrb$
for square-lattice strips of widths $2 \leq L_x \leq 8$
with free boundary conditions in the transverse direction.
We have checked the self-consistency of our finite-lattice results
using the trivial identity
\be
  Z(m_{\rm F} \times n_{\rm F})  \;=\;   Z(n_{\rm F} \times m_{\rm F})
\ee
for all pairs $2 \le m,n \le 8$.

%
%
\subsection{$L_x = 2_{\rm F}$} \label{sec2F}

In this case the transfer matrix is one-dimensional,
and the result is trivial:
\be
  Z(2_{\rm F} \times n_{\rm F}) \;=\;  q (q-1) (q^2 - 3q + 3)^{n-1}
  \;.
\ee
Since there is only one eigenvalue, there is obviously no crossing,
hence ${\cal B} = \emptyset$.
However, there are zeros for all $n$ at $q=0,1$ (trivially)
and for all $n \ge 2$ at $q = (3 \pm \sqrt{3} \, i)/2$.

%
%
\subsection{$L_x = 3_{\rm F}$} \label{sec3F}

In this case the transfer matrix is two-dimensional.
The allowed partitions are given by
${\bf P} = \left\{ 1, \delta_{13}\right\}$.
In this basis the transfer matrix is equal to\footnote{
   As was found a quarter of a century ago by
   Biggs and Meredith \cite[p.~11]{Biggs_76}.
}
\be
T(3_{\rm F}) \;=\; \left( \begin{array}{cc}
                             q^3 - 5 q^2 + 10 q -8 & q^2 - 4 q + 5 \\
                             1                     & q -2
                          \end{array} \right)
  \;,
\ee
and the partition function is equal to
\be
Z(3_{\rm F} \times n_{\rm F})  \;=\; q(q-1) \,
   \left(\! \begin{array}{c}
             q-1 \\
             1
          \end{array} \!\right) ^{\! {\rm T}}
   \cdot
   T(3_{\rm F})^{n-1} \cdot
   \left(\! \begin{array}{c}
             1 \\
             0
          \end{array} \!\right)
\ee
where ${}^{\rm T}$ denotes transpose.

We can rewrite the above expression for the partition function
as in \reff{partitionfunction2x2}.
The polynomials $P_1$, $P_2$, $P_3$ and $P_4$
entering the definitions of the eigenvalues and amplitudes
\reff{eigenvalues2x2}/\reff{amplitudes2x2}/\reff{amplitudeproduct2x2}
are given by
\begin{subeqnarray}
P_1(q)  &=& (q^2-3q+5)(q-2) \\
P_2(q)  &=& (q^2 - 5q +7)(q^4 - 5 q^3 + 11 q^2 - 12 q + 8)
  \slabel{P_2_3F} \\
P_3(q)  &=& q(q-1)(q^4 - 6q^3 + 14 q^2 -15 q + 8)  \\
P_4(q)  &=& q^2 (q-1)^2 (q-2)
\end{subeqnarray}

\begin{figure}
  \centering
  \epsfxsize=400pt\epsffile{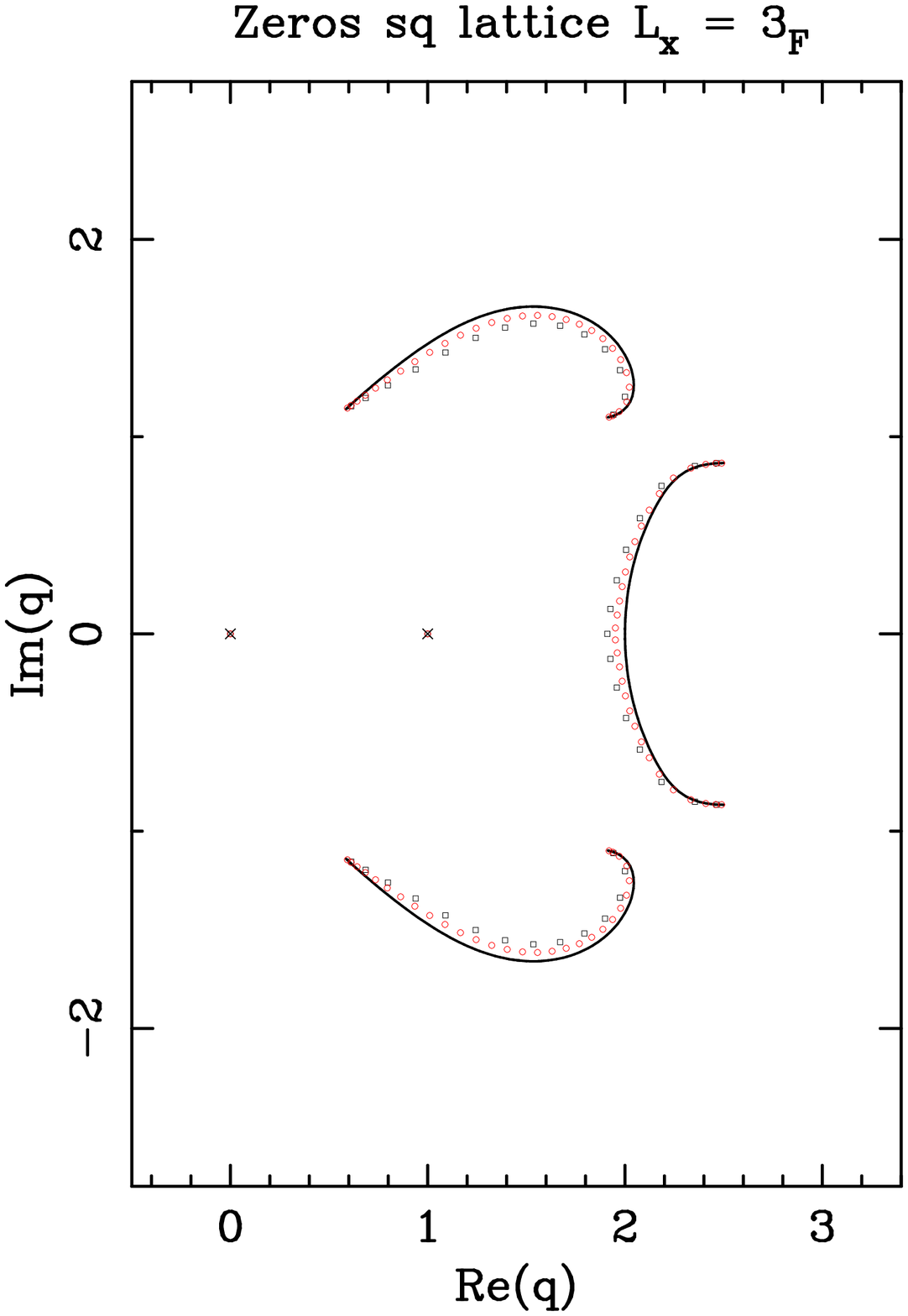}
  \caption{
  Zeros of the partition function of the $q$-state Potts antiferromagnet
  on the square lattices $3_F \times 15_F$ (squares),
  $3_F \times 30_F$ (circles) and $3_F\times\infty_F$ (solid line).
  The isolated limiting zeros are depicted by a $\times$.
  The limiting curve was computed using the resultant method.
  }
\label{Figure_sq_3FxInftyF}
\end{figure}

The limiting curve $\scrb$ (see Figure~\ref{Figure_sq_3FxInftyF})
consists of three disjoint arcs.\footnote{
   This curve is also depicted in \cite[Figure 3(a)]{Shrock_98a}.
}
One of them crosses the real axis at $q_0=2$,
and is invariant under complex conjugation;
the other two lie in the first and fourth quadrants, respectively,
and are complex conjugates of each other.
There are six endpoints:
\begin{subeqnarray}
q &\approx& 0.5865699800 \pm 1.1400627519\,i \\
q &\approx& 1.9134300200 \pm 1.0979688996\,i \\
q &=      & {5 \pm \sqrt{3}\,i \over  2 }  \;\approx\; 2.5 \pm 0.8660254038\,i
\end{subeqnarray}
These endpoints are the six roots of the $t=0$ resultant
$\widetilde{R}_0(q) = -P_2(q)$ given by \reff{P_2_3F}.

The zeros of the amplitudes can be found by solving the equation $P_4(q) = 0$.
There are two trivial zeros $q=0,1$, where both amplitudes vanish
simultaneously; both these zeros lie in the region where the eigenvalue
$\lambda_{-}$ is dominant. The non-trivial zero $q=B_4=2$ is a zero of
$\alpha_{-}$,
but not of $\alpha_{+}$ [$\alpha_{+}(q=2) = 2$].
However, this point happens to lie on an dominant-eigenvalue-crossing curve
[$\lambda_{+}(q=2) = - \lambda_{-}(q=2) = 1$],
so it actually belongs to case (b) of the Beraha--Kahane--Weiss theorem.
Indeed, from Table~\ref{table_zeros_free} we see that
the first non-trivial real zero of the partition function
seems to converge to $q=2$, but at the slow (roughly $1/n$) rate
characteristic of non-isolated limiting points
rather than at the fast (exponential) rate characteristic of
isolated limiting points.

%
%
\subsection{$L_x = 4_{\rm F}$} \label{sec4F}

The transfer matrix is three-dimensional. In the basis
${\bf P} = \{ 1, \delta_{13}+\delta_{24},\delta_{14} \}$ it can be
written as
\begin{eqnarray}
 & & T(4_{\rm F}) = \nonumber \\
 & & \quad \left( \begin{array}{ccc}
q^4 - 7 q^3 + 21 q^2 - 32 q + 21  & 2 (q^3 - 6 q^2 + 14 q -12) &
                                                   q^3 - 7 q^2 + 19 q -20 \\
q -2                              & q^2 - 4 q +5               & 3 - q    \\
-1                                & -2(q-2)                    & q^2 -5 q + 7
                      \end{array} \right) \;, \nonumber \\[2mm]
 & &
\end{eqnarray}
and the partition function is equal to
\be
Z(4_{\rm F} \times n_{\rm F})   \;=\;  q(q-1) \,
   \left(\! \begin{array}{c}
               (q-1)^2 \\
               2(q-1) \\
               q-2
            \end{array} \!\right) ^{\! {\rm T}}
   \cdot T(4_{\rm F})^{n-1} \cdot
   \left(\! \begin{array}{c}
               1 \\
               0 \\
               0
            \end{array} \!\right)
   \;.
\ee

\begin{figure}
  \centering
  \epsfxsize=400pt\epsffile{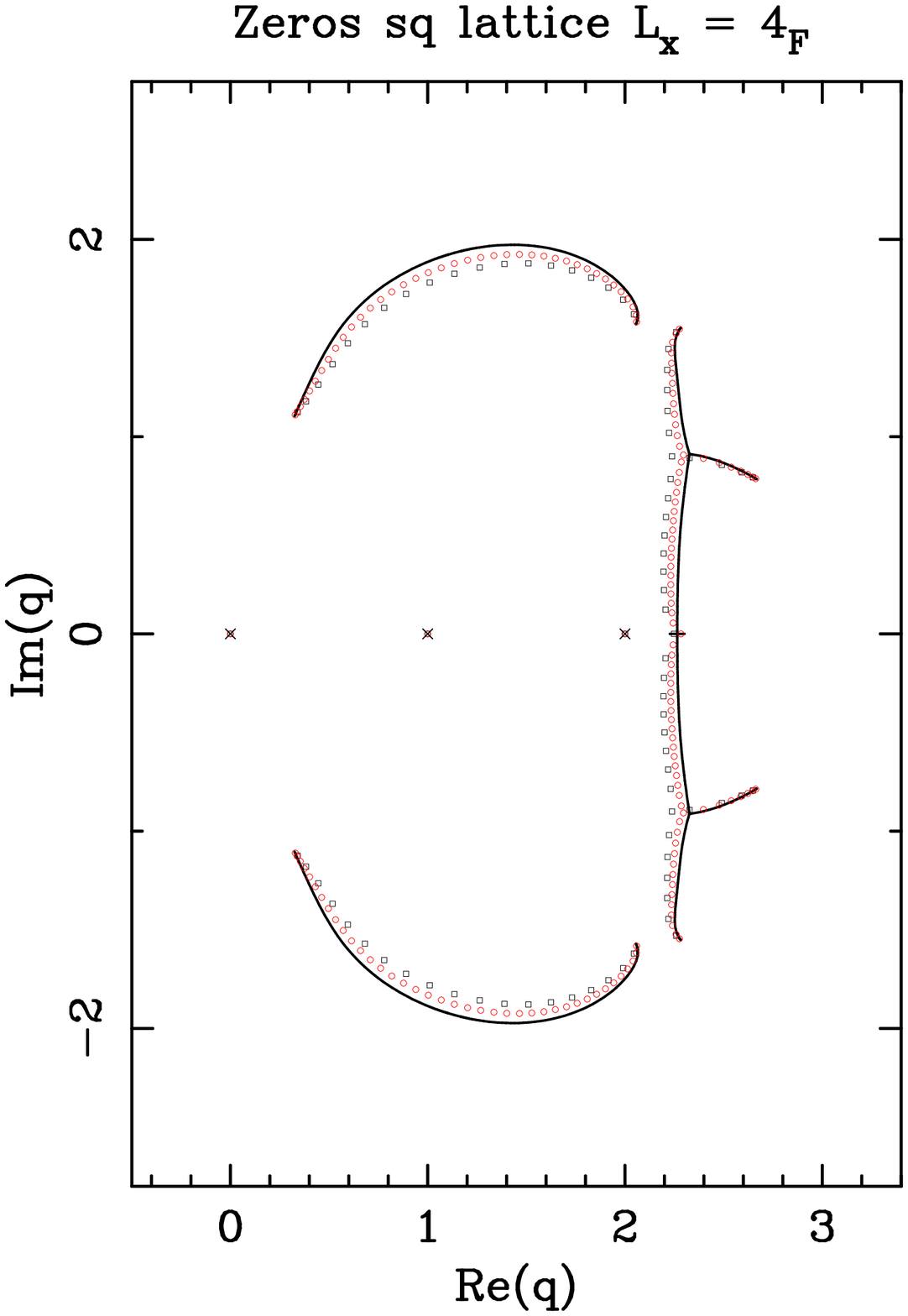}
  \caption{
  Zeros of the partition function of the $q$-state Potts antiferromagnet
  on a square lattices $4_F \times 20_F$ (squares),
  $4_F \times 40_F$ (circles) and $4_F\times\infty_F$ (solid line).
  The isolated limiting zeros are depicted by a $\times$.
  The limiting curve was computed using the resultant method.
  }
\label{Figure_sq_4FxInftyF}
\end{figure}

The limiting curve $\scrb$ (see Figure~\ref{Figure_sq_4FxInftyF})
has three connected components:
again, one crosses the real axis and is self-conjugate,
while the other two stay away from the real axis and are a pair of
mutually conjugate arcs.\footnote{
   This curve is also depicted in \cite[Figure 3(b)]{Shrock_98a}.
}
This time, however, the component that crosses the real axis
is rather complicated:
there is a pair of T points at $q \approx 2.327\pm0.9113\,i$,
and there is a double point on the real axis at $q \approx 2.2649418565$.

There are ten endpoints:
\begin{subeqnarray}
q &\approx& 0.3254743549 \pm 1.1048503376\,i \\
q &\approx& 2.0555822564 \pm 1.5703029256\,i \\
q &\approx& 2.2283590792                     \\
q &\approx& 2.2823594125 \pm 1.5512247035\,i \\
q &\approx& 2.3014157308                     \\
q &\approx& 2.6674264726 \pm 0.7845284722\,i
  \label{4F_endpoints}
\end{subeqnarray}
The horizontal segment emerging from the double point ends at
the pair of real endpoints (\ref{4F_endpoints}c,e).

Using Lemma~\ref{lemma2.2} we find that the points where
at least one amplitude vanishes are given by the zeros of the polynomial
\be
\det D(q) \;=\; 2 q^3 (q-1)^3 (q-2)^2 (q^2-3q+1) (2 q^3-13q^2+27q-17)^2
   \;.
 \label{det_D_4F}
\ee
The values $q=0,1$ are trivial zeros where all three amplitudes vanish
simultaneously.
At $q=2$ the amplitude corresponding to the leading eigenvalue vanishes
(as does one of the two subdominant amplitudes).
The other roots $q = (3 \pm \sqrt{5})/2$ and
$q \approx 1.170516, 2.664742 \pm 0.401127\,i$
all correspond to the vanishing of a subdominant amplitude.
Therefore, according to case (a) of the Beraha--Kahane--Weiss theorem,
the isolated limiting points for this strip are $q=0,1,2$.

{}From Table~\ref{table_zeros_free}, we see that the first non-trivial
real zero converges rapidly to the Beraha number $B_4 = 2$.
In addition, there are further real zeros (whose number increases with $n$)
that tend to the segment $[2.2283590792\ldots, 2.3014157308\ldots]$
of the limiting curve and in the limit $n\to\infty$ become dense
on that segment.

It is curious that \reff{det_D_4F} vanishes also at the Beraha number
$B_5 = (3 + \sqrt{5})/2$ and its conjugate $B_5^* = (3 - \sqrt{5})/2$,
even though both of these correspond to the vanishing
of a subdominant amplitude.

%
%
\subsection{$L_x = 5_{\rm F}$} \label{sec5F}

The transfer matrix is seven-dimensional, and can be expressed in the basis
${\bf P} = \{ 1, \delta_{13}+\delta_{35}, \delta_{24},
                  \delta_{14}+\delta_{25}, \delta_{15},
                  \delta_{135}, \delta_{15} \delta_{24} \}$.
We refrain from giving here the full transfer matrix;
instead, we refer the reader to the {\sc Mathematica} file {\tt transfer1.m}
included in the electronic version of this article at xxx.lanl.gov.

The limiting curve $\scrb$ (see Figure~\ref{Figure_sq_5FxInftyF})
has five connected components.
One of them crosses the real axis at $q_0 \approx 2.4284379020$
and is a self-conjugate arc;
there is a pair of mutually conjugate arcs;
and finally, there is a pair of mutually conjugate components exhibiting
T points at $q \approx 2.423\pm 0.1067\,i$ and
$q \approx 2.291\pm1.561\,i$.

There are 14 endpoints:
\begin{subeqnarray}
q &\approx& 0.1708973690 \pm 1.0464583589\,i \\
q &\approx& 1.9065720451 \pm 1.9339587717\,i \\
q &\approx& 1.9748200483 \pm 1.9395387106\,i \\
q &\approx& 2.3024178902 \pm 1.6190810539\,i \\
q &\approx& 2.3990745384 \pm 0.8206408701\,i \\
q &\approx& 2.4983799650 \pm 0.8199051472\,i \\
q &\approx& 2.7692051339 \pm 0.7143320949\,i
\end{subeqnarray}
Please note that the endpoints
$q \approx 1.9065720451 \pm 1.9339587717\,i$ and
$q \approx 1.9748200483 \pm 1.9395387106\,i$
define a pair of small gaps in the limiting curve.
Likewise, the curves emerging from the endpoints
$q \approx  2.3024178902 \pm 1.6190810539\,i$ and
$q \approx  2.3990745384 \pm 0.8206408701\,i$
do not cross, but define another pair of very small gaps
(see Figure~\ref{Figure_sq_5FxInftyF_bis} for detail).

By Lemma~\ref{lemma2.2}, the points where at least one amplitude vanishes
are given by the zeros of the polynomial
\be
   \det D(q)   \;=\;  q^7 (q-1)^{13} (q-2)^6 (q^2 -3q+1)^5 (q-3)^7 P(q)^2
 \label{det_D_5F}
\ee
where $P(q)$ is a polynomial of degree 37 with integer coefficients
that we report in the file {\tt transfer1.m};
as far as we can tell, $P(q)$ cannot be factored further over the integers.
The values $q=0,1$ are trivial zeros where all seven amplitudes vanish
simultaneously.  (Two of the amplitudes have in fact {\em multiple}\/ roots
at $q=1$.)
At $q=2$ the amplitude corresponding to the leading eigenvalue vanishes
(as do the amplitudes of five of the six subdominant eigenvalues).
In addition, one complex-conjugate pair of zeros of $P(q)$,
namely $q \approx 2.2866147868 \pm 1.0116506019\,i$,
corresponds to the vanishing of a dominant amplitude
(shown with a $\times$ in Figure~\ref{Figure_sq_5FxInftyF}).
This is the first time we find a nonreal isolated limiting zero.
The remaining zeros of $\det D(q)$ correspond to the vanishing
of subdominant amplitudes.

The first non-trivial real zero (see Table~\ref{table_zeros_free}) converges
quickly to the Beraha number $B_4=2$.
The next real zero appears to be converging slowly (at a roughly $1/n$ rate)
to the value $q_0 \approx 2.4284379020$ where the limiting curve $\scrb$
intersects the real axis, which is smaller than the next Beraha number $B_5$.
The convergence to the complex isolated limiting points at
$q \approx 2.2866147868 \pm 1.0116506019\,i$ is quite again rapid.
Indeed, if we select for each strip length $n$ the zero closest to
the limiting point, and fit the distance from the limiting point
to an inverse power of $n$, the effective power seems to grow with $n$;
this is compatible with the expected exponential convergence.

It is curious that \reff{det_D_5F} vanishes also at the Beraha numbers
$B_5 = (3 + \sqrt{5})/2$ and $B_6 = 3$ ---
and hence also at the conjugate $B_5^* = (3 - \sqrt{5})/2$ ---
even though all of these correspond to the vanishing of a subdominant amplitude.

%
%
\subsection{$L_x = 6_{\rm F}$} \label{sec6F}

The transfer matrix is 13-dimensional;
it can be found in the {\sc Mathematica} file {\tt transfer1.m}.

The limiting curve $\scrb$ (see Figure~\ref{Figure_sq_6FxInftyF})
has five connected components.
One of the components is self-conjugate and crosses the real axis
with a double point at $q \approx 2.5328721401$,
from which there emerges a small horizontal segment running from
$2.5286467909 \ltapprox q \ltapprox 2.5370979311$
(see Figure~\ref{Figure_sq_6FxInftyF_bis}a for detail).
The other four connected components form two mutually conjugate pairs.
One pair consists of arcs running from
$q \approx 0.0689480595 \pm 0.9874383424\,i$
to $q \approx 1.6648104050 \pm 2.1404062947\,i$.
The other pair exhibits T points at $q \approx 2.039\pm1.964\,i$,
$q \approx 2.332\pm 1.638\,i$ and $q \approx 2.478\pm 1.213\,i$,
the latter of which is the end of a small bulb-like region
enclosed by the limiting curve
(see Figures~\ref{Figure_sq_6FxInftyF_bis}b,c).
There are two small gaps at $q \approx 1.67\pm 2.14\,i$.

There are 14 endpoints:
\begin{subeqnarray}
q &\approx& 0.0689480595 \pm 0.9874383424\,i \\
q &\approx& 1.6648104050 \pm 2.1404062947\,i \\
q &\approx& 1.6870381566 \pm 2.1423501191\,i \\
q &\approx& 2.0370674106 \pm 1.9742433636\,i \\
q &\approx& 2.3334923547 \pm 1.6492963460\,i \\
q &\approx& 2.5286467909                     \\
q &\approx& 2.8373380200 \pm 0.6533586125\,i \\
q &\approx& 2.5370979311
\end{subeqnarray}

The bulb-like region is rather unusual.
The point at which it starts, $q \approx 2.478\pm 1.213\,i$,
really is a T point:
computations show that each of the three curves
of dominant equimodularity arriving at the T point
has a smooth continuation into the region beyond the T point,
where it becomes a curve of {\em subdominant}\/ equimodularity
(see Figure~\ref{Figure_sq_6FxInftyF_bis}c,
 where the dominant curves are shown with solid black lines and the
 subdominant curves with dashed \subdominantcolor\ lines);
moreover, the $t$ value varies continuously along each of these three curves.
One of the subdominant curves lies entirely within the enclosed
bulb-like region and ends at a $t=0$ subdominant endpoint
$q \approx 2.5018915620 \pm 1.1501464506\,i$.
This means that there is a branch cut for these two subdominant eigenvalues,
which become the dominant eigenvalues outside the bulb-like region.

By Lemma~\ref{lemma2.2}, the points where at least one amplitude vanishes
are given by the condition $\det D(q) = 0$. This determinant can be written
as
\be
\det D(q) \;=\; q^{13} (q-1)^{23} (q-2)^{12} (q^2-3q+1)^8 (q-3)^{27}
       (q^3 - 5q^2 +6q-1) P(q)^2
\label{det_D_6F}
\ee
where $P(q)$ is a polynomial of degree 218 with integer coefficients that we
report in the file {\tt transfer1.m}.
There are two trivial zeros $q=0,1$ where all the
amplitudes vanish simultaneously.
There are five non-trivial zeros of $\det D(q)$ that correspond
to the vanishing of the dominant amplitude:
$q=2$, $q\approx 2.0617791396 \pm 1.7315562279\, i$,
$q\approx 2.3406021969 \pm 1.3825644365\, i$
(shown with a $\times$ in Figure~\ref{Figure_sq_6FxInftyF}).

The first non-trivial real zero converges rapidly to the Beraha number $B_4=2$
(see Table~\ref{table_zeros_free}), while the second non-trivial zero
appears to be converging (at a roughly $1/n^2$ rate) to
the real endpoint $q \approx 2.5286467909$.
We expect that there will be further real zeros
(whose number increases with $n$)
that tend to the segment $[2.5286467909\ldots, 2.5370979311\ldots]$
of the limiting curve and in the limit $n\to\infty$ become dense
on that segment.
The convergence to the complex isolated limiting points at
$q \approx  2.0617791396 \pm 1.7315562279\, i$ and
$q\approx 2.3406021969 \pm 1.3825644365\, i$ is in both cases very fast and
is compatible with an exponential rate.

It is curious that \reff{det_D_6F} vanishes also at the Beraha numbers
$B_5 = (3 + \sqrt{5})/2$, $B_6 =3$ and $B_7\approx 3.246979603717$ ---
and hence also at their conjugates $B_5^* = (3 - \sqrt{5})/2$,
$B_7^{(2)} \approx 1.5549581321$ and $B_7^{(3)} \approx 0.1980622642$ ---
though all of these correspond to the vanishing of a subdominant amplitude.

%
%
\subsection{$L_x = 7_{\rm F}$} \label{sec7F}

The transfer matrix is 32-dimensional;
it can be found in the {\sc Mathematica} file {\tt transfer1.m}.

The limiting curve $\scrb$ (see Figure~\ref{Figure_sq_7FxInftyF})
appears to have seven connected components
(but there may be more: we cannot be sure, as we were unable to compute
 the $t=0$ resultant).
One of these components is a self-conjugate arc,
and crosses the real axis at $q_0 \approx 2.6062482130$;
it has endpoints at $q \approx 2.622974 \pm 0.609548\,i$.
The other six components form three mutually conjugate pairs.
One pair consists of arcs running from
$q\approx -0.002412 \pm 0.933080\,i$ to $q \approx 1.425603 \pm 2.248902\,i$.
A very small gap (see Figure~\ref{Figure_sq_7FxInftyF_bis}a for detail)
separates these arcs from the second pair,
which starts at the endpoint $q \approx 1.433184 \pm 2.248834\,i$
and ends in a tiny bulb-like region with a T point at
$q \approx 2.415 \pm  1.497\,i$ (see Figure~\ref{Figure_sq_7FxInftyF_bis}b).
A small gap (see Figure~\ref{Figure_sq_7FxInftyF_bis}c)
separates these components from the third pair,
which exhibits T points at $q \approx 2.577 \pm  1.133\,i$
and endpoints at $q \approx 2.616006 \pm 0.616910\,i$;
the latter is in turn separated by a very small gap from the
self-conjugate arc (see Figure~\ref{Figure_sq_7FxInftyF_bis}d).

Please note that $\scrb$ enters for the first time into the half-plane
$\real q < 0$;  we conjecture that this occurs for {\em all}\/
lattice widths $L_x \ge 7_{\rm F}$.

There are probably 14 endpoints:
\begin{subeqnarray}
q &\approx&           -0.002412 \pm 0.933080\,i \\
q &\approx& \phantom{-}1.425603 \pm 2.248902\,i \\
q &\approx& \phantom{-}1.433184 \pm 2.248834\,i \\
q &\approx& \phantom{-}2.445207 \pm 1.471332\,i \\
q &\approx& \phantom{-}2.616006 \pm 0.616910\,i \\
q &\approx& \phantom{-}2.622974 \pm 0.609548\,i \\
q &\approx& \phantom{-}2.886041 \pm 0.602908\,i
\end{subeqnarray}
We have determined these endpoints by the direct-search method;
therefore, they are less accurate than endpoints computed 
by the resultant method, and the list is possibly incomplete.

Due to limitations of CPU time,
we were unable to obtain the explicit expression for $\det D(q)$ as
a polynomial in $q$;  we were therefore unable to obtain all its roots.
Instead, we evaluated $\det D(q)$ numerically at selected values of $q$.
In particular, motivated by the results of the preceding subsections,
we computed $\det D(q)$ numerically at the first 50 Beraha numbers;
and in those cases where $\det D(q) = 0$, we numerically diagonalized
the transfer matrix in order to ascertain {\em which}\/ amplitude(s)
are the one(s) that vanish.
We find two trivial zeros $q=0,1$ where all the amplitudes
vanish simultaneously, and one non-trivial real zero $q=2$ where
the dominant amplitude vanishes (along with others).
In addition, $\det D(q)$ vanishes at the Beraha numbers
$B_5 = (3+\sqrt{5})/2$, $B_6=3$, $B_7\approx 3.246979603717$ and
$B_8 = 2 + \sqrt{2}$ (and hence also at their conjugates);
these all correspond to the vanishing of a subdominant amplitude.
Finally, inspection of Figure~\ref{Figure_sq_7FxInftyF}
suggests the existence of at least two pairs of complex-conjugate
isolated limiting points,
$q \approx 2.48873 \pm 0.75416\,i$ and $q \approx 1.65436 \pm 2.01881\,i$;
and we confirm that the absolute value of the dominant amplitude
is very small in both cases
($2.9 \times 10^{-11}$ and $2.1 \times 10^{-6}$, respectively),
suggesting that this amplitude does indeed have a zero nearby.
There might exist further isolated limiting points not found here.

The first non-trivial real zero converges quickly to the Beraha number
$B_4=2$ (see Table~\ref{table_zeros_free}), and the next
real zero converges (at a roughly $1/n$ rate) to the value of
$q_0 \approx 2.6062482130$ for this lattice,
which is slightly smaller than the Beraha number $B_5$.

%
%
\subsection{$L_x = 8_{\rm F}$} \label{sec8F}

The transfer matrix is 70-dimensional;
it can be found in the {\sc Mathematica} file {\tt transfer1.m}.

The limiting curve ${\cal B}$ (see Figure~\ref{Figure_sq_8FxInftyF})
appears to consist of six connected components
that form three mutually conjugate pairs (but there may be more).
One pair is defined by arcs running from
$q\approx -0.054426 \pm 0.884363\,i$ to   $q\approx 1.211959 \pm 2.301760\,i$.
A very small gap (see Figure~\ref{Figure_sq_8FxInftyF_bis}a)
separates these arcs from the second pair of arcs,
which run from  $q\approx 1.214531 \pm 2.301385\,i$
to $q\approx 2.326565 \pm 1.753667 \,i$.
Another small gap (see Figure~\ref{Figure_sq_8FxInftyF_bis}b)
separates these arcs from the third pair of components,
which run from $q \approx 2.330755  \pm 1.737504 \,i$
to $q\approx 2.660260 \pm 0.001257\,i$
and also have T points at $q\approx 2.640\pm 1.114\,i$
(see Figure~\ref{Figure_sq_8FxInftyF_bis}c).
Note that the limiting curve does {\em not}\/ cross the real axis:
the closest approach is $q\approx 2.660260 \pm 0.001257\,i$
(see Figure~\ref{Figure_sq_8FxInftyF_bis}d).

There are probably 14 endpoints:
\begin{subeqnarray}
q &\approx&           -0.054426  \pm 0.884363 \,i \\
q &\approx& \phantom{-}1.211959  \pm 2.301760 \,i \\
q &\approx& \phantom{-}1.214531  \pm 2.301385 \,i \\
q &\approx& \phantom{-}2.326565  \pm 1.753667 \,i \\
q &\approx& \phantom{-}2.330755  \pm 1.737504 \,i \\
q &\approx& \phantom{-}2.660260  \pm 0.001257 \,i \\
q &\approx& \phantom{-}2.921658  \pm 0.560969 \,i
\end{subeqnarray}
We have again determined these endpoints by the direct-search method.

Once again we were unable to compute the determinant $\det D(q)$
as a polynomial in $q$, so we followed the same numerical method as
in Section~\ref{sec7F} to locate at least some of the isolated limiting points.
There are of course two trivial isolated real zeros at $q=0,1$ where all
the amplitudes vanish simultaneously.
In addition, $\det D(q)$ vanishes at the Beraha numbers
$B_4 = 2$, $B_5=(3+\sqrt{5})/2$, $B_6=3$, $B_7\approx 3.246979603717$,
$B_8 = 2 + \sqrt{2}$ and $B_9\approx 3.5320888862$
(and hence also at their conjugates).
Unfortunately, we were unable to compute the corresponding amplitudes:
even using 4000-digit numerical precision,
the inverse of the change-of-basis matrix was obtained with
no precision at all!
Nevertheless, by inspection of Figure~\ref{Figure_sq_7FxInftyF} 
we can guess that $B_4$ and $B_5$ are indeed isolated limiting points
(see also the discussion below about the rate of convergence to those points),
while $B_6,B_7,B_8,B_9$ are not.
Inspection of Figure~\ref{Figure_sq_7FxInftyF} also suggests that there are
two possible pairs of complex-conjugate isolated limiting points, namely
$q\approx 1.33321 \pm 2.15164\,i$ and $q\approx 2.39242 \pm 1.11180\,i$;
but we are again unable to confirm this by a direct evaluation of the 
amplitudes at those points. 

The convergence of the first non-trivial real zero to $B_4 = 2$ is very rapid
(see Table~\ref{table_zeros_free}).
The second non-trivial real zero converges fairly rapidly
to the Beraha number $B_5 = (3+\sqrt{5})/2$,
which lies slightly below the point $q \approx 2.66$ where the limiting curve
comes close to (but does not cross) the real axis.
This rapid convergence suggests that $B_4$ and $B_5$
are indeed isolated limiting points.
This is the first square-lattice strip with free b.c.\ for which
$B_5$ appears as an isolated limiting point.

\section{Numerical Results for the Square-Lattice Chromatic Polynomial:
   \hfill\break Periodic Transverse Boundary Conditions}
   \label{sec6}

We have also computed the transfer matrix $T(q)$
and the limiting curves $\scrb$
for square-lattice strips of widths $2 \leq L_x \leq 8$
with periodic boundary conditions in the transverse direction.
We have checked our results for widths $2 \le m_{\rm P} \le 8$
and lengths $n_{\rm F} = 2,3$ by comparing to the results of
Biggs--Damerell--Sands \cite{Biggs_72}
(resp.\ Shrock--Tsai \cite{Shrock_99c,Shrock_00a})
for {\em width}\/ $n_{\rm F} = 2$ (resp.\ $n_{\rm F} = 3$)
and {\em length}\/ $m_{\rm P}$, using the trivial identity
\be
   Z(m_{\rm P} \times n_{\rm F})  \;=\;  Z(n_{\rm F} \times m_{\rm P})
   \;.
\ee

\subsection{$L_x = 2_{\rm P}$} \label{sec2P}

This is of course identical to $L_x = 2_{\rm F}$ (Section~\ref{sec2F}).

\subsection{$L_x = 3_{\rm P}$} \label{sec3P}

This case is also trivial, as the transfer matrix is one-dimensional.
The result is
\be
Z(3_{\rm P} \times n_{\rm F}) = q (q-1) (q-2) (q^3 - 6 q^2 + 14 q - 13)^{n-1}
\ee
The dominant-eigenvalue-crossing curve is of course
the empty set ${\cal B} = \emptyset$.
However, there are zeros for all $n$ at $q=0,1,2$ (trivially)
and for all $n \ge 2$ at
$q \approx 1.7733011742 \pm 1.4677115087\,i$ and $q \approx 2.4533976515$.

%
%
\subsection{$L_x = 4_{\rm P}$} \label{sec4P}

The transfer matrix is two-dimensional.
In the basis ${\bf P} = \left\{ 1, \delta_{13}+\delta_{24}\right\}$
it can be written as
\be
T(4_{\rm P}) \;=\; \left( \begin{array}{cc}
                    q^4 - 8 q^3 + 28 q^2 - 51 q + 41 & 2(q^3-6q^2+14q-12) \\
                    2q-5                             & q^2 - 4q +5
                      \end{array} \right)
  \;,
\ee
and the partition function is equal to
\be
Z(4_{\rm P} \times n_{\rm F})  \;=\;  q(q-1) \,
  \left(\! \begin{array}{c}
             q^2-3q+3 \\
             2(q-1)
          \end{array} \!\right) ^{\! {\rm T}}
   \cdot T(4_{\rm P})^{m-1} \cdot
   \left(\! \begin{array}{c}
             1 \\
             0
          \end{array} \!\right)
   \;.
\ee

The eigenvalues of the transfer matrix $T(4_{\rm P})$ and their
corresponding amplitudes are given by
\reff{eigenvalues2x2}/\reff{amplitudes2x2}/\reff{amplitudeproduct2x2}
with\footnote{
  Our results are in agreement with those of
  \protect\cite[eqns.~(7.1)--(7.4)]{Shrock_98c}.
  In particular, we have $b_{sq(4),c,1} = - \tr T$ and
  $b_{sq(4),c,2} = \det T$.
}
\begin{subeqnarray}
P_1(q) &=& q^4 - 8 q^3 + 29q^2 - 55q + 46 \\
P_2(q) &=& q^8 -16 q^7 + 118 q^6 - 526 q^5 + 1569 q^4 - 3250 q^3 +
                  4617 q^2 \nonumber \\
       & & \quad - 4136 q + 1776 \\
P_3(q) &=& q(q-1)(q^6 - 11q^5 + 54 q^4 -152 q^3 + 266 q^2 -277 q + 128)  \\
P_4(q) &=& 2 q^2 (q-1)^2 (q-2) (q^2 - 3q +1) (2q-5)^2
\end{subeqnarray}

The limiting curve $\scrb$ (see Figure~\ref{Figure_sq_4PxInftyF})
contains three pieces.\footnote{
   This curve is also depicted in \cite[Figure 3(a)]{Shrock_98c}.
}
It crosses the real axis at
the double point $q \approx 2.3026282864$ (with $t \approx 0.092$),
from which there emerges a horizontal segment running from
$q \approx 2.2533697671$ to $q \approx 2.3516882809$.

There are eight endpoints given by the zeros of the resultant at $t=0$
[$= -P_2(q)$]:
\begin{subeqnarray}
q &\approx& 0.7098031013 \pm 2.0427103451\,i \\
q &\approx& 1.9923366166 \pm 1.5941556425\,i \\
q &\approx& 2.2533697671                     \\
q &\approx& 2.3516882809                     \\
q &\approx& 2.9953312581 \pm 1.4266372190\,i
\end{subeqnarray}

The zeros of the amplitudes can be found by solving $P_4(q)=0$.
There are trivial zeros $q=0,1$, where both amplitudes vanish simultaneously;
both these zeros lie in the region where the eigenvalue $\lambda_+$ is dominant.
There are five non-trivial zeros: $q=2$, $5/2$ (double) and
$(3\pm \sqrt{5})/2$.  All of them lie in regions where there is
a unique dominant eigenvalue; but only for $q=2$ does the amplitude
corresponding to the dominant eigenvalue vanish.
The Beraha--Kahane--Weiss theorem thus implies that the isolated
limiting points of zeros are precisely $q=0,1,2$.
{}From  Table~\ref{table_zeros_cyl} we see that the first
non-trivial real zero does indeed converge rapidly to the Beraha number $B_4=2$.
In addition, there are further real zeros (whose number increases with $n$)
that tend to the segment $[2.2533697671\ldots, 2.3516882809\ldots]$
of the limiting curve and in the limit $n\to\infty$ become dense
on that segment.

It is again curious that the amplitude corresponding to the subdominant
eigenvalue vanishes at the Beraha number $B_5 = (3 + \sqrt{5})/2$
and its conjugate $B_5^* = (3 - \sqrt{5})/2$.

%
%
\subsection{$L_x = 5_{\rm P}$} \label{sec5P}

The transfer matrix is again two-dimensional.
In the basis ${\bf P} = \{ 1,\
  \delta_{13}+\delta_{24}+\delta_{35}+\delta_{41}+\delta_{52} \}$
it can be written as
\begin{eqnarray}
 & & T(5_{\rm P}) \;=\; \nonumber \\
 & & \quad \left( \begin{array}{cc}
  q^5 - 10q^4 + 45q^3 - 115q^2 + 169q -116 & 5(q^4- 9q^3+ 34q^2- 63q +47) \\
  q^2 - 6 q + 10                           & q^3 - 9 q^2 + 29 q  -32
                      \end{array} \right)  \;, \nonumber \\[2mm]
       & &
\end{eqnarray}
and the partition function is equal to
\be
Z(5_{\rm P} \times m_{\rm F}) \;=\; q(q-1)(q-2)\,
   \left(\!  \begin{array}{c}
                  q^2-2q+2 \\
                  5(q-1)
             \end{array} \!\right) ^{\! {\rm T}}
   \cdot T(5_{\rm P})^{m-1} \cdot
   \left(\! \begin{array}{c}
                  1 \\
                  0
            \end{array} \!\right)
   \;.
\ee

The eigenvalues of the transfer matrix $T(5_{\rm P})$ and their
corresponding amplitudes are given by
\reff{eigenvalues2x2}/\reff{amplitudes2x2}/\reff{amplitudeproduct2x2}
with\footnote{
  Our results are in agreement with those of
  \protect\cite[eqns.~(3.8)--(3.10)]{Shrock_00d}.
}
\begin{subeqnarray}
P_1(q) &=& q^5 - 10 q^4 + 46q^3 - 124q^2 + 198q - 148 \\
P_2(q) &=& q^{10} -20 q^9 + 188 q^8 - 1092 q^7 + 4356 q^6 - 12596 q^5 +
           27196 q^4 \nonumber \\
       & & \quad - 44212 q^3 + 52708 q^2 - 41760 q + 16456 \\
P_3(q) &=& q(q-1)(q-2)(q^7 - 12q^6 + 66 q^5 -214 q^4 + 450 q^3 \nonumber \\
       & & \quad -646 q^2 + 608q - 268)  \\
P_4(q) &=& 5 q^2 (q-1)^2 (q-2)^2 (q^2 - 3 q +1) (q-3) (q^2 - 6 q + 10)^2
\end{subeqnarray}

The limiting curve contains five pieces, and it
crosses the real axis at $q_0 \approx 2.6916837012$.\footnote{
   This curve is also depicted in \cite[Figure 2]{Shrock_00d}.
}
There are ten endpoints given by the zeros of the resultant at $t=0$
[$= -P_2(q)$]:
\begin{subeqnarray}
q &\approx& 0.1650212134 \pm 1.9190897717\,i \\
q &\approx& 2.0895893895 \pm 1.9436539472\,i \\
q &\approx& 2.5034648023 \pm 2.0851731765\,i \\
q &\approx& 2.5680063227 \pm 0.4886738235\,i \\
q &\approx& 2.6739182721 \pm 0.5983324603\,i
\end{subeqnarray}

The zeros of the amplitudes can be found by solving $P_4(q) = 0$.
There are trivial zeros at $q=0,1,2$, where both amplitudes
vanish simultaneously;  all of them lie in the region where
the eigenvalue $\lambda_{-}$ is dominant.
The non-trivial zeros of $P_4$ are $q=3$, $3\pm i$ (each of them double)
and $(3\pm\sqrt{5})/2$. Only for $q=(3+\sqrt{5})/2 = B_5$ does the amplitude
corresponding to the leading eigenvalue vanish. So the isolated zeros
are expected to converge when the strip length goes to infinity to the
first four Beraha numbers $B_2,\ldots,B_5=0,1,2,(3+\sqrt{5})/2$.

{}From Table~\ref{table_zeros_cyl} we see that the third real zero is
trivially equal to $B_4=2$ (a cylindrical square lattice of odd width
is not 2-colorable).  The first non-trivial zero converges rapidly
to the Beraha number $B_5$, while the last real zero seems to be converging
slowly (at a roughly $1/n$ rate) to the value $q_0 \approx 2.6916837012$
where the limiting curve $\scrb$ crosses the real axis.

The next Beraha number $B_6=3$ is also a zero of $P_4(q)$,
but this corresponds to the vanishing of the subdominant amplitude.

%
%
\subsection{$L_x = 6_{\rm P}$} \label{sec6P}

The transfer matrix is five-dimensional;
it can be found in the {\sc Mathematica} file {\tt transfer1.m}.

The limiting curve $\scrb$ (see Figure~\ref{Figure_sq_6PxInftyF})
has three connected components.
Two of them form a pair of mutually conjugate arcs,
which extend for the first time into the half-plane $\real q <0$.
The third component is self-conjugate and crosses the real axis
with a double point at $q \approx 2.6110856839$,
from which there emerges a small horizontal segment running from
$2.6089429411\ltapprox q \ltapprox 2.6132283584$
(see Figure~\ref{Figure_sq_6PxInftyF_bis}a for detail).
Note that this crossing lies slightly {\em below}\/
the Beraha number $B_5 = (3+\sqrt{5})/2 \approx 2.6180339887$.
This component also has T points at $q \approx 2.650 \pm 1.240\,i$
(see Figures~\ref{Figure_sq_6PxInftyF_bis}b,c for detail).

Inspection of Figure~\ref{Figure_sq_6PxInftyF} might lead one to think
that the self-conjugate component also exhibits a cusp at
$q \approx 2.568 \pm 1.395\,i$:
indeed, the curve $\scrb$ seems to have a discontinuous derivative there.
Moreover, there is no nearby dominant endpoint
(which rules out the alternative hypothesis of a T point from which
there emerges a very short curve terminating at a dominant endpoint);
and anyway the value of $t$ is also reasonably continuous around the
alleged cusp (providing further evidence against the idea of a T point).
However, by zooming on this region we can see that it is actually
a single smooth curve (see Figure~\ref{Figure_sq_6PxInftyF_bis}b,c,d).
What happens is that there is a {\em subdominant}\/ endpoint very close
to this curve
(the subdominant curve is shown with dashed \subdominantcolor\ lines);
moreover, the subdominant eigenvalues are very close in modulus
to the dominant ones. In particular, at the subdominant endpoint we have
\begin{subeqnarray}
\lambda_{\rm dom} &=& 10.4372144110 \, \exp(-0.9719979634\,i) \\
\lambda_{\rm sub} &=& 10.4046337888 \, \exp(-2.8500677201\,i)
\end{subeqnarray}
Near the subdominant endpoint, the two subdominant eigenvalues
are very rapidly changing (as they have a square-root branch point),
so that their crossing in modulus with the dominant eigenvalue
occurs on a curve $\scrb$ of rapidly changing slope.

There are ten endpoints:
\begin{subeqnarray}
q &\approx&           -0.1318891429 \pm 1.7132242811\,i \\
q &\approx& \phantom{-}1.9257517021 \pm 2.2876287010\,i \\
q &\approx& \phantom{-}2.0571168133 \pm 2.3885607275\,i \\
q &\approx& \phantom{-}2.6089429411                     \\
q &\approx& \phantom{-}2.6132283584                     \\
q &\approx& \phantom{-}3.1711921718 \pm 0.8639071723\,i
\end{subeqnarray}

By Lemma~\ref{lemma2.2}, the points where at least one amplitude vanishes
are given by the condition $\det D(q) = 0$. This determinant can be written as
\be
\det D(q) \;=\; q^5 (q-1)^5 (q-2)^4 (q^2-3q+1)^2 (q-3)^3 (q^3-5q^2+6q-1) P(q)^2
\label{det_D_6P}
\ee
where $P(q)$ is a polynomial of degree 28 with integer coefficients
that we report in the file {\tt transfer1.m}.
We find two trivial zeros
$q=0,1$ where all the amplitudes vanish simultaneously.
There are three points where a dominant amplitude vanishes:
$q=2$ and $q \approx 2.4813444277 \pm 1.7147613188\,i$
(shown with a $\times$ in Figure~\ref{Figure_sq_6PxInftyF}).
The complex-conjugate pair of isolated limiting zeros lies
extremely near, {\em but not on}\/, the limiting curve $\scrb$
(namely, about 0.006 to the left of $\scrb$). The rate of convergence to
the complex isolated limiting points is roughly $1/n$ instead of the
expected exponential rate. This may be due to the fact that they are
extremely close to the limiting curve ${\cal B}$.

{}From Table~\ref{table_zeros_cyl} we see that the first non-trivial real zero
converges rapidly to the Beraha number $B_4=2$,
while the second non-trivial zero appears to be converging slowly
(at a rate somewhere between $1/n$ and $1/n^2$)
to the real endpoint $q \approx 2.6089429411$.
We expect that there will be further real zeros
(whose number increases with $n$)
that tend to the segment $[2.6089429411\ldots, 2.6132283584\ldots]$
of the limiting curve and in the limit $n\to\infty$ become dense
on that segment.

It is curious that \reff{det_D_6P} vanishes also at the Beraha numbers
$B_5 = (3 + \sqrt{5})/2$, $B_6 =3$ and $B_7\approx 3.246979603717$
and their conjugates, even though all of these correspond to the vanishing
of a subdominant amplitude.

%
%
\subsection{$L_x = 7_{\rm P}$} \label{sec7P}

The transfer matrix is six-dimensional;
it can be found in the {\sc Mathematica} file {\tt transfer1.m}.

The limiting curve $\scrb$ (see Figure~\ref{Figure_sq_7PxInftyF})
has seven connected components.
One of them is a self-conjugate arc that crosses the real axis at
$q_0 \approx 2.7883775115$,
which lies for the first time {\em above}\/ the
Beraha number $B_5 = (3+\sqrt{5})/2 \approx 2.6180339887$;
this arc has endpoints at $q\approx 2.7618995071 \pm 0.4693560083\,i$.
The other six components form three pairs of mutually conjugate components.
The first pair are arcs running from
$q \approx -0.2962497164 \pm 1.5256077564\,i$ to
$q \approx 1.6542262925 \pm 2.4866235231\,i$,
and thus have support on $\real q <0$.
The second pair are also arcs, running from
$q\approx 1.6947007027 \pm 2.5327609879\,i$ to
$q\approx 2.6589962013 \pm 1.5245516751\,i$.
The last pair of components has endpoints at
$q\approx 2.7275004011 \pm 1.4172937300\,i$,
$q\approx 2.7873170476 \pm 0.4754613769\,i$ and
$q\approx 2.8390155832 \pm 1.3872842928\,i$,
and T points at $q \approx 2.737 \pm 1.405\,i$
(see Figure~\ref{Figure_sq_7PxInftyF_bis}a,b);
it is separated from the self-conjugate arc by a very small gap
(see Figure~\ref{Figure_sq_7PxInftyF_bis}c).

There are 16 endpoints:
\begin{subeqnarray}
q &\approx&           -0.2962497164 \pm 1.5256077564\,i \\
q &\approx& \phantom{-}1.6542262925 \pm 2.4866235231\,i \\
q &\approx& \phantom{-}1.6947007027 \pm 2.5327609879\,i \\
q &\approx& \phantom{-}2.6589962013 \pm 1.5245516751\,i \\
q &\approx& \phantom{-}2.7275004011 \pm 1.4172937300\,i \\
q &\approx& \phantom{-}2.7618995071 \pm 0.4693560083\,i \\
q &\approx& \phantom{-}2.7873170476 \pm 0.4754613769\,i \\
q &\approx& \phantom{-}2.8390155832 \pm 1.3872842928\,i
\end{subeqnarray}

By Lemma~\ref{lemma2.2}, the points where at least one amplitude vanishes
are given by the condition $\det D(q) = 0$. This determinant can be written as
\begin{eqnarray}
\det D(q) &=& q^6 (q-1)^6 (q-2)^6 (q^2-3q+1)^4 (q-3)^3 (q^3-5q^2+6q-1) \times
          \nonumber \\
        & & \qquad (q^2-4q+2) P(q)^2
\label{det_D_7P}
\end{eqnarray}
where $P(q)$ is a polynomial of degree 56 with integer coefficients. This
polynomial can be found in the file {\tt transfer1.m}.
The trivial isolated limiting points are $q=0,1,2$, where all the
amplitudes vanish simultaneously. The non-trivial isolated limiting
points are the Beraha number $B_5 = (3+\sqrt{5})/2$ and the pair of
complex-conjugate roots $q \approx 2.1027473746 \pm 2.2083820861\,i$.
This is the first square-lattice strip with cylindrical b.c.\ for which
$B_5$ appears as an isolated limiting point.

There is trivially a real zero at $q=2$ (because the strip width is odd).
The first non-trivial real zero (see Table~\ref{table_zeros_cyl})
converges rapidly to the Beraha number $B_5 = (3+\sqrt{5})/2$,
while the next real zero appears to converge slowly
(at a roughly $1/n$ rate) to the value $q_0 \approx 2.7883775115$
where the limiting curve $\scrb$ crosses the real axis.
The rate of convergence to the complex isolated limiting points
$q \approx 2.1027473746 \pm 2.2083820861\,i$ is very fast and is compatible
with an exponential rate.

It is curious that \reff{det_D_7P} vanishes also at the Beraha numbers
$B_6 =3$, $B_7\approx 3.246979603717$ and $B_8=2+\sqrt{2}$
and their conjugates, even though all of these correspond to the vanishing
of a subdominant amplitude.

%
%
\subsection{$L_x = 8_{\rm P}$} \label{sec8P}

The transfer matrix is 14-dimensional;
it can be found in the {\sc Mathematica} file {\tt transfer1.m}.

The limiting curve $\scrb$ (see Figure~\ref{Figure_sq_8PxInftyF})
has six connected components,
which define three pairs of mutually conjugate components.
The first pair is defined by arcs running from
$q\approx -0.3908638747 \pm 1.3698634697\,i$ to
$q\approx 1.3863697070 \pm 2.5801346584\,i$,
which thus have support on $\real q <0$.
A small gap (see Figure~\ref{Figure_sq_8PxInftyF_bis}a)
separates these components from the second pair,
which consists of arcs running from
$q\approx 1.3989312933 \pm 2.5988401222\,i$ to
$q\approx 2.5297861557 \pm 1.8426263238\,i$.
The last pair has endpoints at
$q\approx 2.5810431815 \pm 1.8106192070\,i$,
$q\approx 2.7515311636 \pm 0.0025313231\,i$,
$q\approx 2.7812812528 \pm 1.0876657311\,i$ and
$q\approx 3.2111321566 \pm 0.6498638896\,i$,
and T points at $q \approx 2.801\pm 1.043\,i$
and $q \approx 2.783 \pm 1.088\,i$.
Note that the T points at $q \approx 2.783 \pm 1.088\,i$
might look like cusps if we fail to use enough magnification
(see Figure~\ref{Figure_sq_8PxInftyF_bis}b).
But if we magnify the region sufficiently,
we observe that it is indeed an ordinary T point
(see Figure~\ref{Figure_sq_8PxInftyF_bis}c).
Note also that $\scrb$ does not cross the real axis at any point;
rather, there is a very tiny gap between
$q \approx 2.7515311636 \pm 0.0025313231\,i$
(see Figure~\ref{Figure_sq_8PxInftyF_bis}d).

There are 16 endpoints:
\begin{subeqnarray}
q &\approx&           -0.3908638747 \pm 1.3698634697\,i \\
q &\approx& \phantom{-}1.3863697070 \pm 2.5801346584\,i \\
q &\approx& \phantom{-}1.3989312933 \pm 2.5988401222\,i \\
q &\approx& \phantom{-}2.5297861557 \pm 1.8426263238\,i \\
q &\approx& \phantom{-}2.5810431815 \pm 1.8106192070\,i \\
q &\approx& \phantom{-}2.7515311636 \pm 0.0025313231\,i \\
q &\approx& \phantom{-}2.7812812528 \pm 1.0876657311\,i \\
q &\approx& \phantom{-}3.2111321566 \pm 0.6498638896\,i
\end{subeqnarray}

By Lemma~\ref{lemma2.2}, the points where at least one amplitude vanishes
are given by the condition $\det D(q) = 0$. This determinant is given by
\begin{eqnarray}
\det D(q) &=& q^{14}(q-1)^{20}(q-2)^{13}(q^2-3q+q^2)^{11}(q-3)^{18}
       (q^3-5q^2+6q-1)^4 \times \nonumber \\
          & & \qquad (q^2-4q+2) (q^3-6q^2+9q-1) P(q)^2
\label{det_D_8P}
\end{eqnarray}
where $P(q)$ is a polynomial of degree 396 with integer coefficients
(see file {\tt transfer1.m}).
The trivial isolated limiting points are $q=0,1$:
at these points all the amplitudes vanish simultaneously.
The non-trivial isolated limiting points are the
Beraha numbers  $B_4 = 2$ and $B_5 = (3+\sqrt{5})/2$
and the complex-conjugate pairs
$q \approx 1.6836371202 \pm 2.4533856271\,i$ and
$q \approx 2.6775096551 \pm 1.2084144891\,i$
(see Figure~\ref{Figure_sq_8PxInftyF}).

The first non-trivial real zero converges rapidly to the Beraha number
$B_4=2$, and the next real zero converges rapidly to the Beraha number
$B_5=(3+\sqrt{5})/2$ (see Table~\ref{table_zeros_cyl}).
The convergence to the complex isolated limiting points
$q \approx 1.6836371202 \pm 2.4533856271\,i$ and
$q \approx 2.6775096551 \pm 1.2084144891\,i$ is also very fast
and is compatible with an exponential rate.

It is curious that \reff{det_D_8P} vanishes also at the Beraha numbers
$B_6 =3$, $B_7\approx 3.246979603717$, $B_8=2+\sqrt{2}$ and
$B_9\approx 3.532088886238$ and their conjugates,
even though all of these correspond to the vanishing
of a subdominant amplitude.

\section{Discussion and Open Questions}   \label{sec7}

\subsection{Behavior of dominant-eigenvalue-crossing curves $\scrb$}
   \label{sec7.1}

In this paper we have computed the chromatic polynomials
(= zero-temperature antiferromagnetic Potts-model partition functions)
$P_G(q)$ and their zeros for square-lattice strips of width
$2 \leq L_x \leq 8$ and arbitrary length $L_y$
with free and cylindrical boundary conditions.
In particular, we have extracted the limiting curves $\scrb$
of partition-function zeros when the length $L_y$ goes to infinity
at fixed width $L_x$.
By studying the finite-width limiting curves and their behavior
as we increase the width $L_x$,
we hope to shed light on the thermodynamic limit $L_x, L_y \to \infty$.

In Table~\ref{table_summary} we summarize the main properties
of the limiting curves ${\cal B}$ (and of the isolated limiting points)
for all the lattices studied in the previous sections.
Note the identity
\begin{eqnarray}
   \hbox{endpoints}   & = &
   (2 \times \hbox{components})  \,+\,
   (2 \times \hbox{double points})  \,+\,
   (\hbox{T points})  \,-\,
   \nonumber \\
   & & \qquad
   (2 \times \hbox{enclosed regions})
   \;,
\end{eqnarray}
which can be derived by simple topological/graph-theoretic arguments.

Our first conclusion is that the limiting curves become in general
more complicated as the strip width $L_x$ grows.
In particular, the number of connected components,
the number of endpoints, and the number of T points
all tend to increase with the width $L_x$.
(Note that our counts for $L_x = 7_{\rm F}, 8_{\rm F}$
are only lower bounds on the true values.)
Moreover, the size of the gaps between connected components,
and the lengths of the protruding arcs associated to some of the T points,
both seem to decrease with the strip width $L_x$.
The approach to the thermodynamic limit thus appears to be
rather complicated.

A second point deals with the existence or not of enclosed regions.
Shrock \cite[Section III, point 3]{Shrock_99g}
conjectured that for families of graphs with a well-defined lattice
structure, a {\em sufficient}\/ condition for $\scrb$ to separate
the $q$-plane into two or more regions is that the graphs contain
at least one ``global circuit'', defined as a route following
a lattice direction which has the topology of $S^1$ and a length that goes
to infinity as $L_y\rightarrow \infty$.
(For strip graphs, this condition is equivalent to having
 periodic boundary conditions in the longitudinal direction.)
Our results for $L_x = 6_{\rm F}, 7_{\rm F}$ show that this condition,
whether or not it is in fact sufficient for the existence of
enclosed regions, is in any case not {\em necessary}\/:
enclosed regions can arise also with {\em free}\/ longitudinal b.c.\footnote{
   For earlier examples showing that enclosed regions can arise also with
   free longitudinal b.c.,
   see \cite[Figures 2(b), 3(b) and 4(a,b)]{Shrock_98c}.
}
In both these cases, the enclosed regions are small bulb-like regions
located at the end of one of the components of $\scrb$.
Enclosed regions do, however, seem to be atypical for
square-lattice strips with free longitudinal b.c.

A third point concerns the existence of chromatic zeros with $\real q  < 0$,
and more specifically of limiting curves $\scrb$ that intersect the half-plane
$\real q  < 0$.
In 1980, Farrell \cite{Farrell_80} conjectured,
based on computations with small graphs,
that all chromatic roots have $\real q \ge 0$.
We now know that this conjecture is false \cite{Read_91};
indeed, there are families of graphs whose chromatic roots,
taken together, are dense in the whole complex $q$-plane
\cite{Sokal_hierarchical}.
Nevertheless, chromatic roots do seem to have a tendency to avoid
the left half-plane, and it would be interesting to know why.
Our computations show that
for $L_x = 7_{\rm F}, 8_{\rm F}$ and $L_x = 6_{\rm P}, 7_{\rm P}, 8_{\rm P}$,
the locus $\scrb$ does intersect the half-plane $\real q  < 0$.
Indeed, we conjecture that this happens for square-lattice strips of
{\em all}\/ widths $L_x \ge 7$ (free b.c.) or $L_x \ge 6$ (cylindrical b.c.).

Although the limiting curves get more complicated as the strip width
$L_x$ grows, they do exhibit some regularities, as can be seen in
Table~\ref{table_summary} and
Figures~\ref{Figure_limit_F_all} and \ref{Figure_limit_P_all}.
In Figure~\ref{Figure_limit_F_all} we superpose the limiting curves
for all the square-lattice strips with free boundary conditions.
In the leftmost part of the plot ($\real q \ltapprox 2$),
we see that the arcs behave monotonically:
as the width $L_x$ increases, the corresponding arc moves outwards.
In particular, $\min \real q$ decreases monotonically with the  width $L_x$
(see Table~\ref{table_summary}).
A similar behavior is observed in the right part of the plot for $L_x \ge 4$:
the limiting curves have similar shapes and
they move monotonically to the right as $L_x$ grows.
In particular, the point $q_0$ where $\scrb$ crosses the real axis
increases monotonically with the strip width,
as does $\max \real q$ (see again Table~\ref{table_summary}).
In general, the shapes of the limiting curves look roughly similar
to those obtained by Baxter \cite[Figures 5 and 6]{Baxter_87}
for the triangular lattice.
We conjecture that the rightmost endpoints of $\scrb$,
which lie at $q \approx 2.92 \pm 0.56\,i$ for $L_x = 8$,
will tend (slowly) to close up at the critical value $q_c = 3$
as $L_x \to\infty$.
We also conjecture, in analogy with the triangular lattice,
that the values $q_0$ will tend to a number strictly less than 3,
probably somewhere around $2.9$.
However, our strip widths are still too small to give
unambiguous evidence for or against these conjectures.

In Figure~\ref{Figure_limit_P_all} we show the limiting curves for all
the square-lattice strips with cylindrical boundary conditions.
In the leftmost part of the plot ($\real q \ltapprox 2$),
the behavior of the arcs is again monotonic in the strip width;
in particular, $\min \real q$ is again a decreasing function of $L_x$
(see Table~\ref{table_summary}).
However, the qualitative behavior of the limiting curves
on the right side of the plot ($\real q \gtapprox 2.3$)
is clearly {\em not}\/ monotonic:  there is a notorious difference
between strips with even width and those with odd width.
This difference is, in fact, to be expected:
with periodic transverse boundary conditions,
odd widths are in some sense ``unnatural'' as they introduce
frustration in the antiferromagnetic Ising system
(i.e.\ they make the chromatic number 3 rather than 2,
 thereby forcing a chromatic zero at $q=2$).
It is curious that the difference between even and odd widths is
significant only on the rightmost part of the limiting curve
(namely, the part nearest $q=2$ and $q=3$).
In any case, if we consider the even and odd subsequences separately,
then $q_0$ and $\max \real q$ are again monotonically increasing
functions of $L_x$ (see Table~\ref{table_summary}).
Moreover, the limiting curves for
the square-lattice strips with {\em even}\/ width
and cylindrical boundary conditions have,
apart from the case $L_x=4$, a qualitative shape in agreement with
that for free boundary conditions.

Finally, let us compare the limiting curves for
free and cylindrical boundary conditions
(Figures~\ref{Figure_limit_F_all} and \ref{Figure_limit_P_all}).
If we focus on the leftmost part of the plots,
we see that both sets of curves tend to larger values of $|\!\imag q|$
as $L_x$ grows, but the curves for cylindrical boundary conditions reach
large values of $|\!\imag q|$ much faster.
Likewise, on the rightmost part of the plots, both sets of curves
tend to larger values of $\real q$ as $L_x$ grows
(modulo the even-odd oscillation for cylindrical b.c.),
but the curves for cylindrical boundary conditions
do so somewhat faster.  This suggests that the thermodynamic limit
is achieved faster with cylindrical boundary conditions than with
free boundary conditions.

Let us conclude by mentioning briefly the work of Bakaev and Kabanovich
\cite{Bakaev_94}, who computed the large-$q$ series for the
infinite-volume limiting chromatic polynomial through order $z^{36}$
[where $z=1/(q-1)$], using the finite-lattice method.
Tony Guttmann (private communication) has kindly analyzed their series
using differential approximants.
He finds that the nearest singularity to $z=0$ lies at
$z \approx (0.155 \pm 0.005) \pm (0.37 \pm 0.03)i$,
corresponding to $q \approx (1.96 \pm 0.15) \pm (2.30 \pm 0.15)i$.
Curiously enough, this corresponds quite closely to a gap
(pair of nearby endpoints) for width $6_{\rm P}$ (see Section~\ref{sec6P}),
where the limiting chromatic polynomial is indeed singular.
But this may be a coincidence, as there is no corresponding endpoint
for width $8_{\rm P}$,
and the nearby endpoints for strips with free b.c.\ do not seem to be
converging to this value.\footnote{
   For $5_{\rm F}$ there is a gap around $z \approx 0.20 \pm 0.41\,i$,
   but as the width increases, the real part decreases {\em beyond}\/
   the predicted singularity
   (0.13 for $6_{\rm F}$, 0.08 for $7_{\rm F}$, 0.04 for $8_{\rm F}$).
}
The large-$q$ series shows no hint of singularity at $q_c = 3$,
which is very likely a weak essential singularity.

\subsection{Behavior of amplitudes and the Beraha conjecture}  \label{sec7.2}

For all the lattices we studied (up to width $L=8$),
we observed empirically that there is at least one vanishing amplitude
$\alpha_i(q)$ at each of the Beraha numbers up to $B_{L+1}$.
It is reasonable to conjecture that this holds for all $L$:

\begin{conjecture}
 \label{conj7.1}
For a square-lattice strip of width $L$ with free or cylindrical boundary
conditions, at each Beraha number $q = B_2,\ldots,B_{L+1}$
there is at least one vanishing amplitude $\alpha_i(q)$.
That is, $\det D(q) = 0$ for $q = B_2,\ldots,B_{L+1}$.
\end{conjecture}

For all the cases we studied except $7_{\rm F}$ and $8_{\rm F}$
[where we were unable to compute an explicit expression for $\det D(q)$],
we also verified that none of the roots of $\det D(q)$
correspond to Beraha numbers {\em beyond}\/ $B_{L+1}$.
We conjecture that this holds for all $L$:

\begin{conjecture}
 \label{conj7.2}
For a square-lattice strip of width $L$ with either free or cylindrical
boundary conditions, $\det D(q) \neq 0$ for all $q=B_k$ with $k > L+1$.
[We assume, of course, that the chromatic polynomial is written
 in such a way that there are no {\em identically}\/ vanishing amplitudes.]
\end{conjecture}

\noindent
In some cases we found real roots of $\det D(q)$ that are
{\em very close to}\/ Beraha numbers:
for instance, the strip with $L=5_{\rm F}$ (resp. $L=6_{\rm F}$)
has a zero of $\det D(q)$ very close to $B_7$  (resp. $B_9$ and $B_{29}$).
Moreover, $\det D(q)$ does in general have many real roots with $q > B_{L+1}$
(and indeed with $q>4$), as well as real roots with $q<0$.

We can strengthen the preceding conjecture to assert that
$\det D(q)$ is {\em strictly positive}\/
at the Beraha numbers beyond $B_{L+1}$:

\begin{conjecture}
 \label{conj7.3}
For a square-lattice strip of width $L$ with either free or cylindrical
boundary conditions, $\det D(q) >0$ for all $q=B_k$ with $k > L+1$.
\end{conjecture}

\noindent
We have verified this conjecture for all square-lattice strips
with $L\leq 7_{\rm F}$ and $L \le 8_{\rm P}$ up to $k=50$.
Furthermore, the function $f(k) = \det D(B_k)$ seems to be
a monotonically increasing function of $k$ for $k > L+1$.

\bigskip

Conjecture~\ref{conj7.1} asserts that at each of the Beraha numbers
$B_2,\ldots,B_{L+1}$ there is at least one vanishing amplitude $\alpha_i(q)$,
but it says nothing about whether the vanishing amplitude
belongs to a dominant or a subdominant eigenvalue.
Basing ourselves on a suggestion of Baxter \cite[p.~5255]{Baxter_87},
we conjectured that at each Beraha number $q = B_2,\ldots,B_{L+1}$,
the amplitude $\alpha_*(q)$ corresponding to the eigenvalue $\lambda_*(q)$
vanishes:
here $\lambda_*(q)$ is the eigenvalue that is dominant at small real $q$
(e.g.\ at $q=1$), analytically continued up the real axis
(or, if there is a branch point on the real axis,
then just above or below the real axis).
[Note that $\lambda_*(q)$ remains dominant until the path of analytic
 continuation crosses $\scrb$;  after that, it becomes in general
 subdominant.]
We tested this conjecture numerically as follows:
Choose a path in the complex $q$-plane starting at $q=1$
and proceeding to the right just above or below the real axis
(note that for $8_{\rm F}$ and $8_{\rm P}$ we must keep
$|\!\imag q| \ltapprox 0.001$ in order to avoid going around
an endpoint).\footnote{
   Any two such paths give the same analytic continuation
   provided that there are no endpoints in the region between them.
}
Subdivide this path into very small steps,
and at each point compute the eigenvalues of the transfer matrix,
following $\lambda_*(q)$ ``by continuity''
(that is, starting with $\lambda_*(1)$, choose at each $q$ value the eigenvalue
that is closest to the one chosen at the preceding $q$ value).
Then, when the Beraha number $B_k$ is reached,
diagonalize the transfer matrix, rotate the left and right vectors
$\vec{u}(q)$ and $\vec{v}(q)$,
and test whether the amplitude $\alpha_*(q)$ vanishes.

We were unable to carry out this computation for
$L = 7_{\rm F}$ and $8_{\rm F}$:
in the former case, because of CPU-time limitations,
and in the latter case, because of loss of numerical precision
in diagonalizing the matrices (even when we used 1000-digit arithmetic).
In all but one of the other cases
(namely, for $L \le 6_{\rm F}$ and $L \le 7_{\rm P}$),
we found that $\alpha_*(q)$ does indeed vanish 
(sometimes along with other amplitudes) at all the Beraha numbers up to 
$B_{L+1}$.
The strips with $L = 8_{\rm F}$ and $8_{\rm P}$ are, however,
a different case, as the limiting curve ${\cal B}$ does not cross
the real axis. If we follow a path from $q=1$ to $q=+\infty$
along the real axis, we do not cross ${\cal B}$;
so $\lambda_*(q)$ stays dominant everywhere on the real axis
(and in particular at the Beraha numbers $B_6,\ldots,B_9$).
Therefore, if $\alpha_*(q)$ were to vanish at $B_6,\ldots,B_9$,
as our conjecture asserts,
then those Beraha numbers would be isolated limiting points.
But Figures~\ref{Figure_sq_8FxInftyF} and \ref{Figure_sq_8PxInftyF}
show clearly that they are not!
So our conjecture must be {\em false}\/ for the strips
$8_{\rm F}$ and $8_{\rm P}$.
(Indeed, for $8_{\rm P}$ we confirmed explicitly that $\alpha_*$ vanishes
 at $B_2,\ldots,B_5$ but {\em not}\/ at $B_6,\ldots,B_9$;
 rather, it is a {\em subdominant}\/ amplitude that vanishes
 at the latter points.)
More generally, we can expect our conjecture to be false
whenever $\scrb$ does not cross the real axis.\footnote{
   For the same reason, if in the cases $L \le 6_{\rm F}$ and
   $L \le 7_{\rm P}$ we analytically continue the dominant eigenvalue
   at $q=1$ (namely, $\lambda_*$) to the region $q> q_0(L)$
   by following a path that does {\em not} cross ${\cal B}$
   --- e.g., by going through one of the gaps in ${\cal B}$ ---
   then the analytic continuation of $\lambda_*$  will remain dominant
   everywhere. But then, the analytic continuation of the amplitude
   $\alpha_*$ will {\em not}\/ vanish at any of the Beraha numbers to the
   right of ${\cal B}$, since we know that these are {\em not}\/
   isolated limiting points.  Rather, the analytic
   continuation of one or more {\em sub}\/dominant amplitudes at $q=1$
   will vanish at those Beraha numbers.
   We have numerically tested this behavior of the eigenvalues and
   amplitudes along paths that do not cross ${\cal B}$
   in the same way we did for the paths that do cross ${\cal B}$.
}
We are therefore obliged to modify our conjecture as follows:

\begin{conjecture}
  \label{conj7.4}
For a square-lattice strip of width $L$ with free or cylindrical boundary
conditions,
let $\lambda_*(q)$ be the eigenvalue that is dominant at small real $q$
(e.g.\ at $q=1$),analytically continued up the real axis
(or, if there is a branch point on the real axis,
then just above or below the real axis).
Then, {\em provided that the limiting curve ${\cal B}$ crosses the real axis,}
at each Beraha number $q = B_2,\ldots,B_{L+1}$,
the corresponding amplitude $\alpha_*(q)$ vanishes.
[Other amplitudes may vanish as well.]
\end{conjecture}

\noindent
All our numerical evidence is consistent with
Conjecture~\ref{conj7.4}.

\medskip 

\noindent 
{\bf Remark.} 
In the case $L = 8_{\rm P}$, suppose that we follow a path 
that {\em does}\/ cross the limiting curve ${\cal B}$
slightly above the endpoint at $q \approx 2.7515 + 0.0025\,i$.
Then the eigenvalue $\lambda_*$ will cease 
to be dominant to the right of ${\cal B}$;
and we find that the corresponding amplitude $\alpha_*$ 
{\em does}\/ vanish at all the Beraha numbers $B_6,\ldots,B_9$!
So a variant of Conjecture~\ref{conj7.4}
applies in this case as well.
But we are unable to see what general principle might be at work.

\bigskip 

Conjecture~\ref{conj7.4} ``explains''
why the first few Beraha numbers
--- but {\em only}\/ the first few ---
arise as limiting points of chromatic roots,
at least in those cases where the limiting curve ${\cal B}$
crosses the real axis at some point $q_0(L)$.
Indeed, those Beraha numbers that satisfy both $q \le B_{L+1}$ and $q < q_0(L)$
correspond to the vanishing of the dominant amplitude,
hence are isolated limiting points of chromatic roots.
By contrast, the remaining Beraha numbers correspond either
to the vanishing of a subdominant amplitude
(in case $q_0(L) < q \le B_{L+1}$)
or do not correspond to the vanishing of any amplitude (in case $q > B_{L+1}$,
assuming the validity of Conjecture~\ref{conj7.2}).\footnote{
   In case $q_0(L)$ is itself a Beraha number $\le B_{L+1}$
   (as happens e.g.\ for $L=3_{\rm F}$),
   it corresponds to case (b) of the Beraha--Kahane--Weiss theorem,
   hence to a non-isolated limiting point.
   More generally, if ${\cal B}$ intersects the real axis in an
   interval $[q_{0,-}(L), q_{0,+}(L)]$,
   then all the points in this interval correspond to
   non-isolated limiting points,
   even if a dominant amplitude should happen to vanish.
}
As the strip width $L$ grows,
the limiting curve ${\cal B}$ moves to the right
and ``uncovers'' more Beraha numbers;
$q_0(L)$ presumably tends to a limiting value $q_0(\infty)$.
For the triangular lattice, Baxter's \cite{Baxter_87} analytic solution
predicts that $q_0(\infty) \approx 3.81967$,
which lies between $B_{14}$ and $B_{15}$
and in particular lies strictly below the critical point $q_c = 4$.
For the square lattice, an analytic solution is lacking,
but our results in Table~\ref{table_summary}
suggest (assuming monotonicity in $L$) that $q_0(\infty) > 2.788$,
and analogy with the triangular lattice suggests that
$q_0(\infty) < q_c = 3$.
It follows that $q_0(\infty)$ lies between $B_5$ and $B_6$,
so that the first four Beraha numbers $B_2,\ldots,B_5$ ---
{\em but only these}\/ --- will be isolated limiting points.

\subsection{Upper zero-free interval for bipartite planar graphs}

Let $G$ be a loopless planar graph.
Then it is not hard to prove that $P_G(q) > 0$
for all {\em integers}\/ $q \ge 5$;\footnote{
   This is the Five-Color Theorem, which goes back to Heawood in 1890.
   For a proof, see e.g.\ \cite[Theorem V.8, pp.~154--155]{Bollobas_98};
   or for an elegant alternate proof of an even stronger result,
   see \cite[Theorem V.12, pp.~161--163]{Bollobas_98}.
}
moreover, one of the most famous theorems of graph theory
--- the Four-Color Theorem
\cite{Appel_77a,Appel_77b,Appel_89,Robertson_97,Thomas_98} ---
asserts that $P_G(q) > 0$ holds in fact for all integers $q \ge 4$.

It is natural to ask whether these results can be extended from
integer $q$ to {\em real}\/ $q$.
The answer is yes, at least in part:
Birkhoff and Lewis \cite{Birkhoff_46} proved in 1946
that if $G$ is a loopless planar graph,
then $P_G(q) > 0$ for all real numbers $q \ge 5$.\footnote{
  See also Woodall \cite[Theorem 1]{Woodall_97} and
  Thomassen \cite[Theorem 3.1 ff.]{Thomassen_97}
  for alternate proofs of a more general result.
}
Furthermore, they conjectured that $P_G(q) > 0$ also for $4 < q < 5$;
and while no one has yet found a proof,
no one has found a counterexample either,
so it seems plausible (in the light of the Four-Color Theorem)
that the conjecture is true.

Now some planar graphs can be colored with three or even two colors;
their chromatic polynomials $P_G(q)$ are strictly positive
for integers $q \ge 3$ or $q \ge 2$, respectively.
Can {\em these}\/ bounds can be extended to real $q$?
That is, if $G$ is a $k$-colorable planar graph,
do we have $P_G(q) > 0$ for all real $q \ge k$?
Woodall \cite[p.~142]{Woodall_97} conjectured that the answer is yes.
For $k=4$, this is the conjecture of Birkhoff and Lewis mentioned above.
For $k=3$, however, Thomassen \cite[pp.~505--506]{Thomassen_97}
has shown that Woodall's conjecture is false:
there exist 3-colorable planar graphs with real chromatic roots
greater than 3.\footnote{
   Start with a graph $K$ and a real number $q_0$
   for which $P_K(q_0) < 0$.
   Then Thomassen \cite[Theorem 3.9]{Thomassen_97}
   constructs a 2-degenerate
   (and hence 3-colorable \cite[Theorem V.1, p.~148]{Bollobas_98})
   graph $K(m)$ such that $P_{K(m)}(q_0) < 0$;
   moreover, $K(m)$ can be chosen to be planar if $K$ is.
   Since there exist planar graphs $K$ with real chromatic roots $q_1$
   greater than 3,
   and since the 3-colorability of $K(m)$
   [or alternatively the Four-Color Theorem]
   implies that $P_{K(m)}(4) > 0$,
   we can take $q_0 = q_1 - \epsilon$ and conclude that
   $K(m)$ has a chromatic root in the interval $q_0 < q < 4$.
   Thus, the upper zero-free interval for 3-colorable planar graphs
   is the same as that for all planar graphs.

   In presenting this result, Thomassen \cite[p.~506]{Thomassen_97}
   further asserted that there exist planar graphs $K$
   with real chromatic roots {\em arbitrarily close to 4}\/;
   but this assertion apparently arises from a misunderstanding of
   the Beraha--Kahane \cite{Beraha_79} theorem that
   $4_{\rm P} \times n_{\rm F}$ triangular lattices
   have {\em complex}\/ chromatic roots arbitarily close to 4.
   In fact we do not know of any planar graphs with real chromatic roots
   arbitarily close to 4.
   In our study of triangular-lattice strips \cite{transfer3}
   we have thus far found chromatic roots up to $\approx 3.51$;
   and Baxter's \cite{Baxter_87} result $q_0(\infty) \approx 3.81967$
   suggests that sufficiently wide and long pieces of the triangular lattice
   will have real chromatic roots up to at least $B_{14} \approx 3.801938$.
   But we do not know of any planar graphs with chromatic roots
   in the interval $[B_{14},\infty)$.

   We thank Carsten Thomassen and Douglas Woodall for correspondence
   concerning these questions.
}
We can now show that Woodall's conjecture is false also for $k=2$:
there exist 2-colorable (i.e.\ bipartite) planar graphs
with real chromatic roots greater than 2.
For example, the $4_{\rm P} \times 6_{\rm F}$ square lattice
has chromatic roots at $q \approx 2.009978$ and $q \approx 2.168344$;
and the same pattern persists for larger lattices,
with both free and periodic transverse boundary conditions
(see Tables~\ref{table_zeros_free} and \ref{table_zeros_cyl}).
Indeed, for the cases $8_{\rm F} \times n_{\rm F}$ and
$8_{\rm P} \times n_{\rm F}$,
we see numerically (Tables~\ref{table_zeros_free} and \ref{table_zeros_cyl})
that there are real chromatic roots tending to
$B_5 = (3+\sqrt{5})/2 \approx 2.618034$ from below as $n \to \infty$.\footnote{
   The $5_{\rm P} \times n_{\rm F}$ and $7_{\rm P} \times n_{\rm F}$
   lattices have chromatic roots larger than $B_5$,
   but graphs of odd strip width with periodic b.c.\ are not bipartite.
}
This leads us to modify Woodall's conjecture as follows:

\begin{conjecture}
  \label{conj7.5}
Let $G$ be a bipartite planar graph.
Then $P_G(q) > 0$ for real $q \ge B_5 = (3+\sqrt{5})/2$.
\end{conjecture}

\noindent
Let us make two remarks:

1) Planarity here is crucial,
as non-planar bipartite graphs can have arbitrarily large
real chromatic roots.
Indeed, the complete bipartite graphs $K_{n_1,n_2}$,
in the limit $n_2 \to \infty$ with $n_1$ fixed,
have real chromatic roots arbitrarily close to all the integers
from 2 through $\lfloor n_1/2 \rfloor$
\cite[Theorem 8]{Woodall_77}.

2) Initially we conjectured that not only $P_G(q)$
but also {\em all its derivatives}\/ are positive for $q \ge B_5$.
But this turns out to be false:
for example, the $8_{\rm P} \times 16_{\rm F}$ lattice
has $P_G''(B_5) < 0$;
%
%
%
and the $8_{\rm P} \times 24_{\rm F}$ lattice
has $P_G'(q) < 0$ for $2.638337 \ltapprox q \ltapprox 2.687058$.
%
%
%

\subsection{Prospects for future work}  \label{sec7.4}

One very interesting extension of this work is the computation of the
chromatic polynomials for strips with {\em periodic}\/ boundary conditions 
in the {\em longitudinal}\/ direction \cite{transfer4}. 
Such computations have been performed for strip widths $m=2,3$
by {\em ad hoc}\/ methods
\cite{Biggs_72,Biggs_99a,Biggs_99b,Shrock_99b,Shrock_99c,%
Shrock_99e,Shrock_99f,Shrock_00a},
but a systematic transfer-matrix formalism has heretofore been lacking.
In fact, the needed formalism can be obtained by a slight extension
of the methods explained in this paper.
Suppose we want to obtain the chromatic polynomial
for a square-lattice strip $m_{\rm F/P}\times n_{\rm P}$
(i.e., either free or periodic transverse b.c.,
 and periodic longitudinal b.c.).
The idea is simple:
instead of just keeping track of the connectivities among the $m$ sites
on the current top row, we keep track of the connectivities among
the $2m$ sites on the current top row {\em and}\/ the bottom row.
Let us call these sites $1,2,\ldots,m$ and $1',2',\ldots,m'$, respectively.
Initially the top and bottom rows are identical.
We then enlarge the lattice one site at a time,
exactly as in Section~\ref{sec3.2};
the join and detach operations act on the sites of the top row,
with those of the bottom row simply ``going along for the ride''.
At the end, when we have obtained a lattice with $n+1$ rows,
we identify the top and bottom rows.
The partition function for periodic longitudinal boundary conditions
can thus be written as
\be
   Z_{G_n^{\rm per}}(q, \{v_e\})   \;=\;
     \widehat{\bf u}^{\rm T} (\V \H)^n \vv_{\rm equiv}
   \;,
   \label{Z_per_FK}
\ee
where ``equiv'' denotes the partition
$\{\, \{1,1'\}, \{2,2'\}, \,\ldots\, \{m,m'\} \,\}$
and $\widehat{\bf u}^{\rm T}$ is defined by
\be
   \widehat{\bf u}^{\rm T} \;=\;
   {\bf u}^{\rm T} \J_{11'} \J_{22'} \cdots \J_{mm'}
\ee
where ${\bf u}^{\rm T}$ is defined in \reff{def_uT}.

In future work in collaboration with Jesper-Lykke Jacobsen,
we will extend the results of the present paper
to wider strip widths \cite{transfer2},
to the triangular lattice \cite{transfer3},
to periodic boundary conditions in the longitudinal direction \cite{transfer4},
and to nonzero temperature \cite{transfer5}.

\section*{Acknowledgments}

We wish to thank Dario Bini for supplying us the MPSolve 2.0 package
and for many discussions about its use;
Tony Guttmann for analyzing the large-$q$ series;
Jesper-Lykke Jacobsen for independently confirming our transfer matrices
and for pointing out some typographical errors in an early draft;
Hubert Saleur for emphasizing the importance of the Beraha numbers;
Dan Segal for helping us with algebra and number theory;
Norman Weiss for suggesting that we study the resultant;
and Robert Shrock for many helpful conversations
throughout the course of this work.

The authors' research was supported in part
by U.S.\ National Science Foundation grant PHY-9900769 (J.S.\ and A.D.S.)\
and CICyT (Spain) grant AEN97-1680 (J.S.).
It was completed while the second author was a Visiting Fellow
at All Souls College, Oxford, where he was supported in part
by Engineering and Physical Sciences Research Council grant GR/M 71626
and aided by the warm hospitality of John Cardy and the
Department of Theoretical Physics.

\clearpage

%
%
%
%
\begin{table}
\hspace*{-2cm}
\begin{tabular}{|r|l|l|l|l|}
\hline\hline
\multicolumn{1}{|c|}{$n$} &
\multicolumn{1}{|c|}{$B_n$ (exact)} &
\multicolumn{1}{|c|}{$B_n$ (num)} &
\multicolumn{1}{|c|}{$p_n(q)$}   &
\multicolumn{1}{|c|}{Other $B_n^{(k)}$} \\
\hline\hline\vphantom{$\overline{\overline{M}}$}
2   & 0                  & 0           & $q$        &     \\
\hline\vphantom{$\overline{\overline{M}}$}
3   & 1                  & 1           & $q-1$        &     \\
\hline\vphantom{$\overline{\overline{M}}$}
4   & 2                  & 2           & $q-2$        &     \\
\hline\vphantom{$\overline{\overline{M}}$}
5   & $(3 + \sqrt{5})/2$ & 2.6180339887& $q^2-3q+1$ &
      $(3 - \sqrt{5})/2$ \\
\hline\vphantom{$\overline{\overline{M}}$}
6   & 3                  & 3           & $q-3$        &     \\
\hline\vphantom{$\overline{\overline{M}}$}
7   &                    & 3.2469796037& $q^3-5q^2+6q-1$ &
      0.1980622642       \\
    &                    &             &            &
      1.5549581321       \\
\hline\vphantom{$\overline{\overline{M}}$}
8   & $2 + \sqrt{2}$     & 3.4142135624& $q^2-4q+2$ &
      $2 - \sqrt{2}$     \\
\hline\vphantom{$\overline{\overline{M}}$}
9   &                    & 3.5320888862& $q^3-6q^2+9q-1$ &
      0.1206147584       \\
    &                    &             &            &
      2.3472963553       \\
\hline\vphantom{$\overline{\overline{M}}$}
10  & $(5 + \sqrt{5})/2$ & 3.6180339887& $q^2-5q+5$ &
      $(5 - \sqrt{5})/2$ \\
\hline\vphantom{$\overline{\overline{M}}$}
11  &                    & 3.6825070657& $q^5-9q^4+28q^3-35q^2+15q-1$ &
      0.0810140528       \\
    &                    &             &            &
      0.6902785321       \\
    &                    &             &            &
      1.7153703235       \\
    &                    &             &            &
      2.8308300260       \\
\hline\vphantom{$\overline{\overline{M}}$}
12  & $2 + \sqrt{3}$     & 3.7320508076& $q^2-4q+1$ &
      $2 - \sqrt{3}$     \\
\hline\vphantom{$\overline{\overline{M}}$}
13  &                    & 3.7709120513& $q^6-11q^5+45q^4-84q^3+70q^2$&
      0.0581163651       \\
    &                    &             & $\qquad -21q+1$          &
      0.5029785037       \\
    &                    &             &            &
      1.2907902259       \\
    &                    &             &            &
      2.2410733605       \\
    &                    &             &            &
      3.1361294935       \\
\hline\vphantom{$\overline{\overline{M}}$}
14  &                    & 3.8019377358& $q^3-7q^2+14q-7$ &
      0.7530203963       \\
    &                    &             &            &
      2.4450418679       \\
\hline\vphantom{$\widetilde{\widetilde{M}}$}
15  & $(9+\sqrt{5}+\sqrt{30-6\sqrt{5}})/4$ &
                           3.8270909153& $q^4-9q^3+26q^2-24q+1$ &
      0.0437047985       \\
    &                    &             &            &
      1.7909430735       \\
    &                    &             &            &
      3.3382612127       \\
\hline\vphantom{$\widetilde{\widetilde{M}}$}
16  & $2 +\sqrt{2+\sqrt{2}}$ &
                           3.8477590650& $q^4-8q^3+20q^2-16q+2$ &
      0.1522409350       \\
    &                    &             &            &
      1.2346331353      \\
    &                    &             &            &
      2.7653668647       \\
\hline\hline
\end{tabular}
\vspace{5mm}
\caption{
Beraha numbers $B_n = 4 \cos^2(\pi/n)$ and their minimal polynomials $p_n(q)$.
For each $n$ we give the exact expression of the Beraha number $B_n$
whenever it can be expressed in terms of square roots alone;
its numerical value to 10 decimal places;
the unique irreducible monic polynomial $p_n(q)$
with integer coefficients having $B_n$ as a root;
and the other zeros of $p_n(q)$, which are the generalized Beraha numbers
$B_n^{(k)} = 4 \cos^2(k\pi/n)$ with $k$ relatively prime to $n$.
}
\label{table_monomials}
\end{table}

\clearpage

%
%
\begin{table}
\centering
\begin{tabular}{|r|r|r|r|r|r|r|r|}
\hline\hline
$m$& $B(m)$ & $C_m$ &TriFree$(m)$&SqFree$(m)$& $d_m$ &
                      TriCyl$(m)$ &SqCyl$(m)$ \\
\hline\hline
1  &      1 &      1 &        1 &       1 &      1 &       1 &      1\\
2  &      2 &      2 &        1 &       1 &      1 &       1 &      1\\
3  &      5 &      5 &        2 &       2 &      1 &       1 &      1\\
4  &     15 &     14 &        4 &       3 &      3 &       2 &      2\\
5  &     52 &     42 &        9 &       7 &      6 &       2 &      2\\
6  &    203 &    132 &       21 &      13 &     15 &       5 &      5\\
7  &    877 &    429 &       51 &      32 &     36 &       6 &      6\\
8  &   4140 &   1430 &      127 &      70 &     91 &      15 &     14\\
9  &  21147 &   4862 &      323 &     179 &    232 &      28 &     22\\
10 & 115975 &  16796 &      835 &     435 &    603 &      67 &     51\\
11 & 678570 &  58786 &     2188 &    1142 &   1585 &     145 &     95\\
12 &4213597 & 208012 &     5798 &    2947 &   4213 &     368 &    232\\
13 &   27644437 &  742900 &  15511 &  7889 & 11298 &  870 &  499 \\
14 &  190899322 & 2674440 &  41835 & 21051 & 30537 & 2211 & 1241 \\
\hline\hline
\end{tabular}
\vspace{1cm}
\caption{
   Dimension of the transfer matrix.
   For each strip width $m$ we give
   the number $B(m)$ of all partitions,
   the number $C_m$ of non-crossing partitions,
   the number $\hbox{\rm TriFree}(m) = M_{m-1}$
     of non-crossing non-nearest-neighbor partitions
     with free boundary conditions,
   and the number $\hbox{\rm SqFree}(m)$ of equivalence classes
     of non-crossing non-nearest-neighbor partitions modulo reflection
     in the center of the strip.
   We also give the number $d_m$ ($= R_m$ for $m \ge 2$)
     of non-crossing non-nearest-neighbor partitions
     with periodic boundary conditions,
   the number $\hbox{\rm TriCyl}(m)$ of equivalence classes of
     such partitions modulo translations,
   and the number $\hbox{\rm SqCyl}(m)$ of equivalence classes of
     such partitions modulo translations and reflections.
}
\label{table_dimensions}
\end{table}

\clearpage

%
%
\begin{table}[t]
\vspace*{-1cm}
\centering
\scriptsize
\begin{tabular}{|l|l|l|l|l|}
\hline\hline
 Lattice & 3rd Zero & 4th Zero   & 5th Zero   & 6th Zero   \\
\hline\hline
$ 3_{\rm F}\times 3_{\rm F}$ &  1.646039212420  &   &   &   \\
$ 3_{\rm F}\times 6_{\rm F}$ &   &   &   &   \\
$ 3_{\rm F}\times 9_{\rm F}$ &  1.862295803794  &   &   &   \\
$ 3_{\rm F}\times 12_{\rm F}$ &   &   &   &   \\
$ 3_{\rm F}\times 15_{\rm F}$ &  1.910244567418  &   &   &   \\
$ 3_{\rm F}\times 18_{\rm F}$ &   &   &   &   \\
$ 3_{\rm F}\times 21_{\rm F}$ &  1.932253338339  &   &   &   \\
$ 3_{\rm F}\times 24_{\rm F}$ &   &   &   &   \\
$ 3_{\rm F}\times 27_{\rm F}$ &  1.945103511556  &   &   &   \\
$ 3_{\rm F}\times 30_{\rm F}$ &   &   &   &   \\
\hline
$ 4_{\rm F}\times 4_{\rm F}$ &   &   &   &   \\
$ 4_{\rm F}\times 8_{\rm F}$ &   &   &   &   \\
$ 4_{\rm F}\times 12_{\rm F}$ &  2.000607664038  &  2.183434328589  &   &   \\
$ 4_{\rm F}\times 16_{\rm F}$ &  2.000017521546  &  2.226186181588  &   &   \\
$ 4_{\rm F}\times 20_{\rm F}$ &  2.000000515361  &  2.248253640526  &   &   \\
$ 4_{\rm F}\times 24_{\rm F}$ &  2.000000015170  &  2.261494080470  &   &   \\
$ 4_{\rm F}\times 28_{\rm F}$ &  2.000000000447  &  2.270172437566  &   &   \\
$ 4_{\rm F}\times 32_{\rm F}$ &  2.000000000013  &  2.276213662199  &   &   \\
$ 4_{\rm F}\times 36_{\rm F}$ &  2.000000000000  &  2.280609243979  &   &   \\
$ 4_{\rm F}\times 40_{\rm F}$ &  2.000000000000  &  2.283918256290  &   &   \\
$ 4_{\rm F}\times 100_{\rm F}$ &  2.000000000000  &  2.236070288638  &  2.284202920228  &  2.297805980307  \\
\hline
$ 5_{\rm F}\times 5_{\rm F}$ &  1.955073615801  &   &   &   \\
$ 5_{\rm F}\times 10_{\rm F}$ &  2.000022457863  &  2.243311545349  &   &   \\
$ 5_{\rm F}\times 15_{\rm F}$ &  1.999999994509  &   &   &   \\
$ 5_{\rm F}\times 20_{\rm F}$ &  2.000000000001  &  2.335823711578  &   &   \\
$ 5_{\rm F}\times 25_{\rm F}$ &  2.000000000000  &   &   &   \\
$ 5_{\rm F}\times 30_{\rm F}$ &  2.000000000000  &  2.365828458342  &   &   \\
$ 5_{\rm F}\times 35_{\rm F}$ &  2.000000000000  &   &   &   \\
$ 5_{\rm F}\times 40_{\rm F}$ &  2.000000000000  &  2.381070502769  &   &   \\
$ 5_{\rm F}\times 45_{\rm F}$ &  2.000000000000  &   &   &   \\
$ 5_{\rm F}\times 50_{\rm F}$ &  2.000000000000  &  2.390328275726  &   &   \\
\hline
$ 6_{\rm F}\times 6_{\rm F}$ &  2.001381451484  &  2.196830038914  &   &   \\
$ 6_{\rm F}\times 12_{\rm F}$ &  2.000000000760  &  2.390498998123  &   &   \\
$ 6_{\rm F}\times 18_{\rm F}$ &  2.000000000000  &  2.448434501424  &   &   \\
$ 6_{\rm F}\times 24_{\rm F}$ &  2.000000000000  &  2.475714120608  &   &   \\
$ 6_{\rm F}\times 30_{\rm F}$ &  2.000000000000  &  2.491245543049  &   &   \\
$ 6_{\rm F}\times 36_{\rm F}$ &  2.000000000000  &  2.501126630104  &   &   \\
$ 6_{\rm F}\times 42_{\rm F}$ &  2.000000000000  &  2.507892809644  &   &   \\
$ 6_{\rm F}\times 48_{\rm F}$ &  2.000000000000  &  2.512775536891  &   &   \\
$ 6_{\rm F}\times 54_{\rm F}$ &  2.000000000000  &  2.516440469487  &   &   \\
$ 6_{\rm F}\times 60_{\rm F}$ &  2.000000000000  &  2.519276871603  &   &   \\
$ 6_{\rm F}\times 240_{\rm F}$ &  2.000000000000  &  2.534921463459  &   &  \\
\hline
$ 7_{\rm F}\times 7_{\rm F}$ &  1.999994176430  &   &   &   \\
$ 7_{\rm F}\times 14_{\rm F}$ &  2.000000000000  &  2.451966225086  &   &   \\
$ 7_{\rm F}\times 21_{\rm F}$ &  2.000000000000  &   &   &   \\
$ 7_{\rm F}\times 28_{\rm F}$ &  2.000000000000  &  2.523291736983  &   &   \\
$ 7_{\rm F}\times 35_{\rm F}$ &  2.000000000000  &   &   &   \\
$ 7_{\rm F}\times 42_{\rm F}$ &  2.000000000000  &  2.548814353555  &   &   \\
$ 7_{\rm F}\times 49_{\rm F}$ &  2.000000000000  &   &   &   \\
$ 7_{\rm F}\times 56_{\rm F}$ &  2.000000000000  &  2.562226841180  &   &   \\
$ 7_{\rm F}\times 63_{\rm F}$ &  2.000000000000  &   &   &   \\
$ 7_{\rm F}\times 70_{\rm F}$ &  2.000000000000  &  2.570504933475  &   &   \\
\hline
$ 8_{\rm F}\times 8_{\rm F}$ &  2.000000005426  &  2.391719919086  &   &   \\
$ 8_{\rm F}\times 16_{\rm F}$ &  2.000000000000  &  2.531414423190  &   &   \\
$ 8_{\rm F}\times 24_{\rm F}$ &  2.000000000000  &  2.576747844784  &   &   \\
$ 8_{\rm F}\times 32_{\rm F}$ &  2.000000000000  &  2.597790566370  &   &   \\
$ 8_{\rm F}\times 40_{\rm F}$ &  2.000000000000  &  2.609001821476  &   &   \\
$ 8_{\rm F}\times 48_{\rm F}$ &  2.000000000000  &  2.614901253573  &   &   \\
$ 8_{\rm F}\times 56_{\rm F}$ &  2.000000000000  &  2.617322619879  &   &   \\
$ 8_{\rm F}\times 64_{\rm F}$ &  2.000000000000  &  2.617921204353  &   &   \\
$ 8_{\rm F}\times 72_{\rm F}$ &  2.000000000000  &  2.618018253506  &   &   \\
$ 8_{\rm F}\times 80_{\rm F}$ &  2.000000000000  &  2.618031848556  &   &   \\
\hline
\hline
 Beraha &2   & 2.618033988750  &     &  \\
\hline
\end{tabular}
\caption{
   Real zeros of the chromatic polynomials of finite square-lattice strips
   with free boundary conditions in both directions, to 12 decimal places.
   A blank means that the zero in question is absent.
   The first two real zeros $q=0,1$ are exact on all lattices.
   ``Beraha'' indicates the Beraha numbers $B_4 = 2$ and
   $B_5 = (3+\sqrt{5})/2$.
}
\label{table_zeros_free}
\end{table}

\clearpage
%
%
\begin{table}[t]
\centering
\scriptsize
\begin{tabular}{|l|l|l|l|l|l|l|}
\hline\hline
 Lattice & 3rd Zero & 4th Zero   & 5th Zero   & 6th Zero   & 7th Zero   & 8th Zero   \\
\hline\hline
$ 4_{\rm P}\times 4_{\rm F}$ &   &   &   &   &   &   \\
$ 4_{\rm P}\times 8_{\rm F}$ &  2.000937646653  &  2.233582851404  &   &   &   &   \\
$ 4_{\rm P}\times 12_{\rm F}$ &  2.000011295331  &  2.285151240169  &   &   &   &   \\
$ 4_{\rm P}\times 16_{\rm F}$ &  2.000000139385  &  2.307528225343  &   &   &   &   \\
$ 4_{\rm P}\times 20_{\rm F}$ &  2.000000001721  &  2.319813608989  &   &   &   &   \\
$ 4_{\rm P}\times 24_{\rm F}$ &  2.000000000021  &  2.327431319510  &   &   &   &   \\
$ 4_{\rm P}\times 28_{\rm F}$ &  2.000000000000  &  2.332533058471  &   &   &   &   \\
$ 4_{\rm P}\times 32_{\rm F}$ &  2.000000000000  &  2.336139224928  &   &   &   &   \\
$ 4_{\rm P}\times 36_{\rm F}$ &  2.000000000000  &  2.338792911735  &   &   &   &   \\
$ 4_{\rm P}\times 40_{\rm F}$ &  2.000000000000  &  2.340807853864  &   &   &   &   \\
$ 4_{\rm P}\times 100_{\rm F}$ &  2.000000000000  &  2.257013014819  &  2.270836682396  &  2.325455510831  &  2.341961199927  &  2.349426156978  \\
\hline
$ 5_{\rm P}\times 5_{\rm F}$ &  2   &   &   &   &   &   \\
$ 5_{\rm P}\times 10_{\rm F}$ &  2   &  2.579692798743  &   &   &   &   \\
$ 5_{\rm P}\times 15_{\rm F}$ &  2   &   &   &   &   &   \\
$ 5_{\rm P}\times 20_{\rm F}$ &  2   &  2.615053742246  &   &   &   &   \\
$ 5_{\rm P}\times 25_{\rm F}$ &  2   &  2.618482995587  &  2.643045814623  &   &   &   \\
$ 5_{\rm P}\times 30_{\rm F}$ &  2   &  2.617994992234  &   &   &   &   \\
$ 5_{\rm P}\times 35_{\rm F}$ &  2   &  2.618037696771  &  2.658908973824  &   &   &   \\
$ 5_{\rm P}\times 40_{\rm F}$ &  2   &  2.618033639521  &   &   &   &   \\
$ 5_{\rm P}\times 45_{\rm F}$ &  2   &  2.618034021676  &  2.666710728680  &   &   &   \\
$ 5_{\rm P}\times 50_{\rm F}$ &  2   &  2.618033985646  &   &   &   &   \\
\hline
$ 6_{\rm P}\times 6_{\rm F}$ &  2.000004484676  &  2.407498857052  &   &   &   &   \\
$ 6_{\rm P}\times 12_{\rm F}$ &  2.000000000000  &  2.516516196247  &   &   &   &   \\
$ 6_{\rm P}\times 18_{\rm F}$ &  2.000000000000  &  2.551495362906  &   &   &   &   \\
$ 6_{\rm P}\times 24_{\rm F}$ &  2.000000000000  &  2.568645710453  &   &   &   &   \\
$ 6_{\rm P}\times 30_{\rm F}$ &  2.000000000000  &  2.578747609077  &   &   &   &   \\
$ 6_{\rm P}\times 36_{\rm F}$ &  2.000000000000  &  2.585363032613  &   &   &   &   \\
$ 6_{\rm P}\times 42_{\rm F}$ &  2.000000000000  &  2.590008147965  &   &   &   &   \\
$ 6_{\rm P}\times 48_{\rm F}$ &  2.000000000000  &  2.593435585192  &   &   &   &   \\
$ 6_{\rm P}\times 54_{\rm F}$ &  2.000000000000  &  2.596060266523  &   &   &   &   \\
$ 6_{\rm P}\times 60_{\rm F}$ &  2.000000000000  &  2.598129161537  &   &   &   &   \\
$ 6_{\rm P}\times 240_{\rm F}$&  2.000000000000  &  2.610780621890  &   &   &
 &   \\
\hline
$ 7_{\rm P}\times 7_{\rm F}$ &  2   &   &   &   &   &   \\
$ 7_{\rm P}\times 14_{\rm F}$ &  2   &  2.617937723253  &   &   &   &   \\
$ 7_{\rm P}\times 21_{\rm F}$ &  2   &  2.618034017737  &  2.721810707015  &   &   &   \\
$ 7_{\rm P}\times 28_{\rm F}$ &  2   &  2.618033988741  &   &   &   &   \\
$ 7_{\rm P}\times 35_{\rm F}$ &  2   &  2.618033988750  &  2.748882762812  &   &   &   \\
$ 7_{\rm P}\times 42_{\rm F}$ &  2   &  2.618033988750  &   &   &   &   \\
$ 7_{\rm P}\times 49_{\rm F}$ &  2   &  2.618033988750  &  2.760230036513  &   &   &   \\
$ 7_{\rm P}\times 56_{\rm F}$ &  2   &  2.618033988750  &   &   &   &   \\
$ 7_{\rm P}\times 63_{\rm F}$ &  2   &  2.618033988750  &  2.766499503035  &   &   &   \\
$ 7_{\rm P}\times 70_{\rm F}$ &  2   &  2.618033988750  &   &   &   &   \\
\hline
$ 8_{\rm P}\times 8_{\rm F}$ &  2.000000000001  &  2.551072878420  &   &   &   &   \\
$ 8_{\rm P}\times 16_{\rm F}$ &  2.000000000000  &  2.616714700486  &   &   &   &   \\
$ 8_{\rm P}\times 24_{\rm F}$ &  2.000000000000  &  2.618032009068  &   &   &   &   \\
$ 8_{\rm P}\times 32_{\rm F}$ &  2.000000000000  &  2.618033986108  &   &   &   &   \\
$ 8_{\rm P}\times 40_{\rm F}$ &  2.000000000000  &  2.618033988746  &   &   &   &   \\
$ 8_{\rm P}\times 48_{\rm F}$ &  2.000000000000  &  2.618033988750  &   &   &   &   \\
$ 8_{\rm P}\times 56_{\rm F}$ &  2.000000000000  &  2.618033988750  &   &   &   &   \\
$ 8_{\rm P}\times 64_{\rm F}$ &  2.000000000000  &  2.618033988750  &   &   &   &   \\
$ 8_{\rm P}\times 72_{\rm F}$ &  2.000000000000  &  2.618033988750  &   &   &   &   \\
$ 8_{\rm P}\times 80_{\rm F}$ &  2.000000000000  &  2.618033988750  &   &   &   &   \\
\hline
\hline
 Beraha &2   & 2.618033988750  &     &  &  &  \\
\hline
\end{tabular}
\caption{
   Real zeros of the chromatic polynomials of finite square-lattice strips
   with periodic boundary conditions in the transverse direction
   and free boundary conditions in the longitudinal direction,
   to 12 decimal places.
   A blank means that the zero in question is absent.
   The first two real zeros $q=0,1$ are exact on all lattices;
   the third real zero $q=2$ is exact on all lattices of odd width.
   ``Beraha'' indicates the Beraha numbers $B_4 = 2$ and
   $B_5 = (3+\sqrt{5})/2$.
}
\label{table_zeros_cyl}
\end{table}

\clearpage
%
%
\begin{table}
\small
\hspace*{-2cm}
\begin{tabular}{|c||c|c|c|c|c|c|c|c||c|c|}
\cline{2-11}
\multicolumn{1}{c||}{\mbox{}}&
\multicolumn{8}{|c||}{Eigenvalue-Crossing Curves ${\cal B}$} &
\multicolumn{2}{|c|}{Isolated Points}\\
\hline\hline
Lattice      & \# C & \# E & \# T & \# D & \# ER & $\min \real q$ & 
$q_0$                      & $\max \real q$      & \# RI& \# CI   \\
\hline\hline
$2_{\rm F}$  &   0  &   0  & 0    &   0  &  0    &                & 
                           &                     & 2    & 0       \\
$3_{\rm F}$  &   3  &   6  & 0    &   0  &  0    & 0.586570       & 
                        2  &                 2.5 & 2    & 0       \\
$4_{\rm F}$  &   3  &   10 & 2    &   1  &  0    & 0.325474       & 
$[2.228359, 2.301416]$     &            2.667426 & 3    & 0       \\
$5_{\rm F}$  &   5  &   14 & 4    &   0  &  0    & 0.170897       &
                  2.428438 &            2.769205 & 3    & 0       \\
$6_{\rm F}$  &   5  &   14 & 6    &   1  &  2    & 0.063142       &
$[2.528647, 2.537098]$     &            2.837338 & 3    & 2       \\
$7_{\rm F}$  & $7^\dagger$ &  
              $14^\dagger$ & 
               $4^\dagger$ &  
               $0^\dagger$ &  
               $2^\dagger$ &                      $-0.044443$      & 
                  2.606248 &            2.886041 & $3^\dagger$
                                                 & $2^\dagger$    \\
$8_{\rm F}$  & $6^\dagger$ &  
              $14^\dagger$ & 
               $2^\dagger$ &  
               $0^\dagger$ &  
               $0^\dagger$ &                      $-0.130642$      & 
$2.660260 \pm 0.001257\,i^*$&           2.921658 & $4^\dagger$
                                                 & $2^\dagger$    \\
\hline\hline
$3_{\rm P}$  &   0  &    0 & 0    &   0  &   0   &                 &  
                           &                     &  3   & 0       \\
$4_{\rm P}$  &   3  &    8 & 0    &   1  &   0   & 0.709803        & 
   $[2.253370, 2.351688]$  &            2.995331 &  3   & 0       \\
$5_{\rm P}$  &   5  &   10 & 0    &   0  &   0   & 0.165021        & 
                  2.691684 &            2.691684 &  4   & 0       \\
$6_{\rm P}$  &   3  &   10 & 2    &   1  &   0   & $-0.131889$      & 
   $[2.608943, 2.613228]$  &            3.171192 &  3   & 1       \\
$7_{\rm P}$  &   7  &   16 & 2    &   0  &   0   & $-0.296250$      & 
                  2.788378 &            2.839016 &  4   & 1       \\
$8_{\rm P}$  &   6  &   16 & 4    &   0  &   0   & $-0.390864$      & 
$2.751531 \pm 0.002531\,i^*$&           3.211132 &  4   & 2        \\
\hline\hline
\end{tabular}

\vspace{1cm}
\caption{
   Summary of qualitative results for the eigenvalue-crossing curves $\scrb$
   and for the isolated limiting points of zeros.
   For each square-lattice strip considered in this paper,
   we give the number of connected components of $\scrb$ (\# C),
   the number of endpoints (\# E),
   the number of T points (\# T),
   the number of double points (\# D),
   and the number of enclosed regions (\# ER);
   we give the minimum value of $\real q$ on $\scrb$,
   the value(s) $q_0$ where $\scrb$ intersects the real axis
   (${}^*$ denotes an almost-crossing),
   and the maximum value of $\real q$ on $\scrb$.
   We also report the number of real isolated limiting points of zeros
   (\# RI) [which are always successive Beraha numbers $B_2$, $B_3$, \ldots]
   and the number of complex-conjugate pairs of isolated limiting points
   (\# CI).
   The symbol $^\dagger$ indicates uncertain results.
}
\label{table_summary}
\end{table}

\clearpage

%
%
%
%

%
%

\clearpage
%
%

\clearpage
%
%
\begin{figure}
\begin{tabular}{cc}
  \epsfxsize=200pt\epsffile{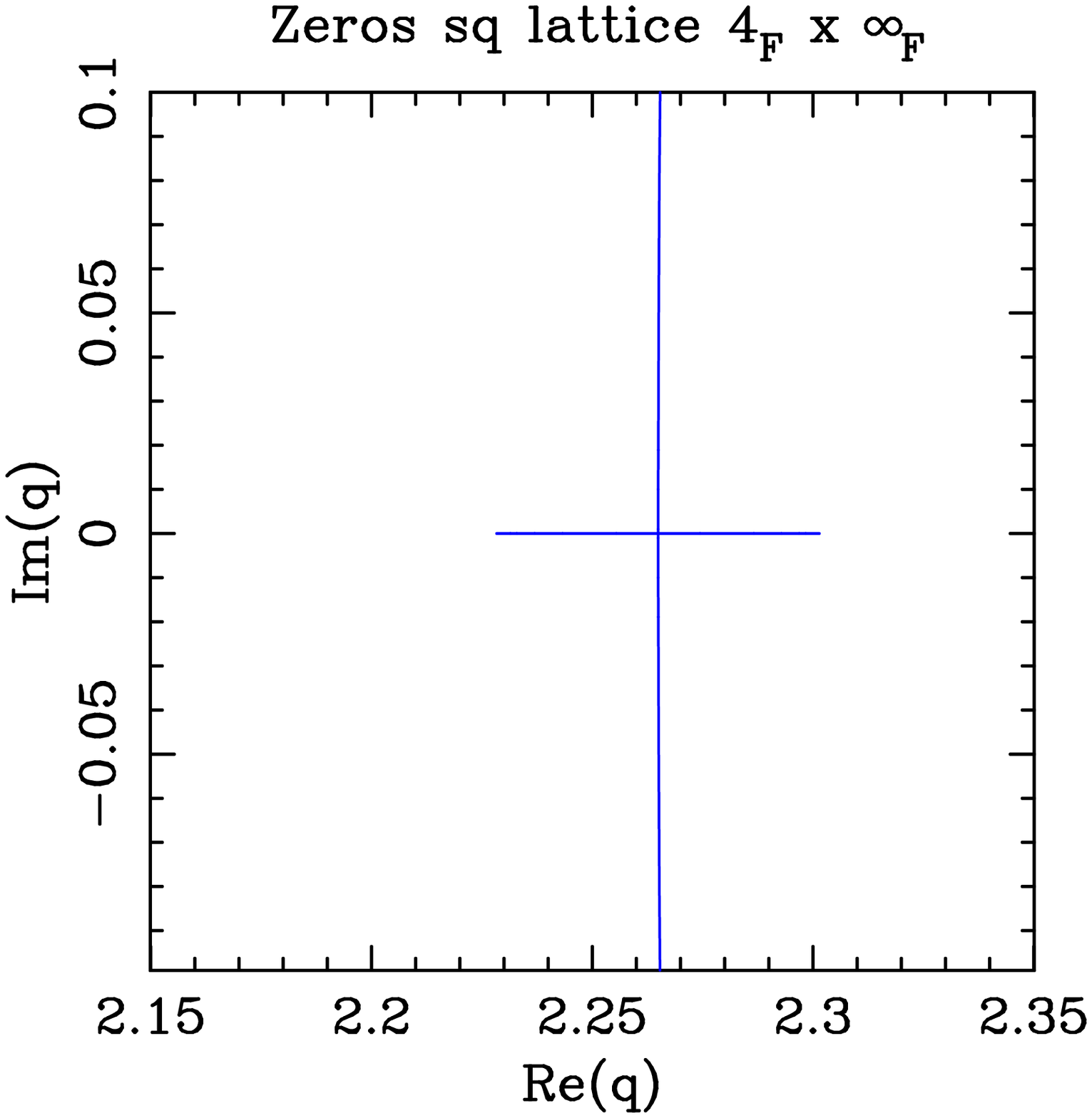} &
  \epsfxsize=200pt\epsffile{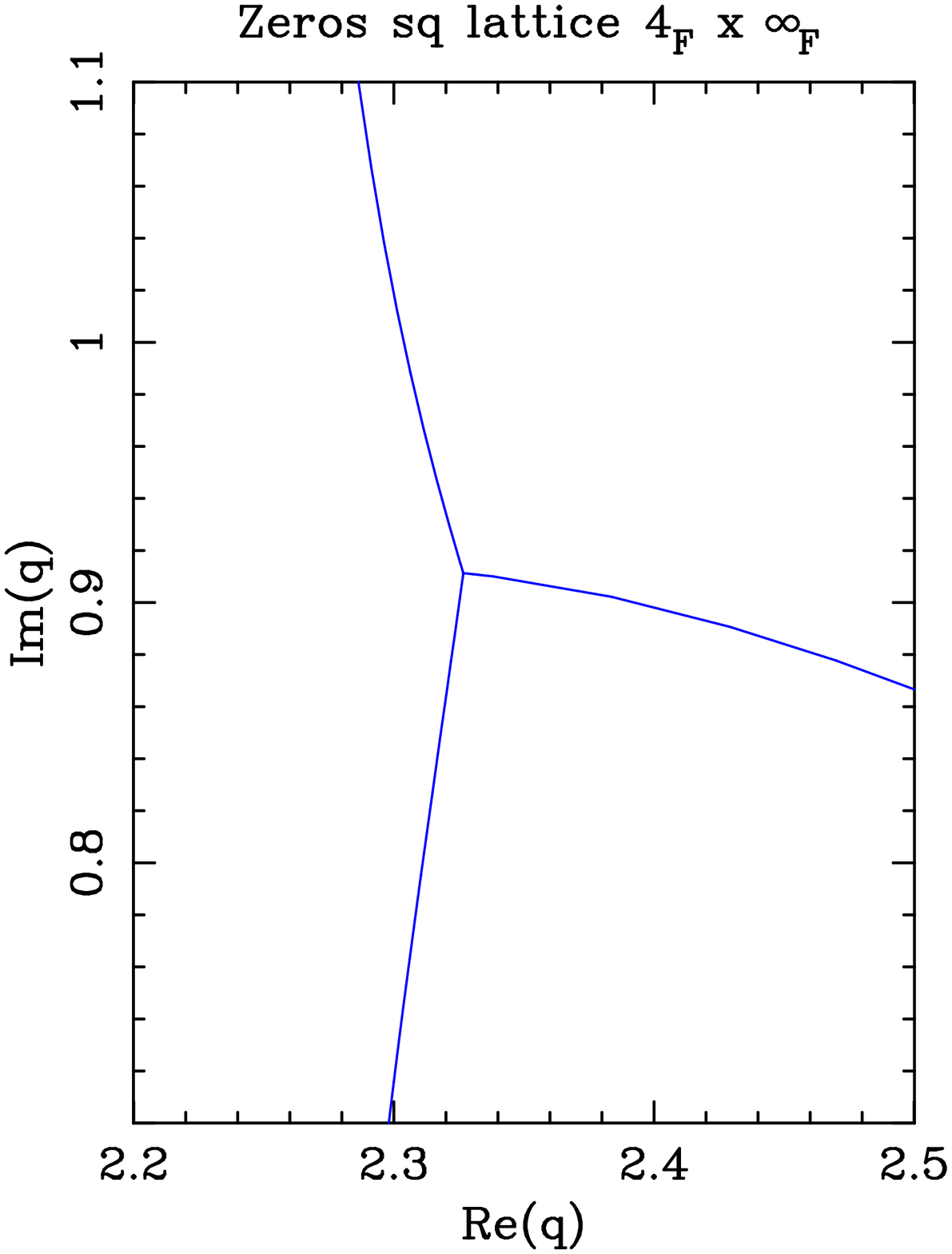} \\
  (a)  & (b)
\end{tabular}
  \caption{
  Detail of the limiting curves $\scrb$ for
  the $q$-state Potts antiferromagnet
  on a square lattice $4_F\times\infty_F$.
  (a) Region near the double point $q \approx 2.2649418565$.
      The value of $t$ is continuous around the double point,
      with $t \approx 0.0621$.
  (b) Region near the T point $q \approx 2.327 + 0.9113\,i$.
      At this T point we have $t \approx (1.818, 12.962, 0.655)$
      and hence $\theta \approx (2.136, 2.988, 1.160)$,
      so that $\sum \theta = 2 \pi$.
  }
\label{Figure_sq_4FxInftyF_bis}
\end{figure}

\clearpage
%
%
\begin{figure}
  \centering
  \epsfxsize=400pt\epsffile{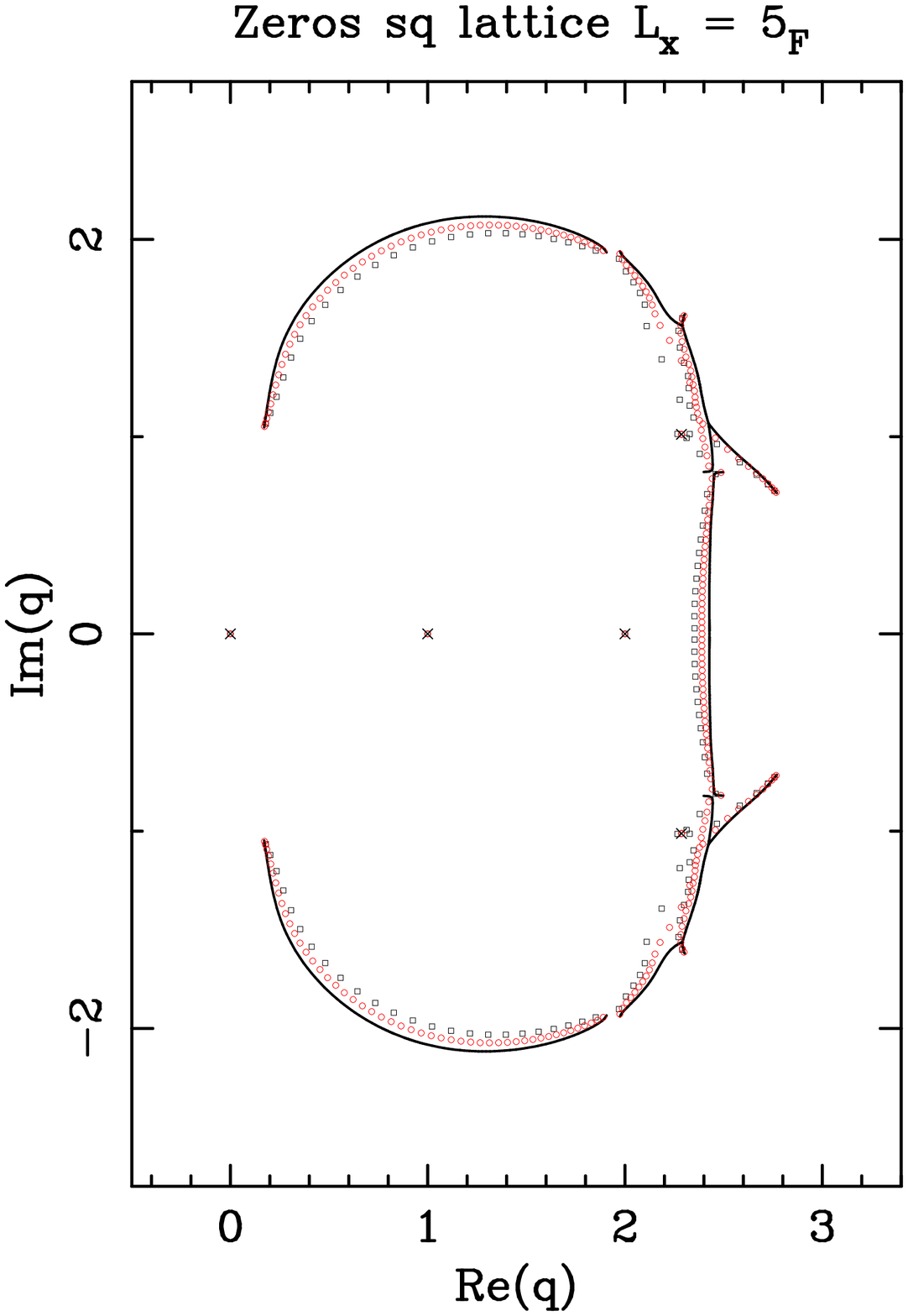}
  \caption{
  Zeros of the partition function of the $q$-state Potts antiferromagnet
  on a square lattices $5_F \times 25_F$ (squares),
  $5_F \times 50_F$ (circles) and $5_F\times\infty_F$ (solid line).
  The isolated limiting zeros are depicted by a $\times$.
  The limiting curve was computed using the resultant method.
  }
\label{Figure_sq_5FxInftyF}
\end{figure}

\clearpage
%
%
\begin{figure}
  \centering
  \begin{tabular}{c}
    \epsfxsize=200pt\epsffile{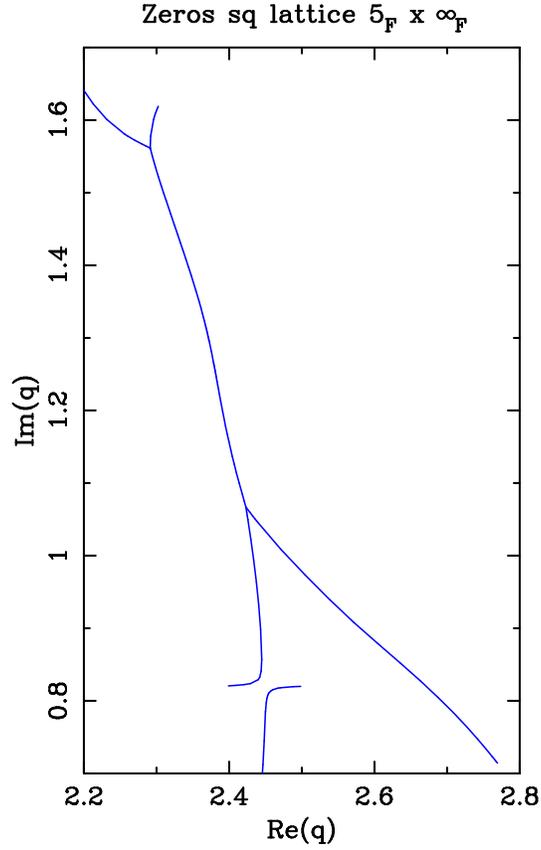}
  \end{tabular}
  \caption{
  Detail of the limiting curves $\scrb$ for the $q$-state Potts antiferromagnet
  on a square lattice $5_F\times\infty_F$. Region near the T points
  of the limiting curve. On the upper T point $q \approx 2.291+1.561\, i$,
  we have $t \approx (0.999,0.179,1.434)$, corresponding to
  $\theta \approx (1.569,0.354,1.924)$. On the lower T point
  $q \approx 2.423 + 0.1067\,i$, we have $t \approx (1.823,0.434,0.774)$ with
  $\theta \approx (2.138,0.820, 1.318)$.
  }
\label{Figure_sq_5FxInftyF_bis}
\end{figure}

\clearpage
%
%
\begin{figure}
  \centering
  \epsfxsize=400pt\epsffile{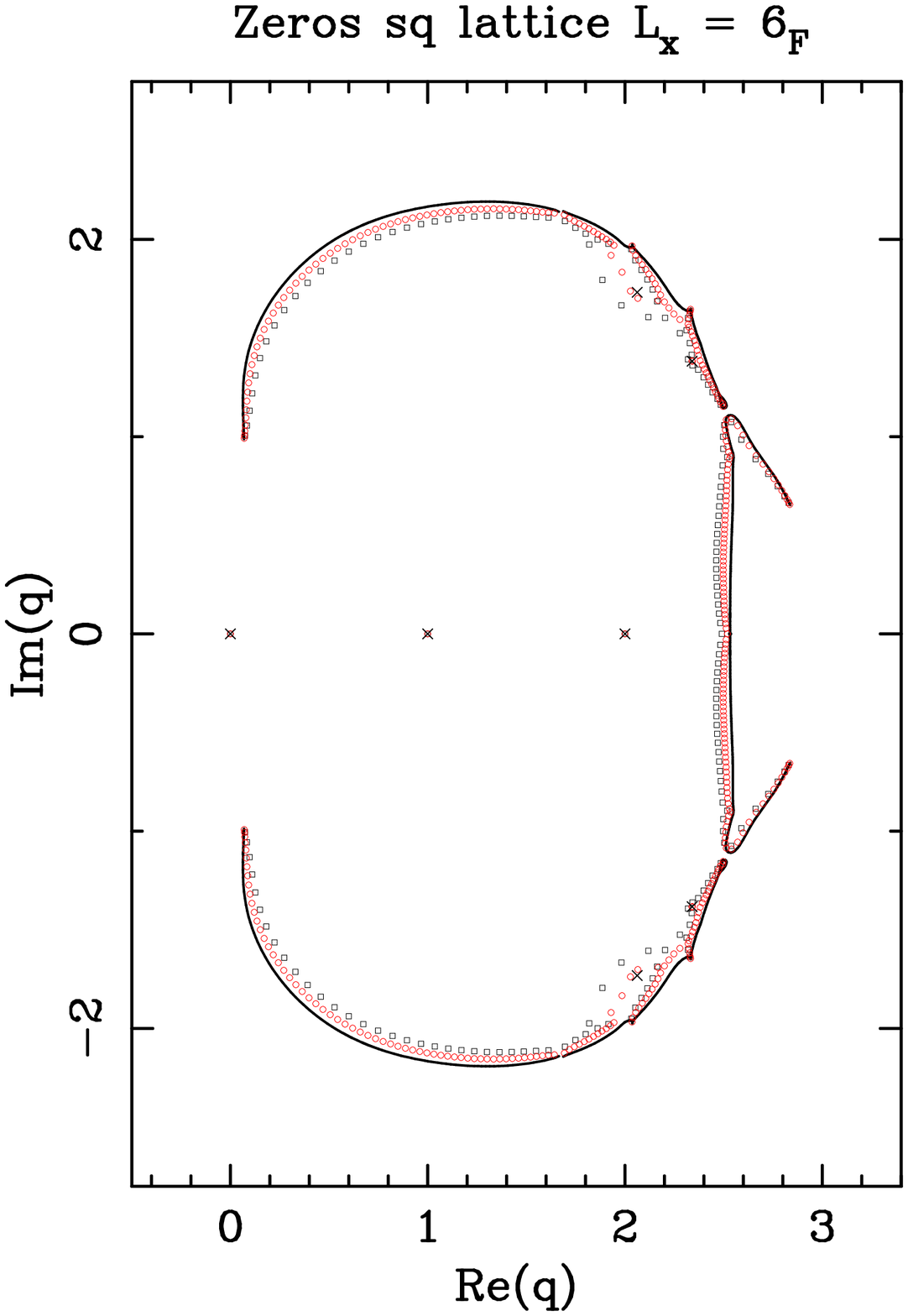}
  \caption{
  Zeros of the partition function of the $q$-state Potts antiferromagnet
  on a square lattices $6_F \times 30_F$ (squares),
  $6_F \times 60_F$ (circles) and $6_F\times\infty_F$ (solid line).
  The isolated limiting zeros are depicted by a $\times$.
  The limiting curve was computed using the resultant method.
  }
\label{Figure_sq_6FxInftyF}
\end{figure}

\clearpage
%
%
\begin{figure}
  \vspace*{-2cm}
  \centering
  \begin{tabular}{cc}
    \multicolumn{2}{c}{\epsfxsize=200pt\epsffile{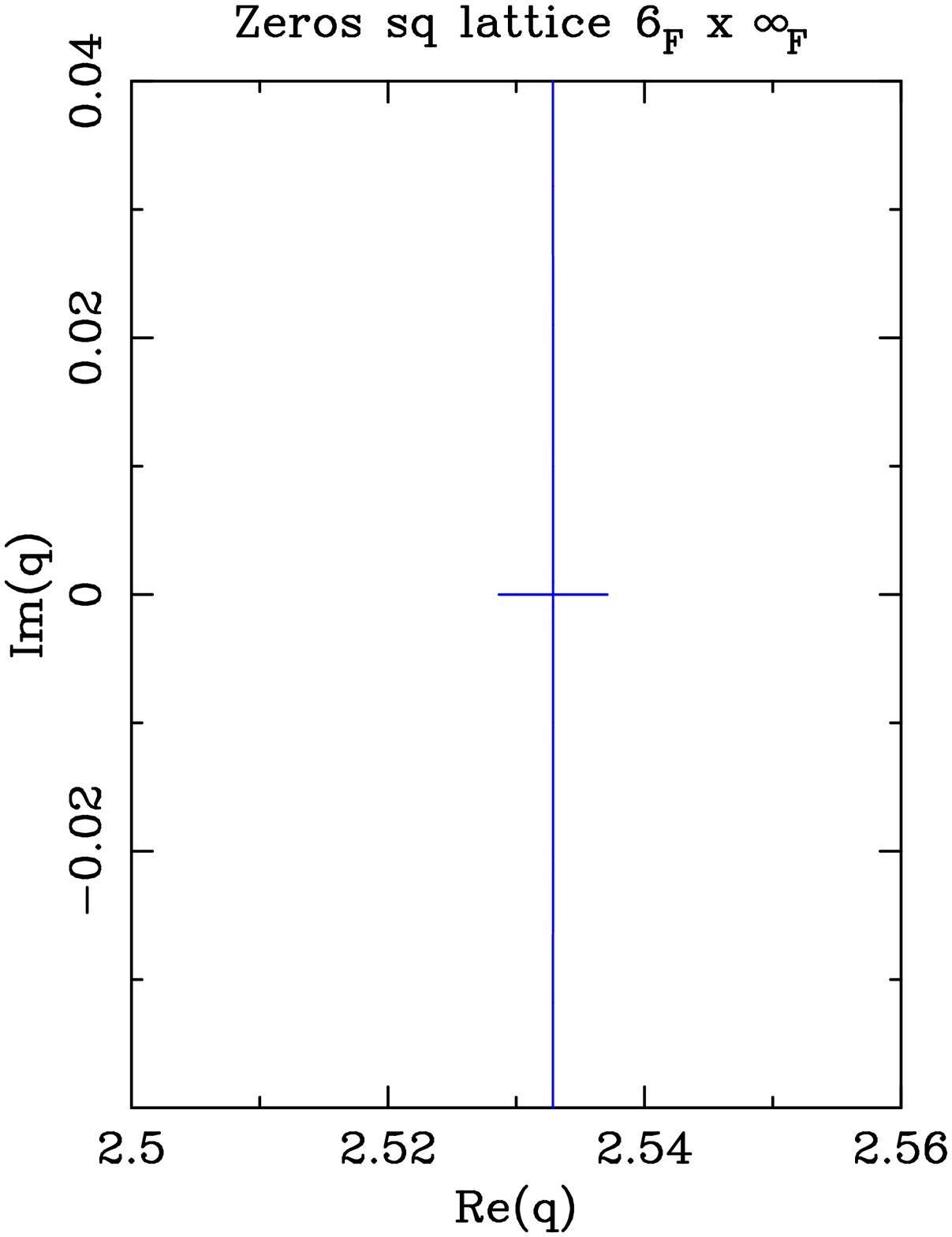}} \\
    \multicolumn{2}{c}{(a)} \\
    \epsfxsize=200pt\epsffile{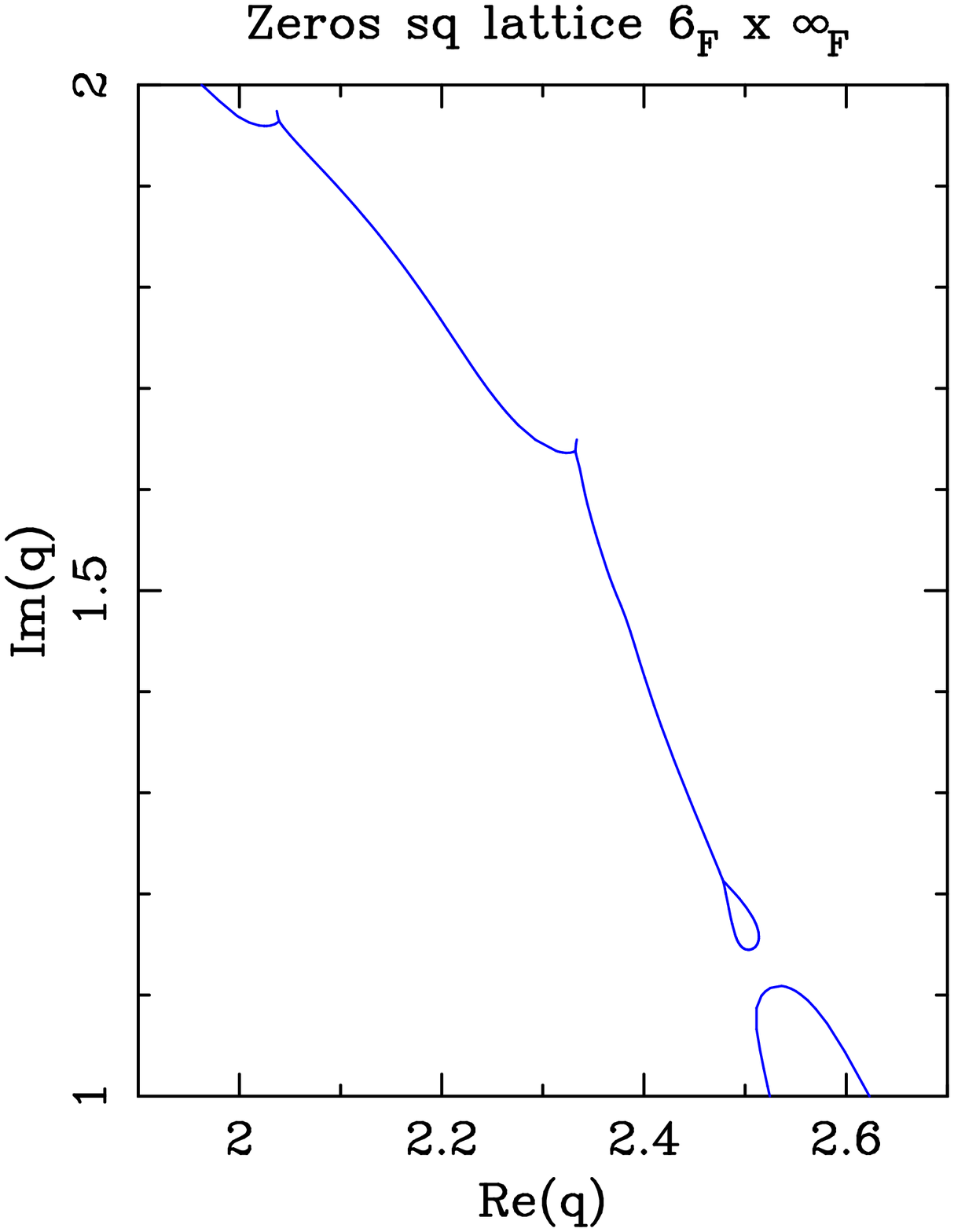} &
    \epsfxsize=200pt\epsffile{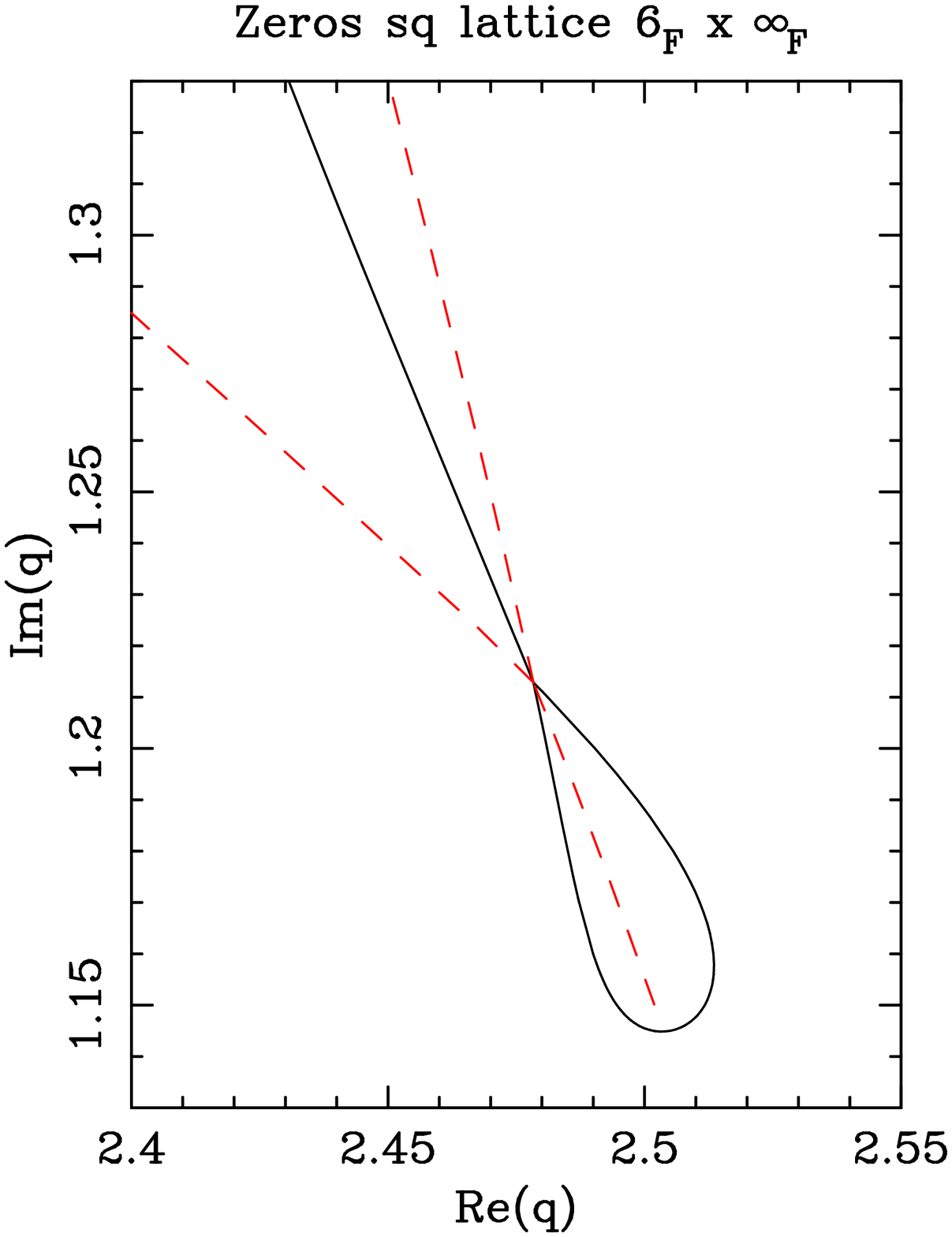}  \\
    (b)  & (c)
  \end{tabular}
  \caption{
  Detail of the limiting curves $\scrb$ for
  the $q$-state Potts antiferromagnet on a square lattice $6_F\times\infty_F$.
  (a) Region around the double point at $q \approx 2.53287$.
      The value of $t$ is continuous around this double point,
      with $t\approx 0.00985$.
  (b) Region containing the three T points.
      On the upper T point $q \approx 2.039 + 1.964\,i$,
      we have $t \approx (0.871,0.0521,0.970)$ with
      $\theta \approx (1.434,0.104,1.540)$;
      on the middle T point $q \approx 2.332 + 1.638\,i$,
      we have $t \approx (18.021,0.0843,7.119)$
      and $\theta \approx (3.031,0.168,2.863)$;
      on the lower T point $q \approx 2.478+1.213\,i$, we have
      $t \approx (0.272,0.618,1.069)$ and $\theta \approx (0.532,1.107,1.638)$.
  (c) Detail of the bulb-like region around $q \approx 2.478+1.213\,i$.
      Dominant crossing curves are depicted in solid black lines, while
      subdominant crossing curves are shown with dashed \subdominantcolor\
      lines.
  }
\label{Figure_sq_6FxInftyF_bis}
\end{figure}

\clearpage
%
%
\begin{figure}
  \centering
  \epsfxsize=400pt\epsffile{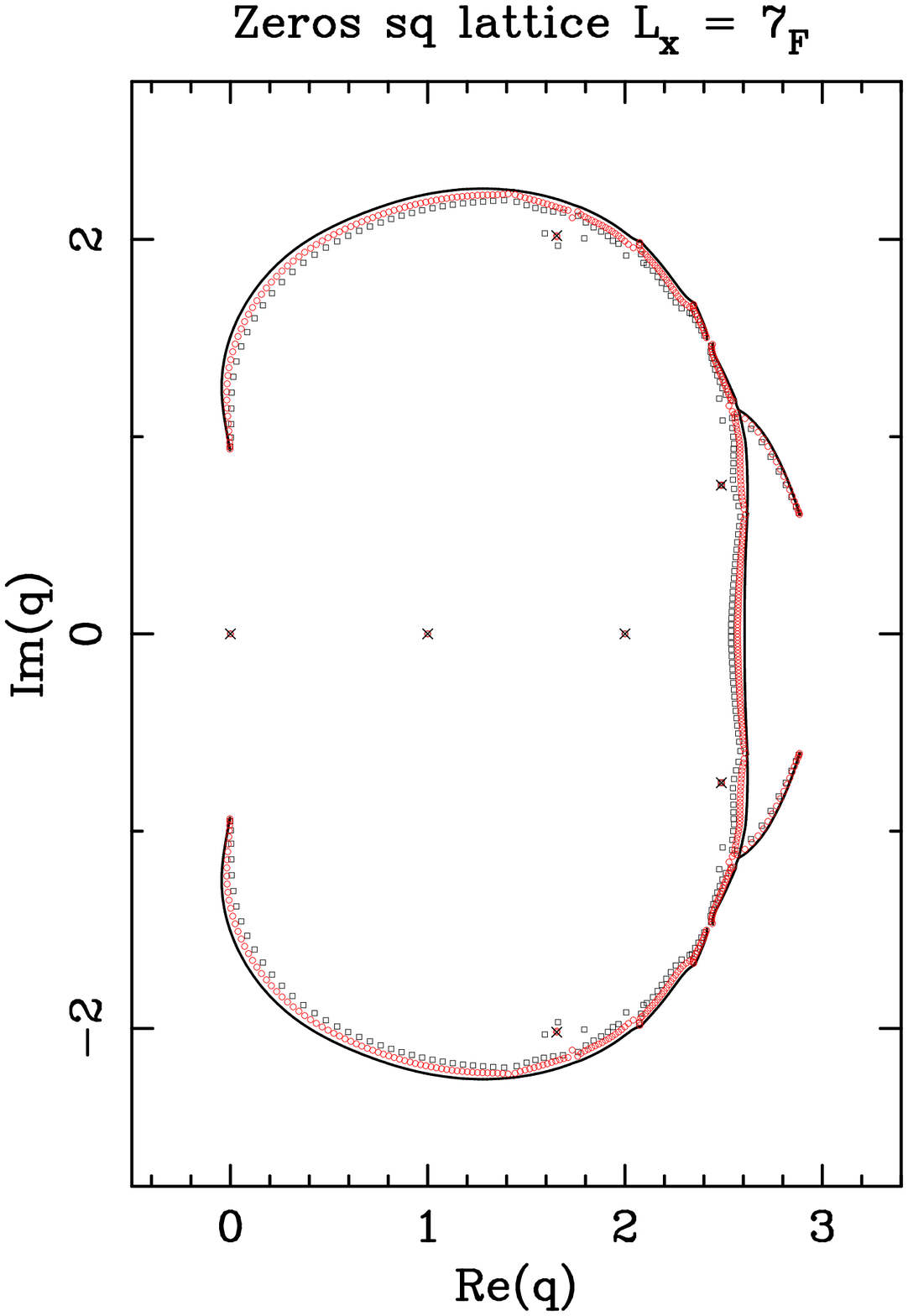}
  \caption{
  Zeros of the partition function of the $q$-state Potts antiferromagnet
  on a square lattices $7_F \times 35_F$ (squares),
  $7_F \times 70_F$ (circles) and $7_F\times\infty_F$ (solid line).
  The isolated limiting zeros are depicted by a $\times$.
  The limiting curve was computed using the direct-search method.
 }
\label{Figure_sq_7FxInftyF}
\end{figure}

\clearpage
%
%
\begin{figure}
  \centering
  \begin{tabular}{cc}
    \epsfxsize=200pt\epsffile{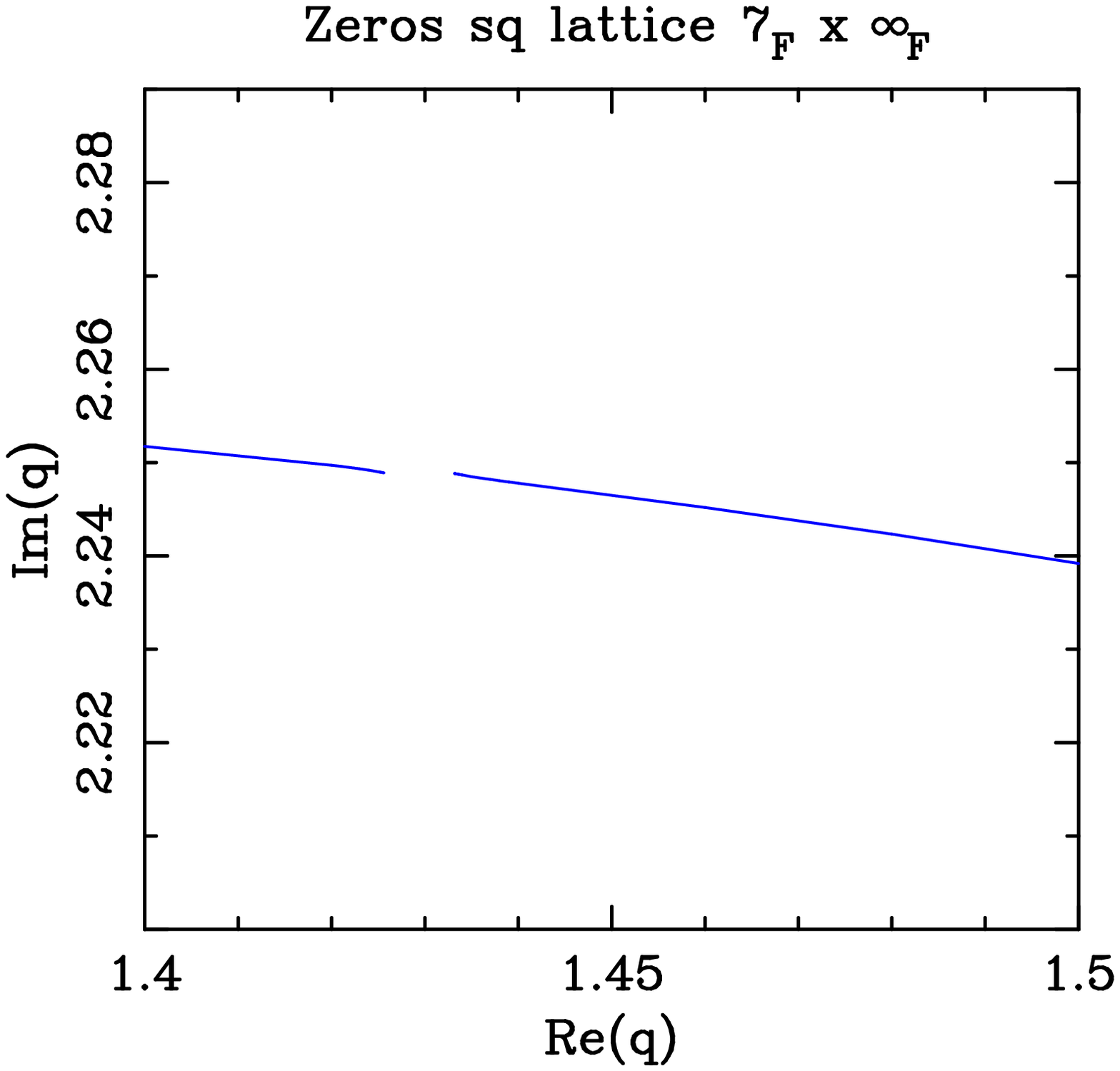} &
    \epsfxsize=200pt\epsffile{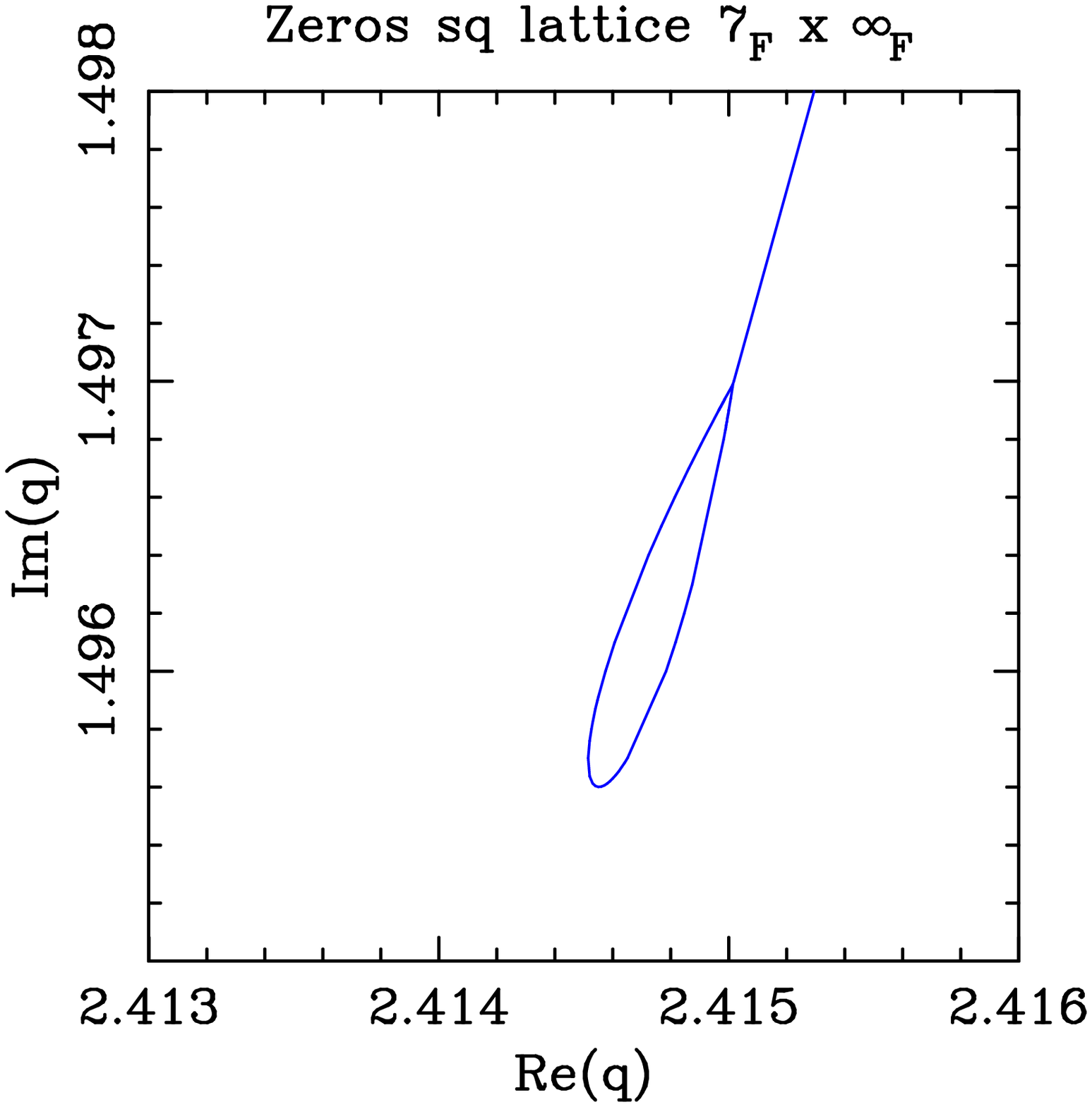} \\
    (a) & (b) \\[2mm]
    \epsfxsize=200pt\epsffile{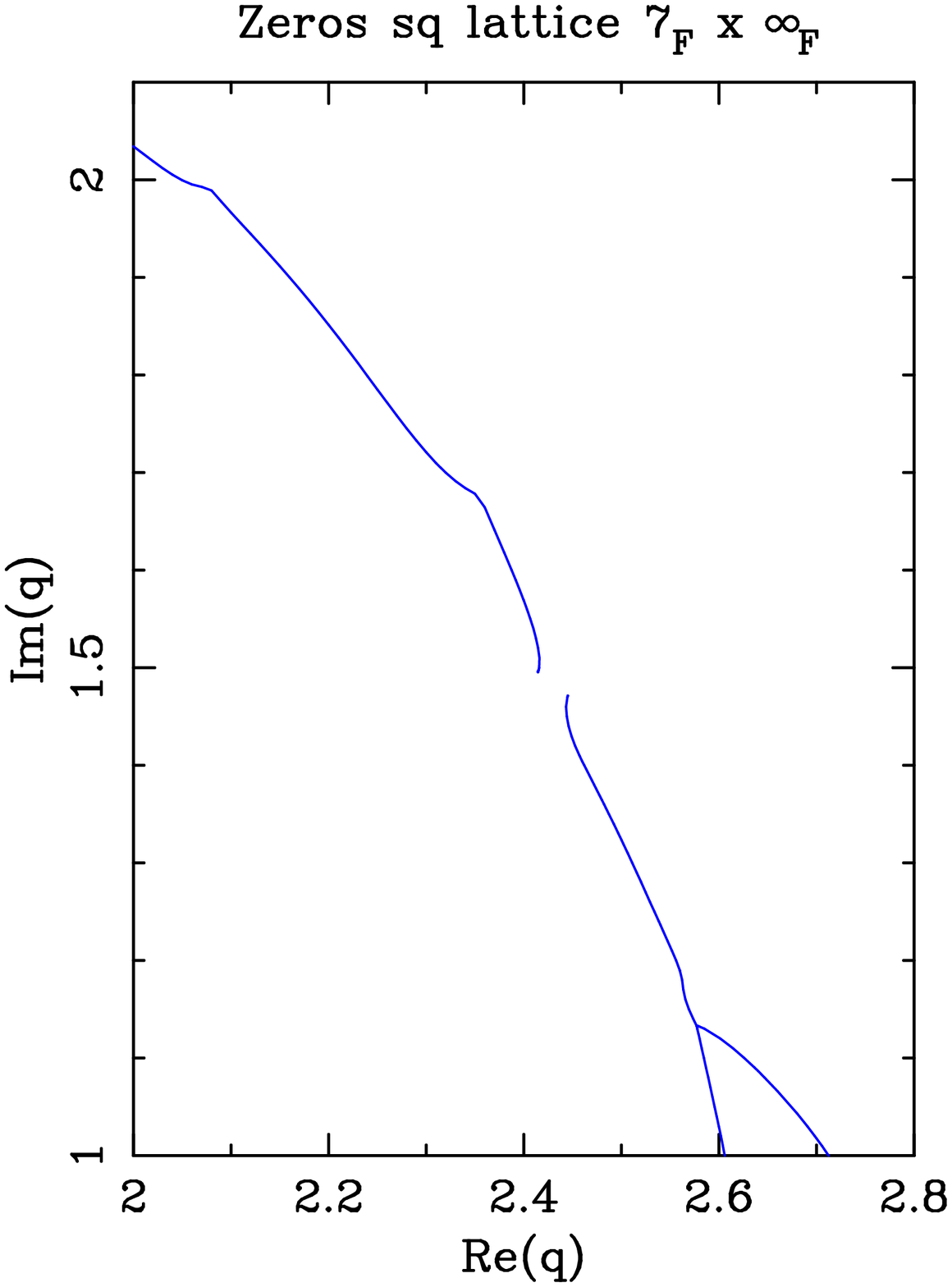} &
    \epsfxsize=200pt\epsffile{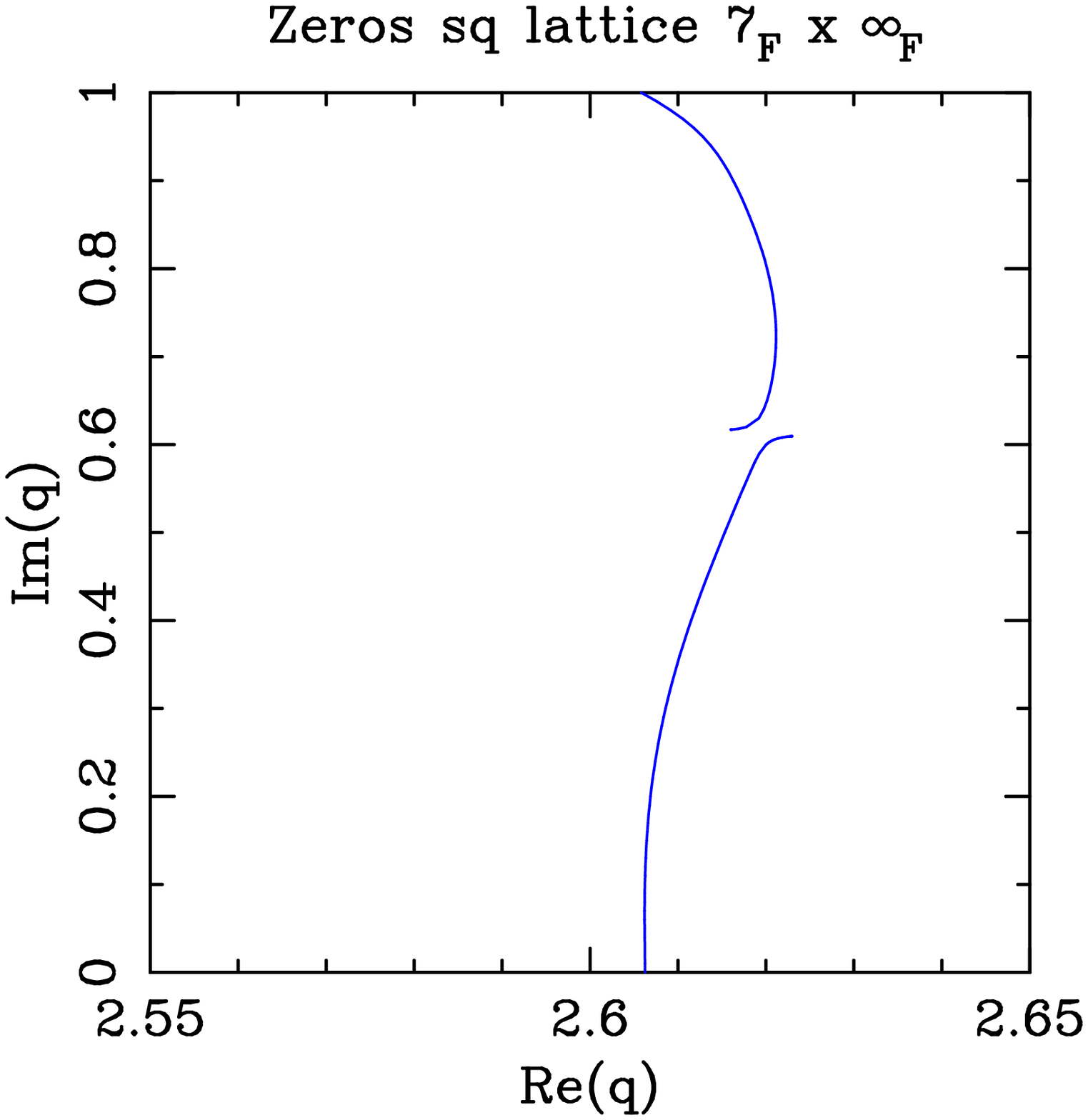} \\
    (c)  & (d)
  \end{tabular}
  \caption{
  Detail of the limiting curves $\scrb$ for
  the $q$-state Potts antiferromagnet on a square lattice $7_F\times\infty_F$.
  (a) Region around the gap between $q\approx 1.425603 +2.248902\,i$ and
      $q\approx 1.433184 +2.248834\,i$.
  (b) Bulb-like region around the T point
      $q \approx 2.415 + 1.497\,i$. At this point we have
      $t \approx (0.023,1.248,1.310)$ and $\theta \approx (0.047,1.791,1.838)$.
  (c) Region around the gap between the bulb-like region at
      $q \approx 2.415 + 1.497\,i$ and the endpoint at
      $q \approx 2.445207 \pm 1.471332\,i$.
      There is also a T point at $q \approx 2.577+1.133\,i$,
      where $t \approx (2.108,2.737,1.016)$
      and $\theta \approx (2.257,2.441,1.586)$.
  (d) Region around the gap between $q\approx 2.616006 + 0.616910\,i$ and
      $q \approx 2.622974+0.609548\,i$.
  }
\label{Figure_sq_7FxInftyF_bis}
\end{figure}

\clearpage
%
%
\begin{figure}
 \centering
  \epsfxsize=400pt\epsffile{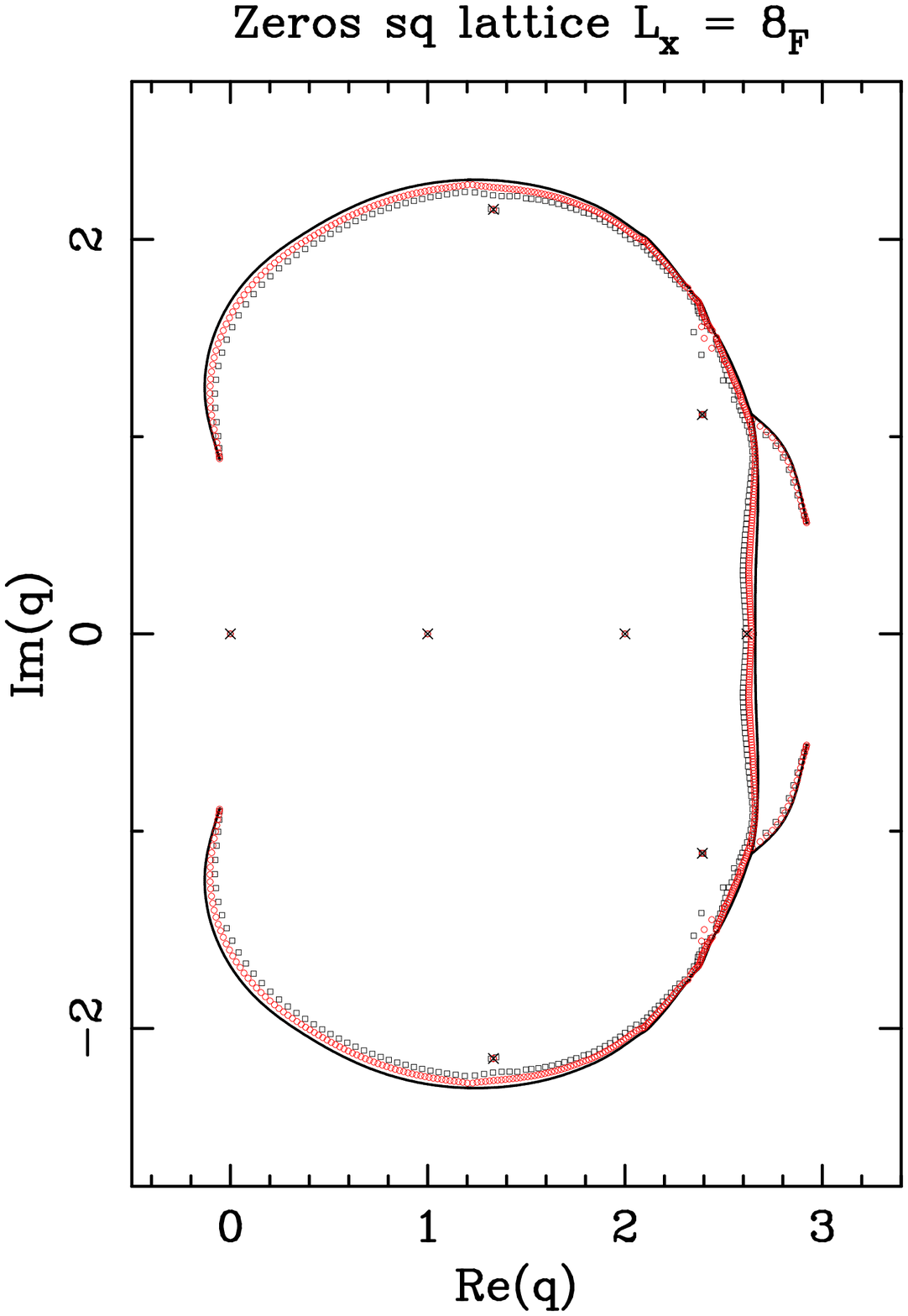}
  \caption{
  Zeros of the partition function of the $q$-state Potts antiferromagnet
  on a square lattices $8_F \times 40_F$ (squares),
  $8_F \times 80_F$ (circles) and $8_F\times\infty_F$ (solid line).
  The isolated limiting zeros are depicted by a $\times$.
  The limiting curve was computed using the direct-search method.
  }
\label{Figure_sq_8FxInftyF}
\end{figure}

\clearpage
%
%
\begin{figure}
  \centering
  \begin{tabular}{cc}
    \epsfxsize=200pt\epsffile{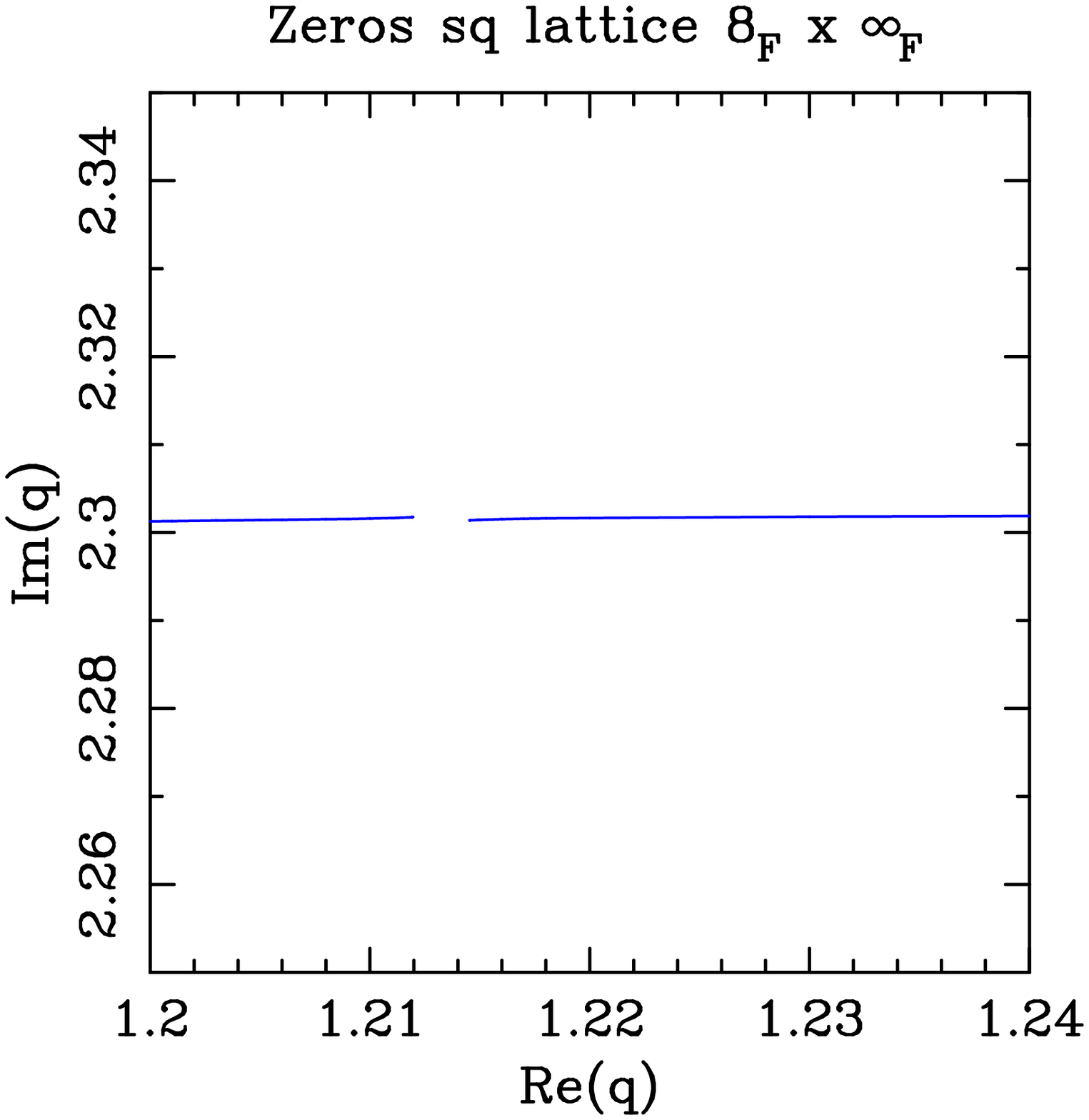} &
    \epsfxsize=200pt\epsffile{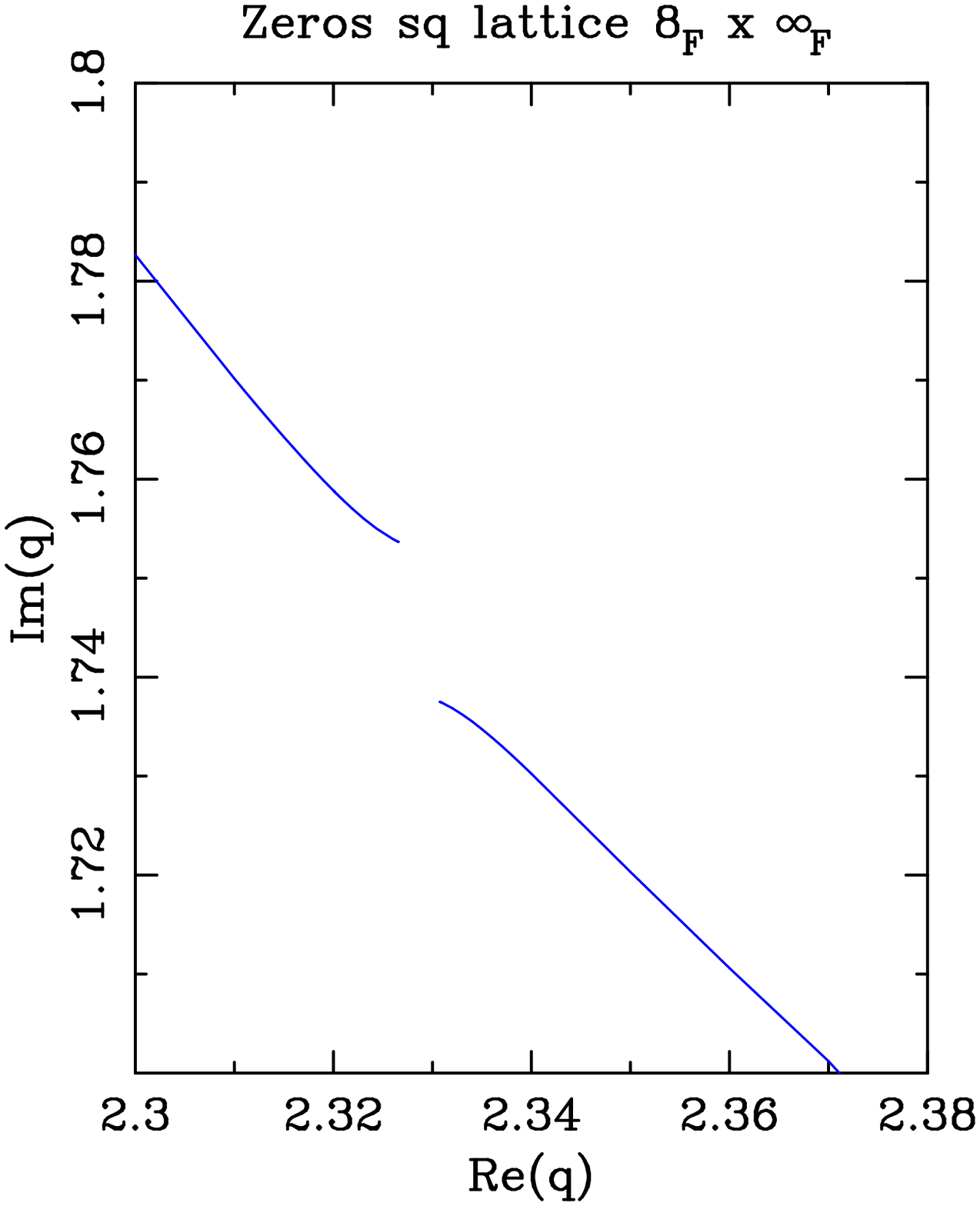}\\
    (a)  & (b) \\[4mm]
    \epsfxsize=200pt\epsffile{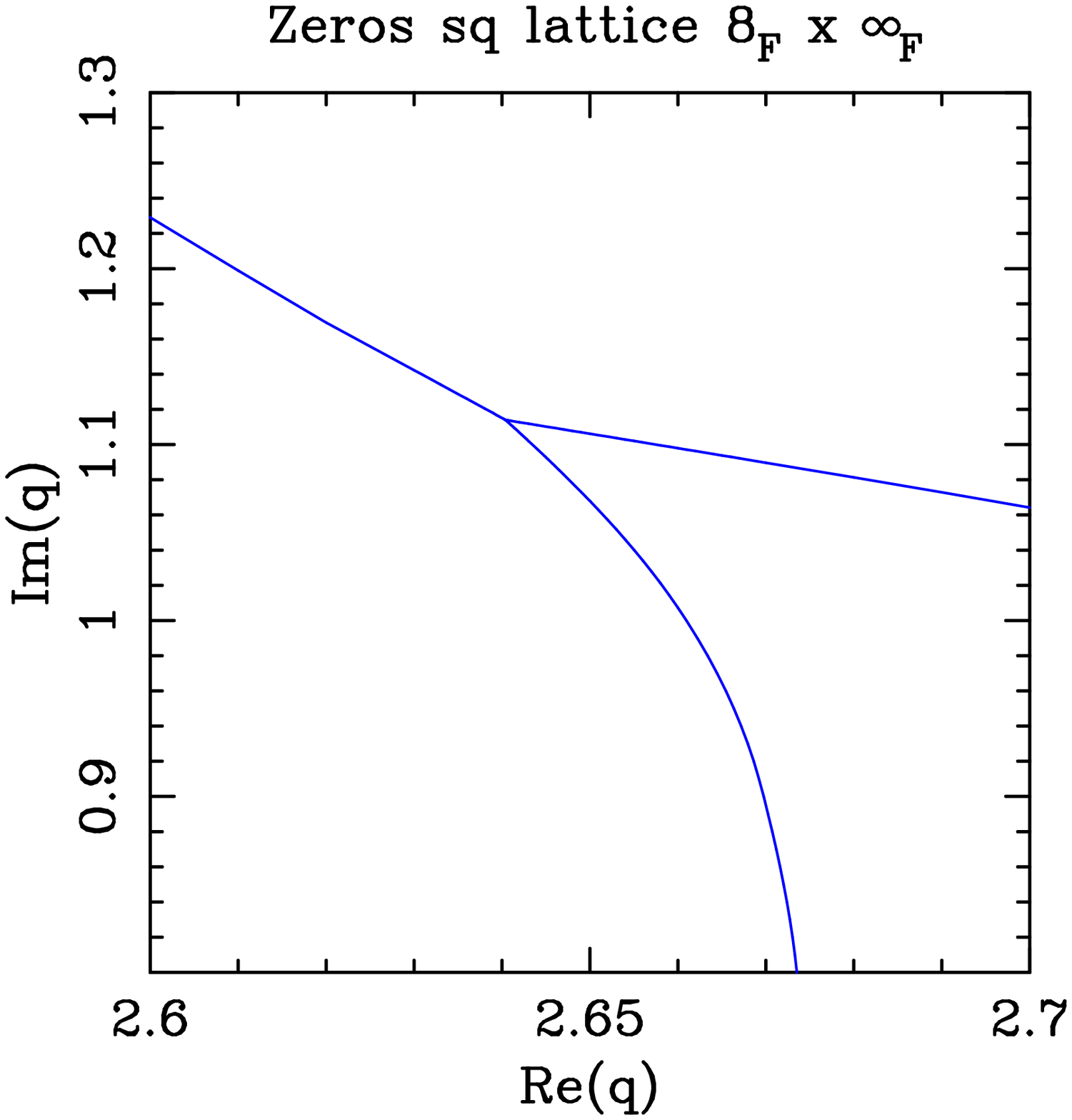} &
    \epsfxsize=200pt\epsffile{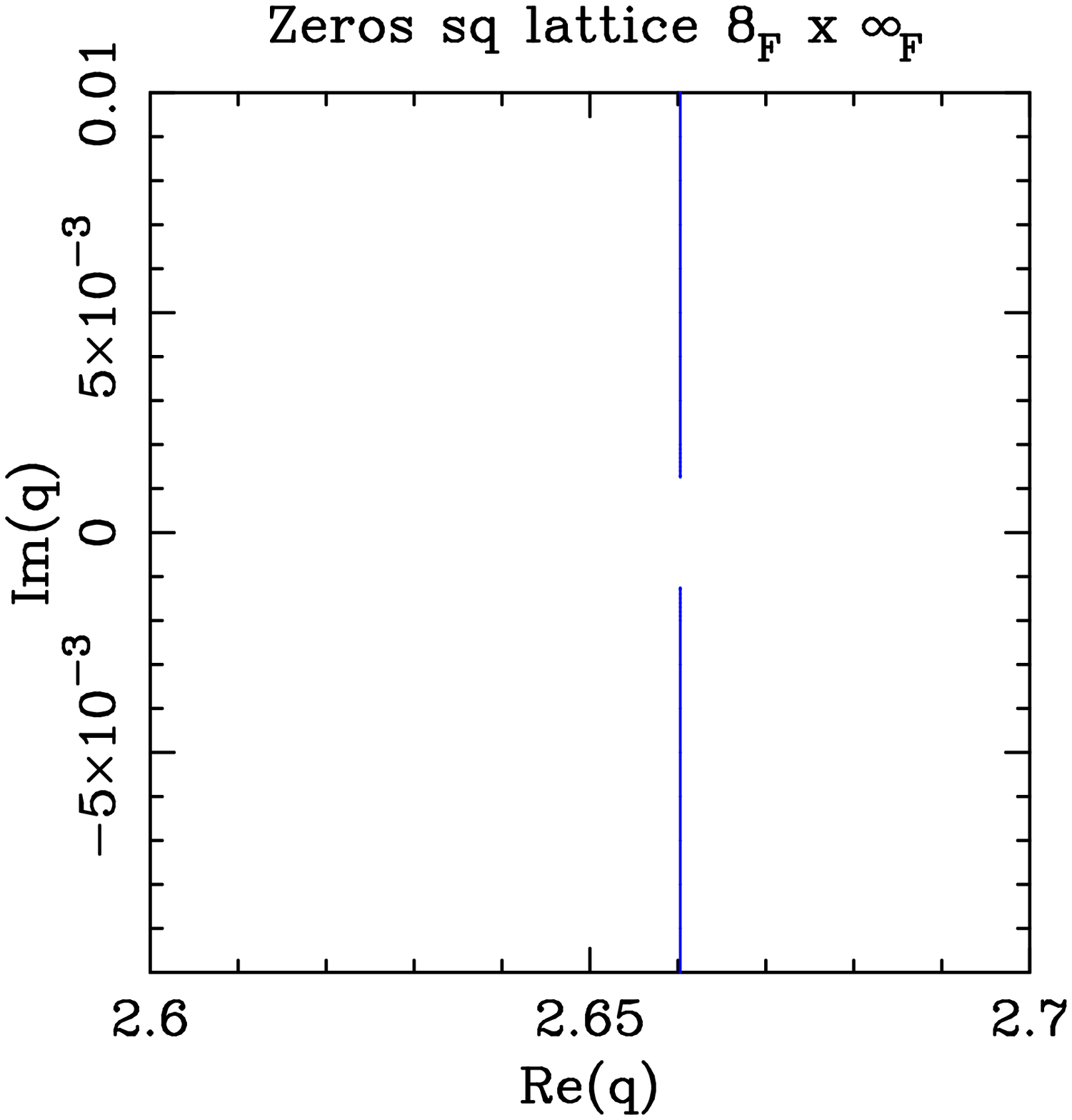} \\
    (c)  & (d)
  \end{tabular}
  \caption{
  Detail of the limiting curves $\scrb$ for
  the $q$-state Potts antiferromagnet on a square lattice $8_F\times\infty_F$.
  (a) Region around the small gap at $q\approx 1.21+2.30\,i$.
  (b) Region around the gap at $q\approx 2.32+1.75\,i$.
  (c) Region around the T point at $q\approx 2.640 + 1.114\,i$. At this point
      we have $t\approx (0.993,1.013,0.0113)$ and
      $\theta\approx(1.563,1.584,0.023)$.
  (d) Region around the tiny gap at $q\approx 2.660260$.
  }
\label{Figure_sq_8FxInftyF_bis}
\end{figure}

\clearpage
%
%
\begin{figure}
  \centering
  \epsfxsize=400pt\epsffile{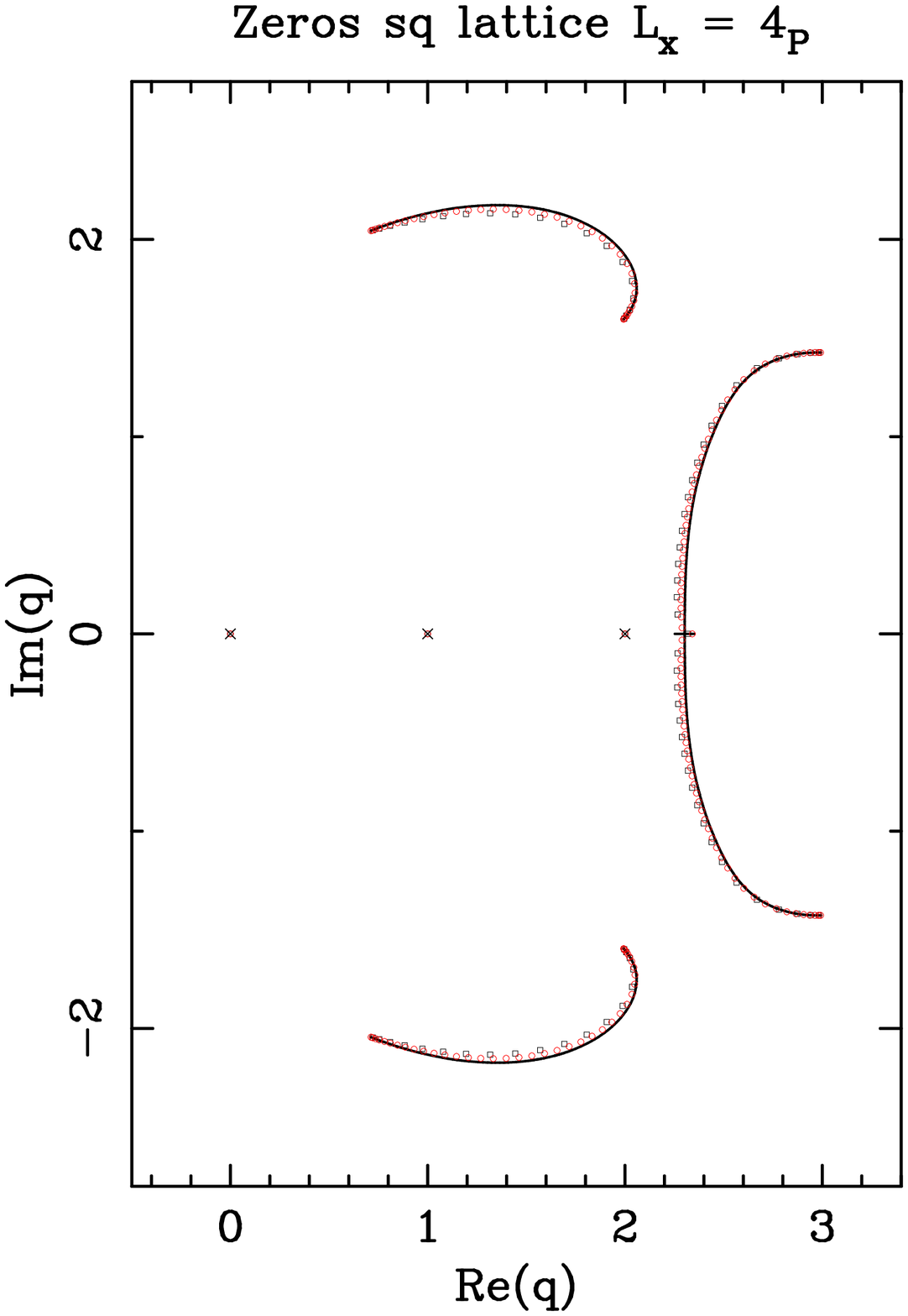}
  \caption{
  Zeros of the partition function of the $q$-state Potts antiferromagnet
  on a square lattices $4_P \times 20_F$ (squares),
  $4_P \times 40_F$ (circles) and $4_P\times\infty_F$ (solid line).
  The isolated limiting zeros are depicted by a $\times$.
  The limiting curve was computed using the resultant method.
  }
\label{Figure_sq_4PxInftyF}
\end{figure}

\clearpage
%
%
\begin{figure}
  \centering
  \epsfxsize=400pt\epsffile{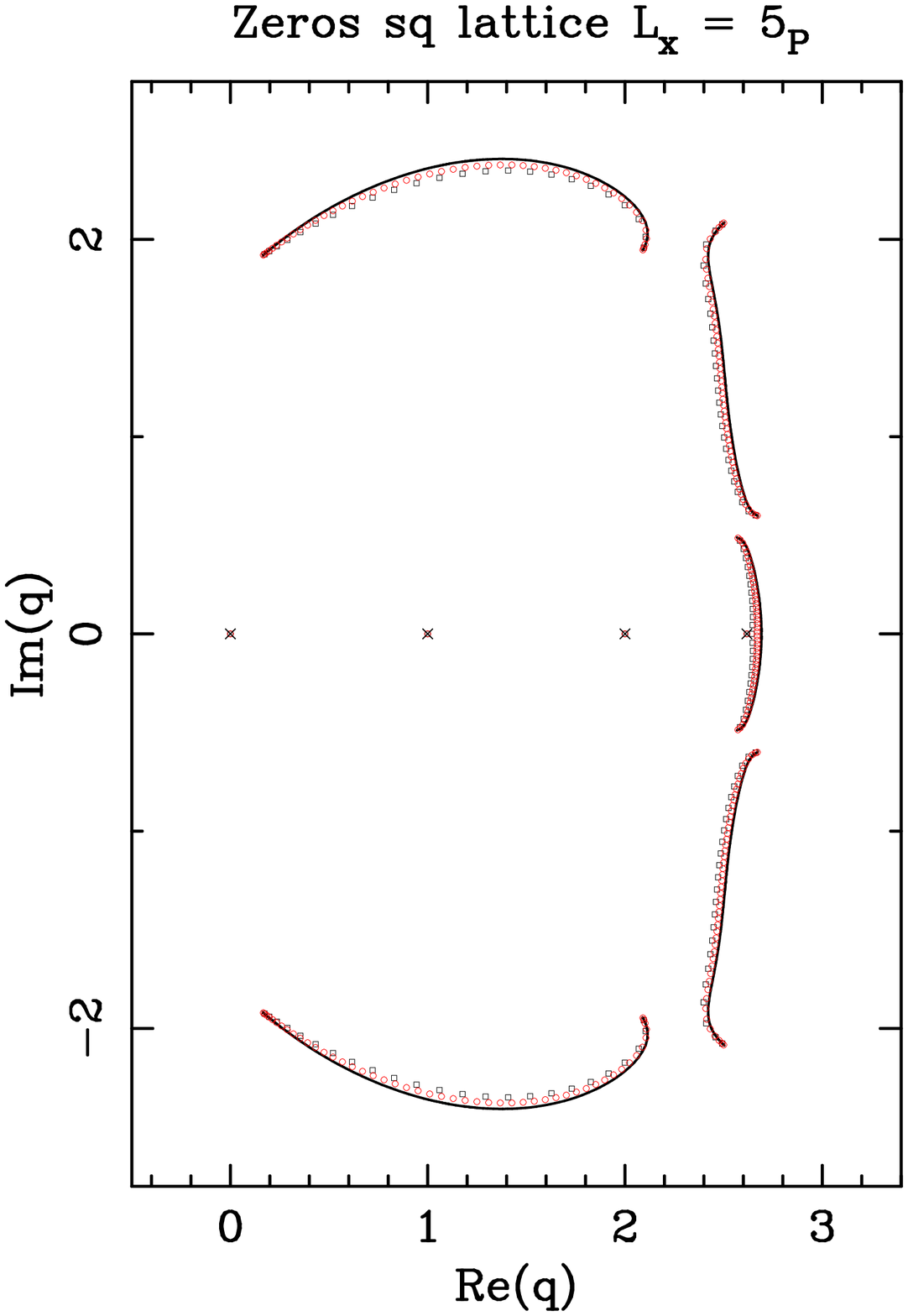}
  \caption{
  Zeros of the partition function of the $q$-state Potts antiferromagnet
  on a square lattices $5_P \times 25_F$ (squares),
  $5_P \times 50_F$ (circles) and $5_P\times\infty_F$ (solid line).
  The isolated limiting zeros are depicted by a $\times$.
  The limiting curve was computed using the resultant method.
  }
\label{Figure_sq_5PxInftyF}
\end{figure}

\clearpage
%
%
\begin{figure}
  \centering
  \epsfxsize=400pt\epsffile{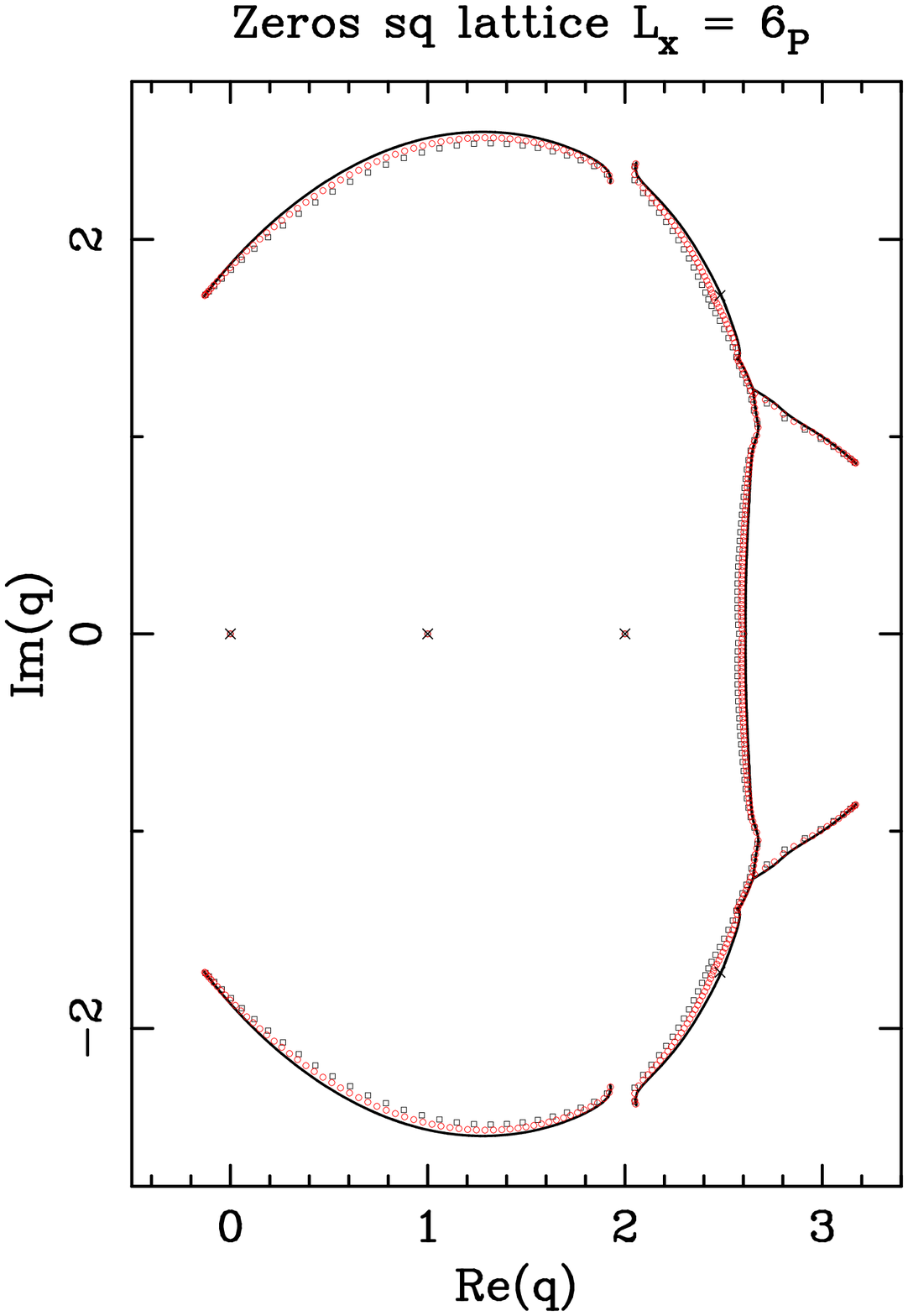}
  \caption{
  Zeros of the partition function of the $q$-state Potts antiferromagnet
  on a square lattices $6_P \times 30_F$ (squares),
  $6_P \times 60_F$ (circles) and $6_P\times\infty_F$ (solid line).
  The isolated limiting zeros are depicted by a $\times$.
  The limiting curve was computed using the resultant method.
  }
\label{Figure_sq_6PxInftyF}
\end{figure}

\clearpage
%
%
\begin{figure}
  \centering
\begin{tabular}{cc}
  \epsfxsize=200pt\epsffile{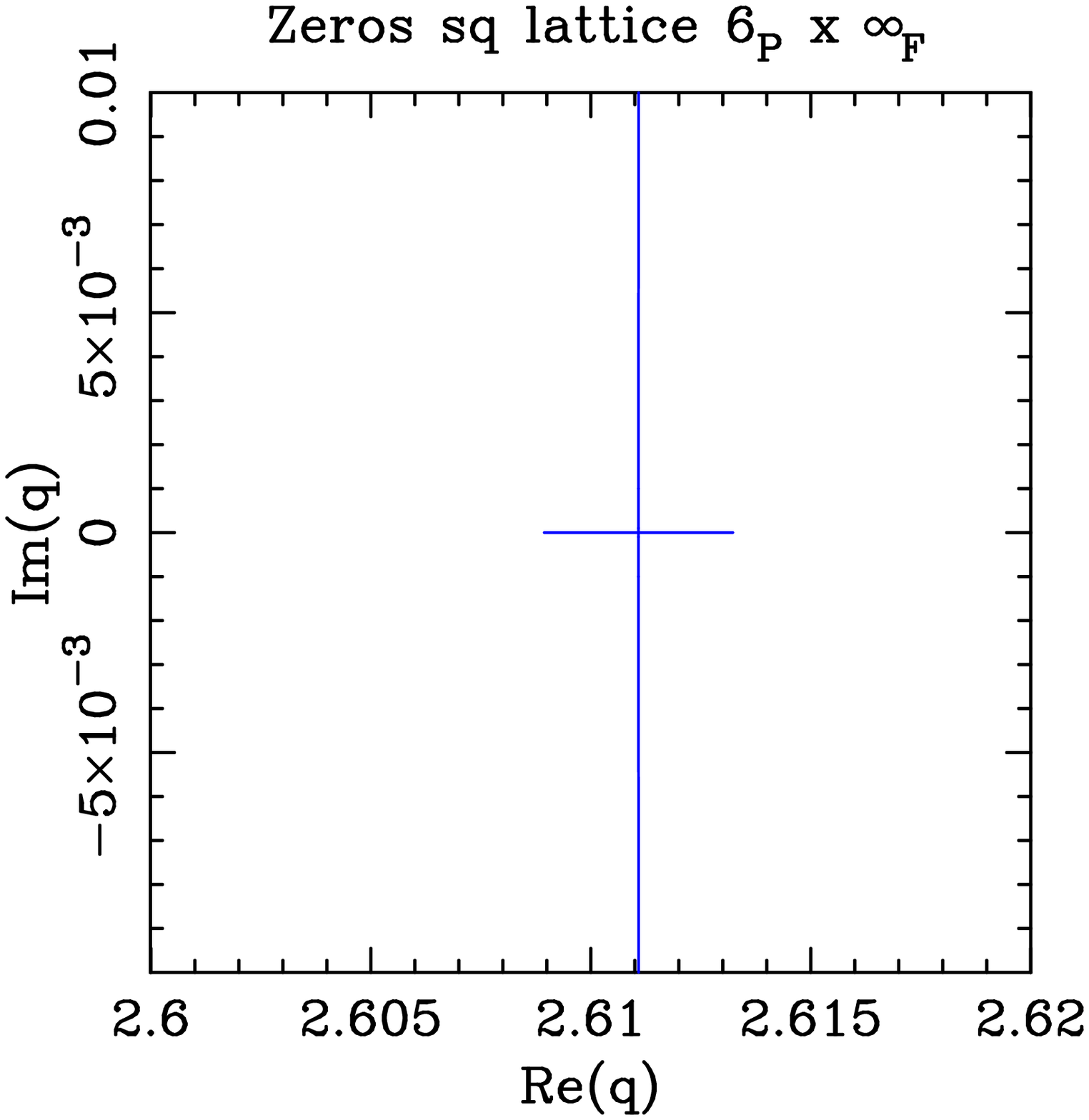} &
  \epsfxsize=200pt\epsffile{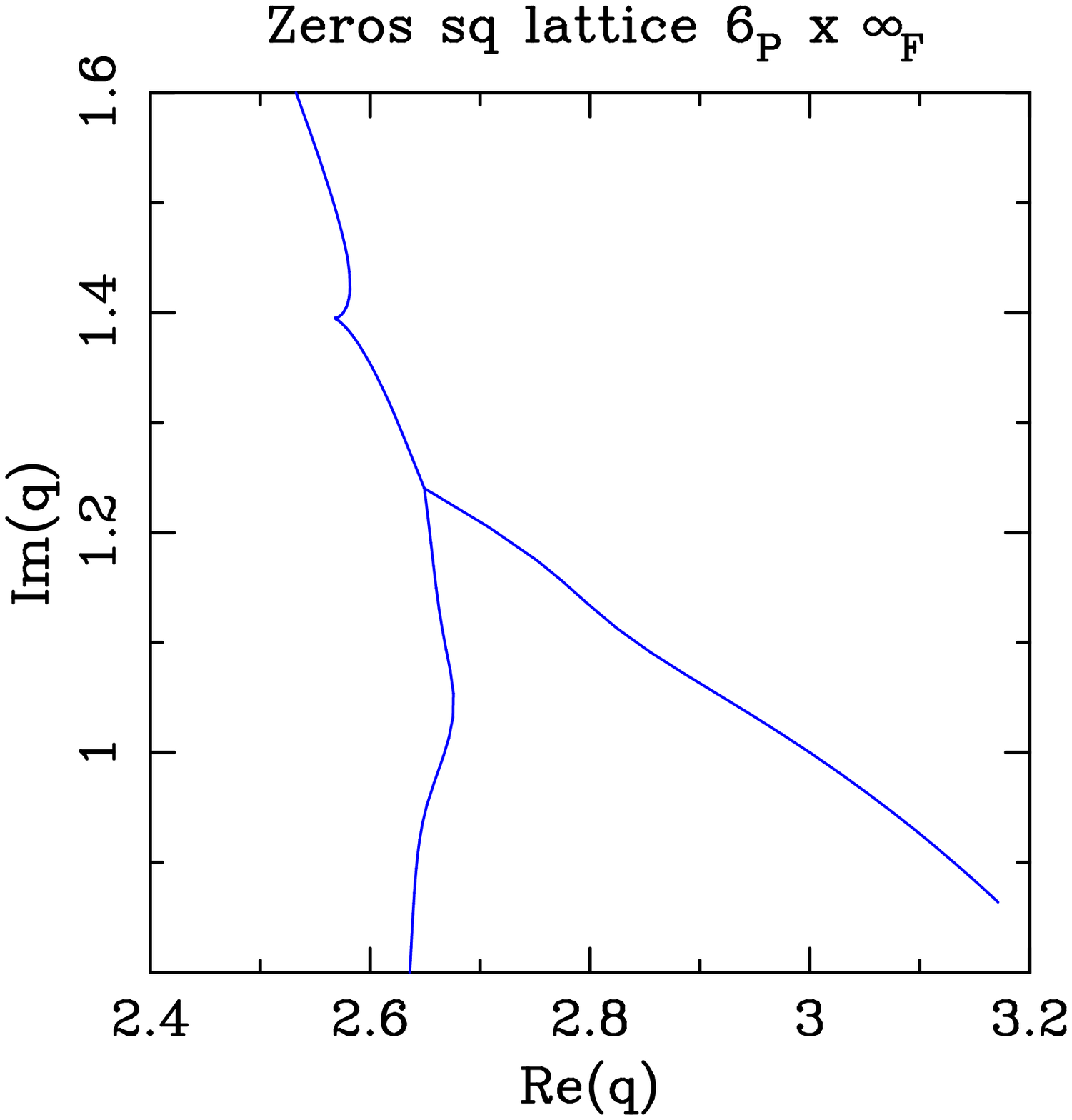} \\
  (a)  & (b) \\[4mm]
  \epsfxsize=200pt\epsffile{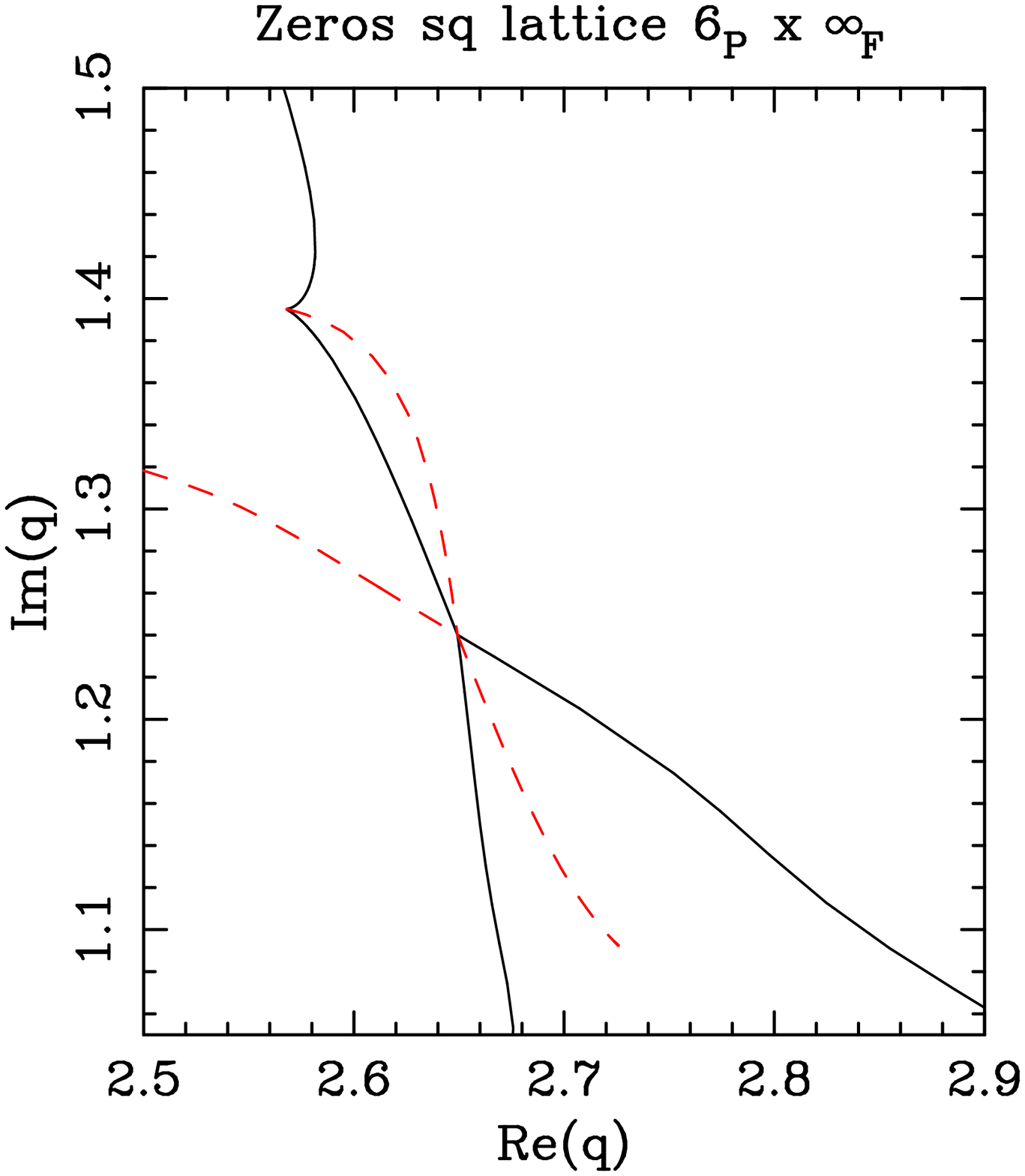} &
  \epsfxsize=200pt\epsffile{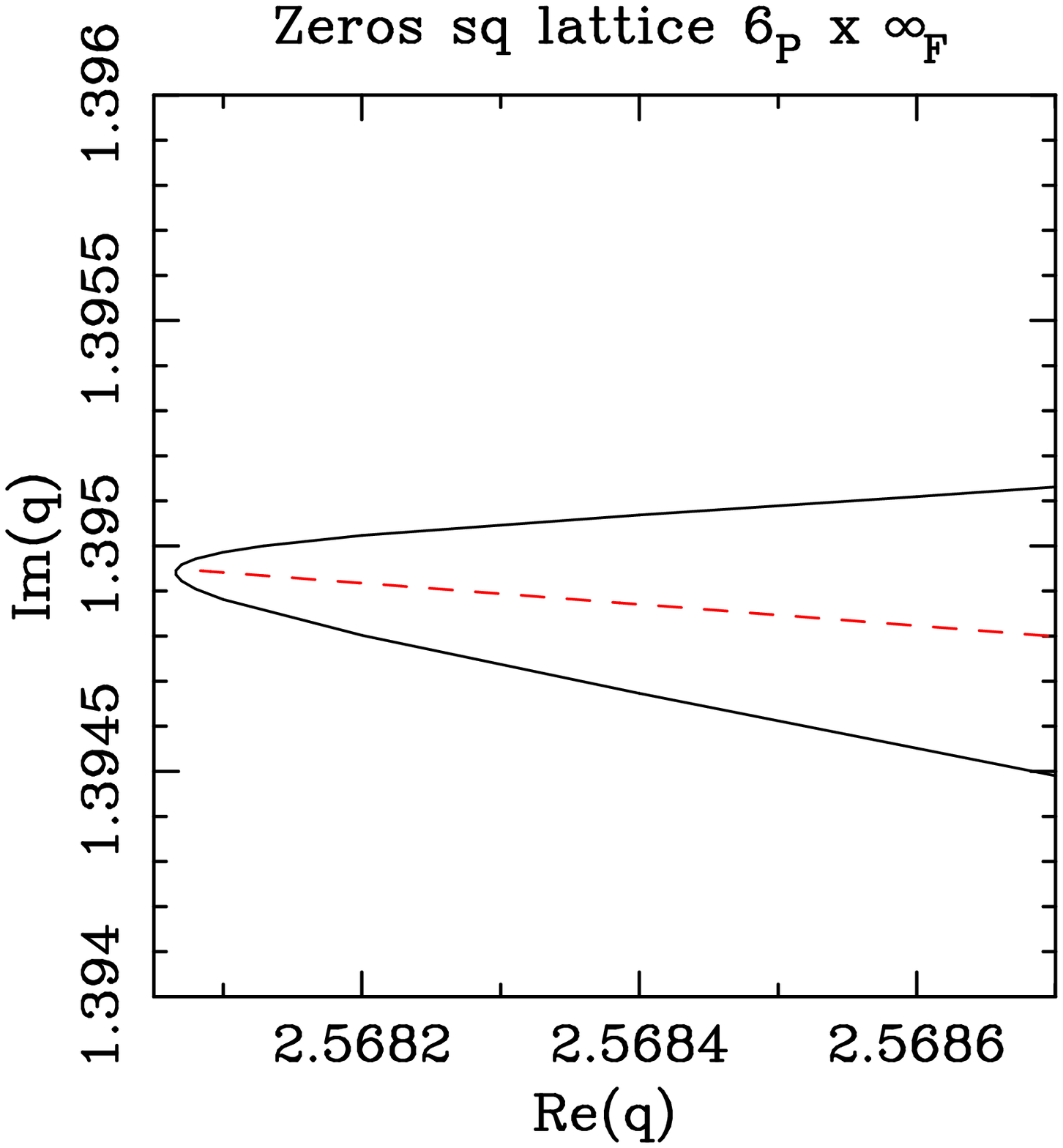} \\
  (c)  & (d) \\
\end{tabular}
  \caption{
  Detail of the limiting curves $\scrb$ for
  the $q$-state Potts antiferromagnet on a square lattice $6_P\times\infty_F$.
  (a) Region around the double point $q \approx 2.6110857$.
      At this double point we have $t \approx 0.0053$.
  (b) Region around the T point at $q\approx 2.650+1.240\,i$.
      At this T point we have
      $t \approx (0.392,1.224,0.562)$ and $\theta\approx(0.748,1.771,1.024)$.
  (c) The same as in (b), but we show the dominant (solid black line)
      and subdominant (dashed \subdominantcolor\ line) crossing curves.
  (d) Blow-up of region around the quasi-cusp at $q\approx 2.568+1.398\,i$.
  }
\label{Figure_sq_6PxInftyF_bis}
\end{figure}

\clearpage
%
%
\begin{figure}
  \centering
  \epsfxsize=400pt\epsffile{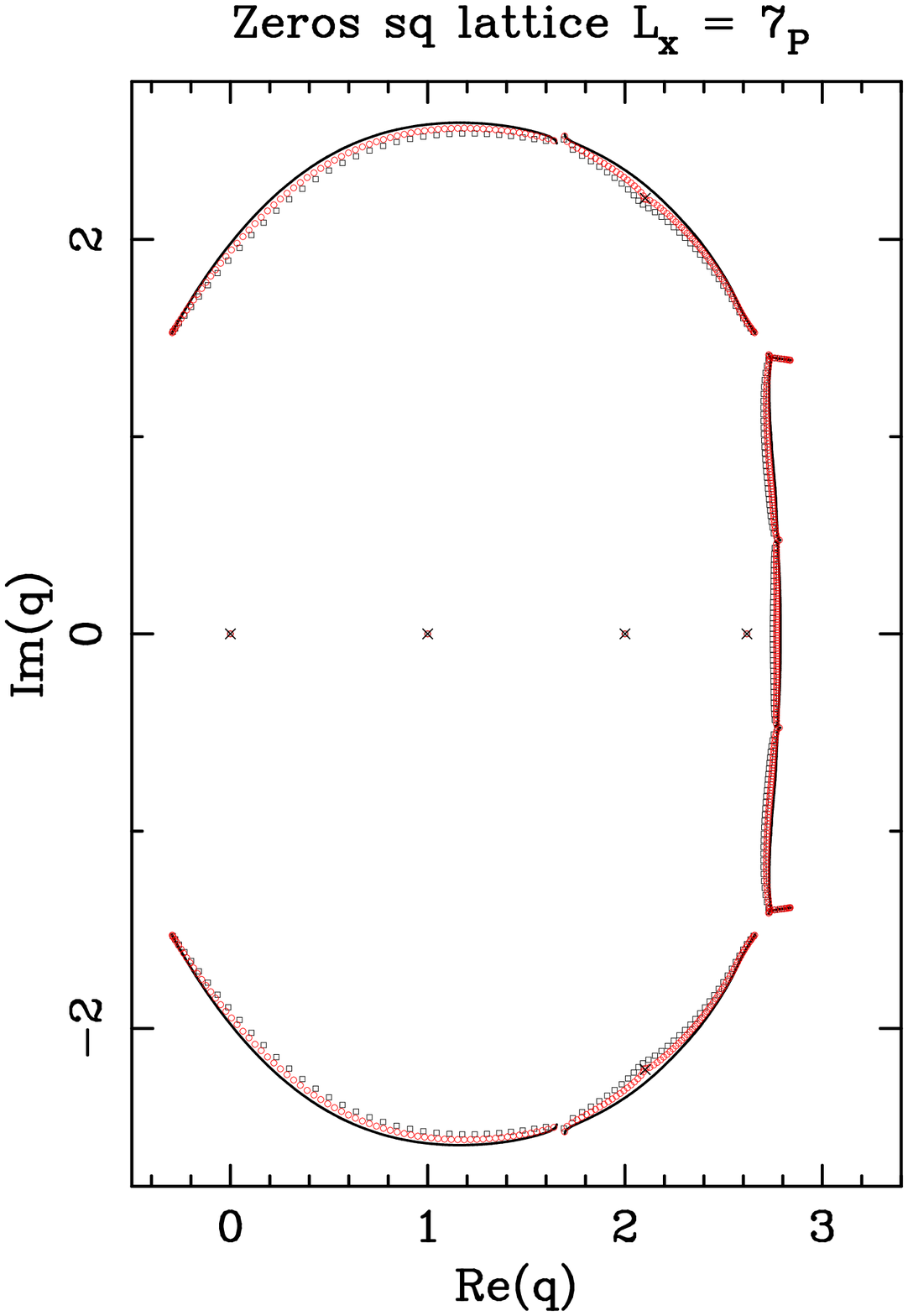}
  \caption{
  Zeros of the partition function of the $q$-state Potts antiferromagnet
  on a square lattices $7_P \times 35_F$ (squares),
  $7_P \times 70_F$ (circles) and $7_P\times\infty_F$ (solid line).
  The isolated limiting zeros are depicted by a $\times$.
  The limiting curve was computed using the resultant method.
  }
\label{Figure_sq_7PxInftyF}
\end{figure}

\clearpage
%
%
\begin{figure}
  \centering
  \begin{tabular}{cc}
    \epsfxsize=200pt\epsffile{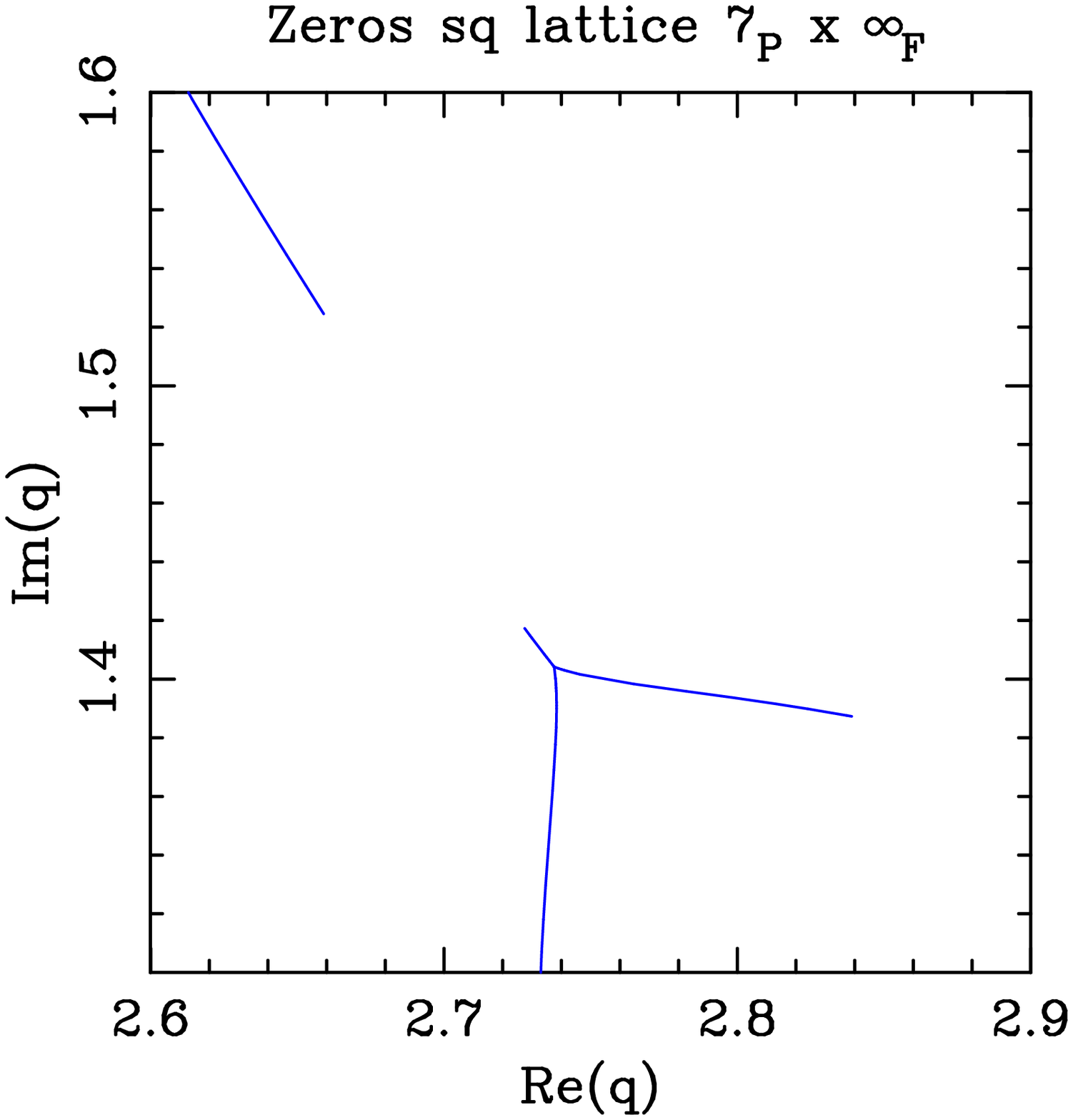} &
    \epsfxsize=200pt\epsffile{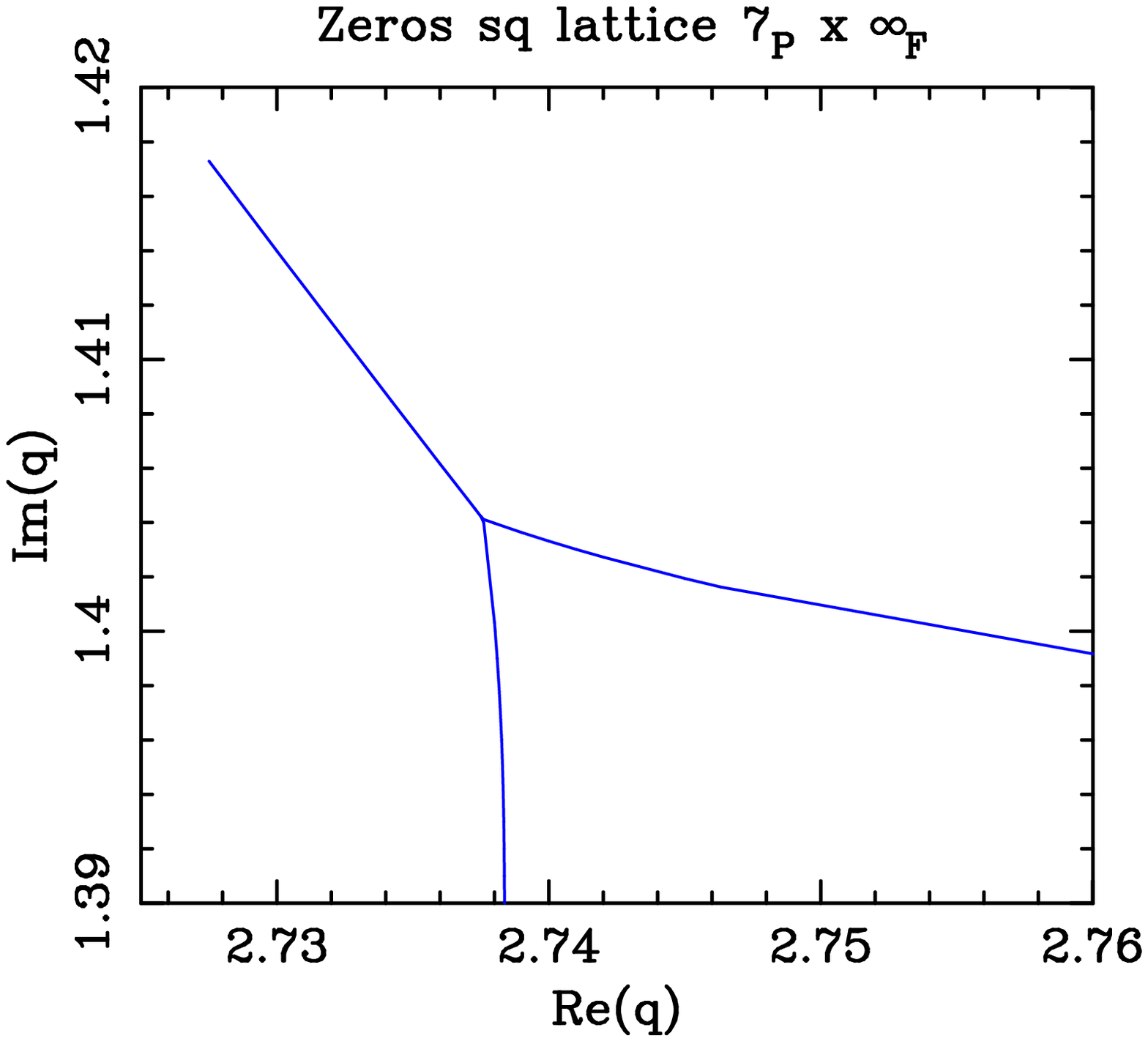} \\
    (a)  & (b) \\[5mm]
    \multicolumn{2}{c}{\epsfxsize=200pt\epsffile{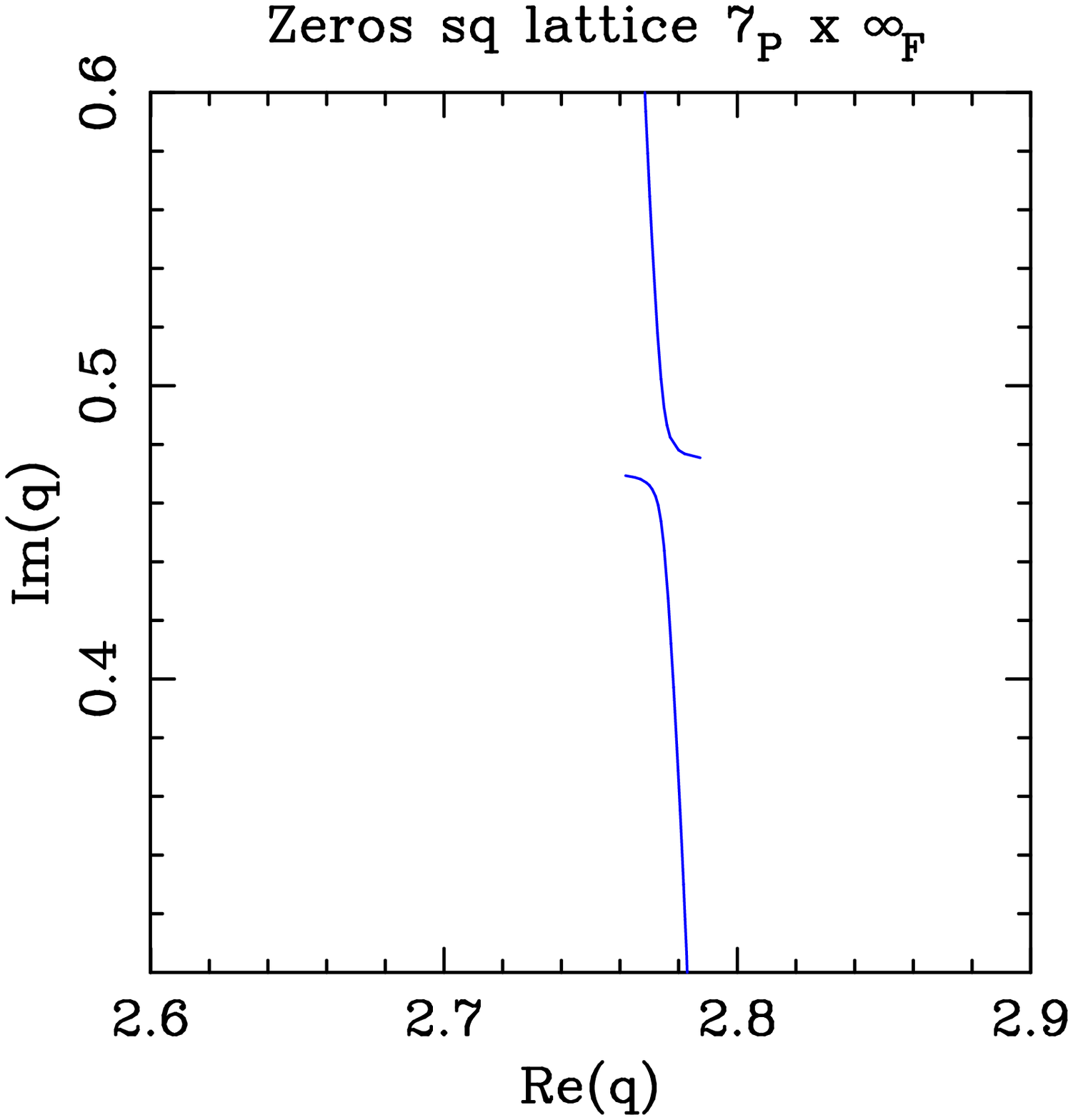}} \\
    \multicolumn{2}{c}{(c)}
  \end{tabular}
  \caption{
  Detail of the limiting curves $\scrb$ for
  the $q$-state Potts antiferromagnet on a square lattice $7_P\times\infty_F$.
  (a) Region around the T point at $q\approx 2.737+1.405\,i$ and the
      gap between $q\approx 2.6590+1.525\,i$ and $q\approx2.7275+1.4173\,i$.
  (b) Detail of the region around the T point. At this point we have
      $t \approx (0.125,0.581,0.425)$ and $\theta\approx (0.248,1.053,0.804)$.
  (c) Region around the gap between the points $q\approx 2.7619+0.46936\,i$ and
      $q\approx 2.7873+0.47546\,i$.
  }
\label{Figure_sq_7PxInftyF_bis}
\end{figure}

\clearpage
%
%
\begin{figure}
  \centering
  \epsfxsize=400pt\epsffile{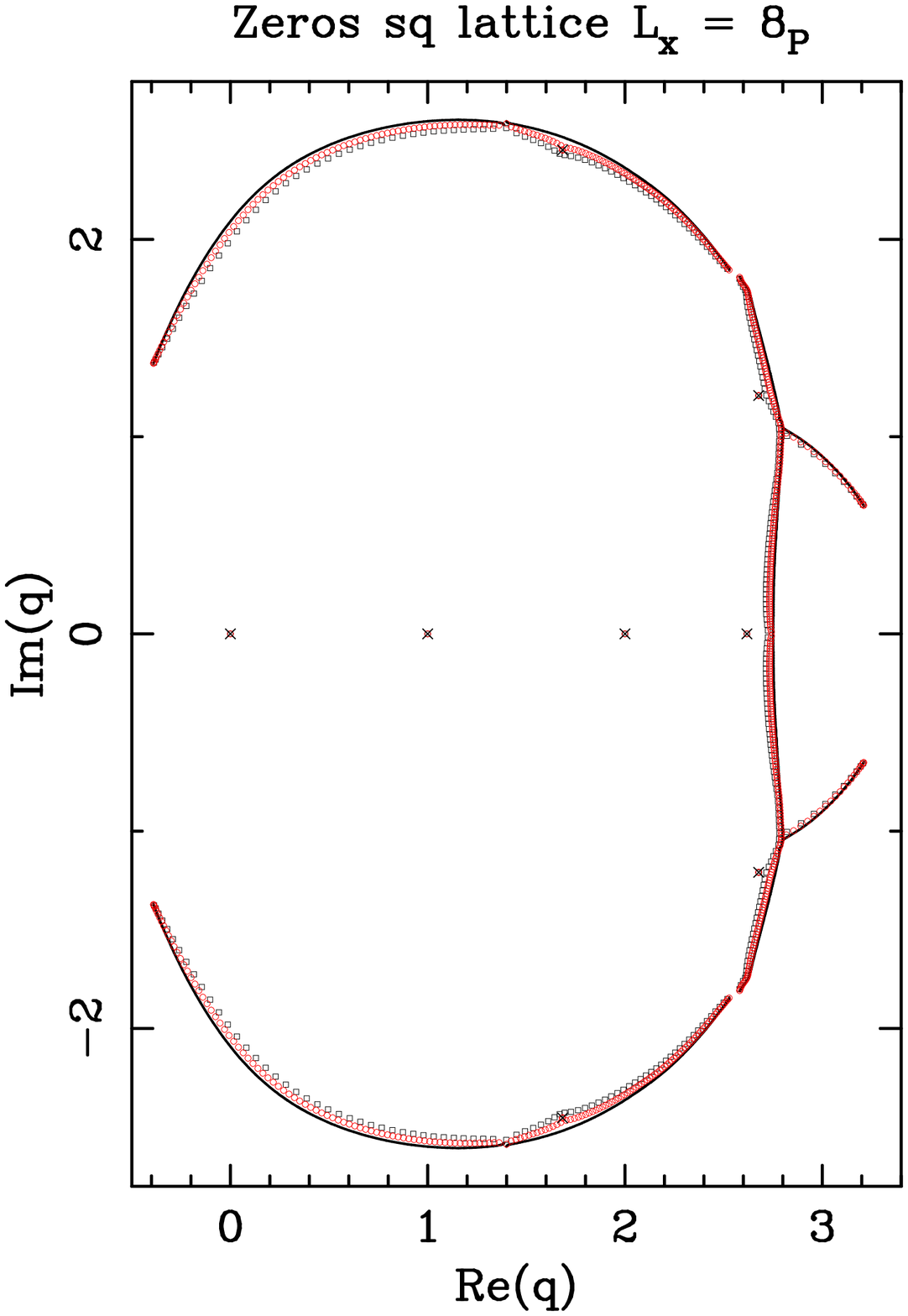}
  \caption{
  Zeros of the partition function of the $q$-state Potts antiferromagnet
  on a square lattices $8_P \times 40_F$ (squares),
  $8_P \times 80_F$ (circles) and $8_P\times\infty_F$ (solid line).
  The isolated limiting zeros are depicted by a $\times$.
  The limiting curve was computed using the resultant method.
  }
\label{Figure_sq_8PxInftyF}
\end{figure}

\clearpage
%
%
\begin{figure}
\begin{tabular}{cc}
  \epsfxsize=200pt\epsffile{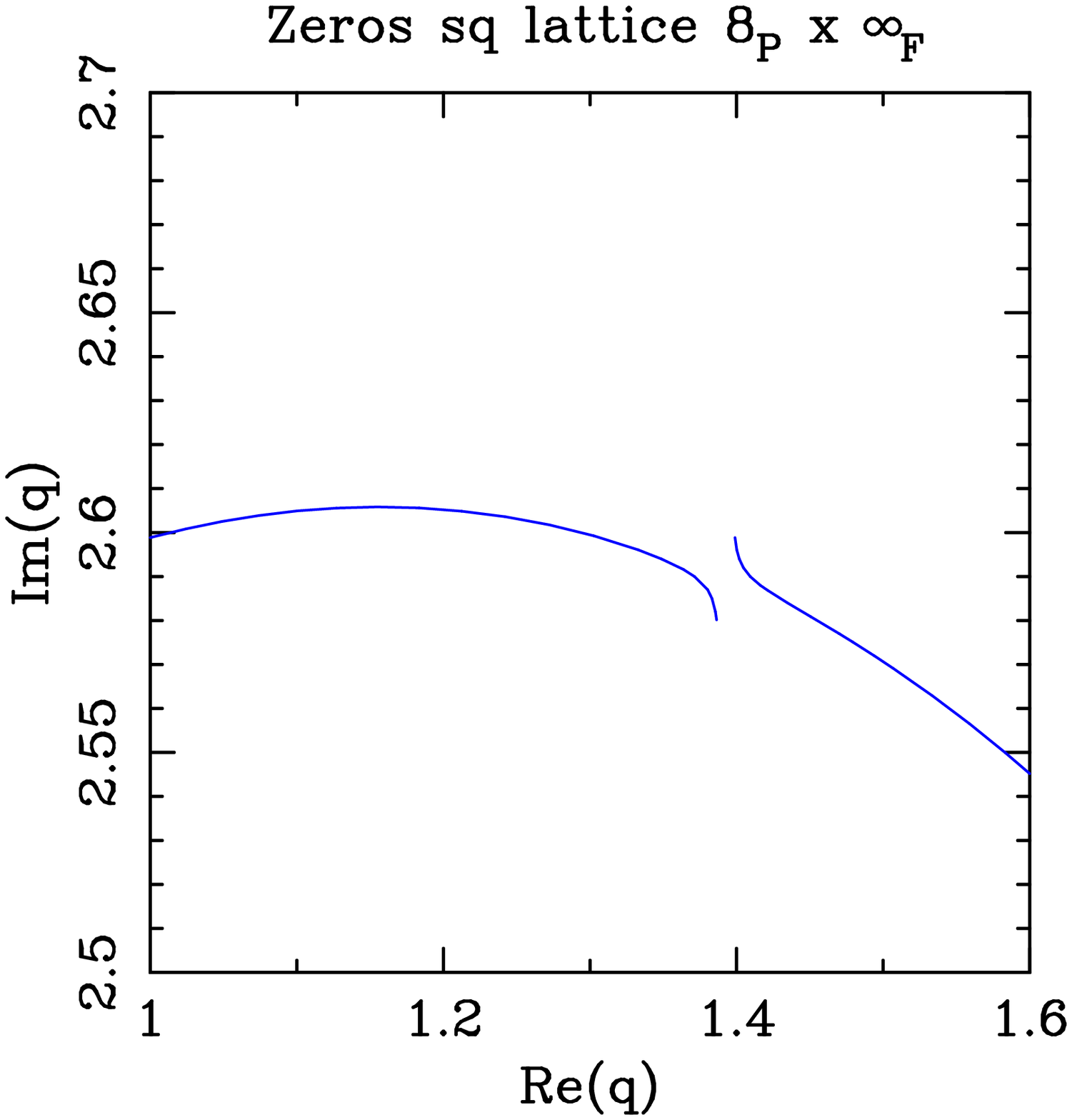} &
  \epsfxsize=200pt\epsffile{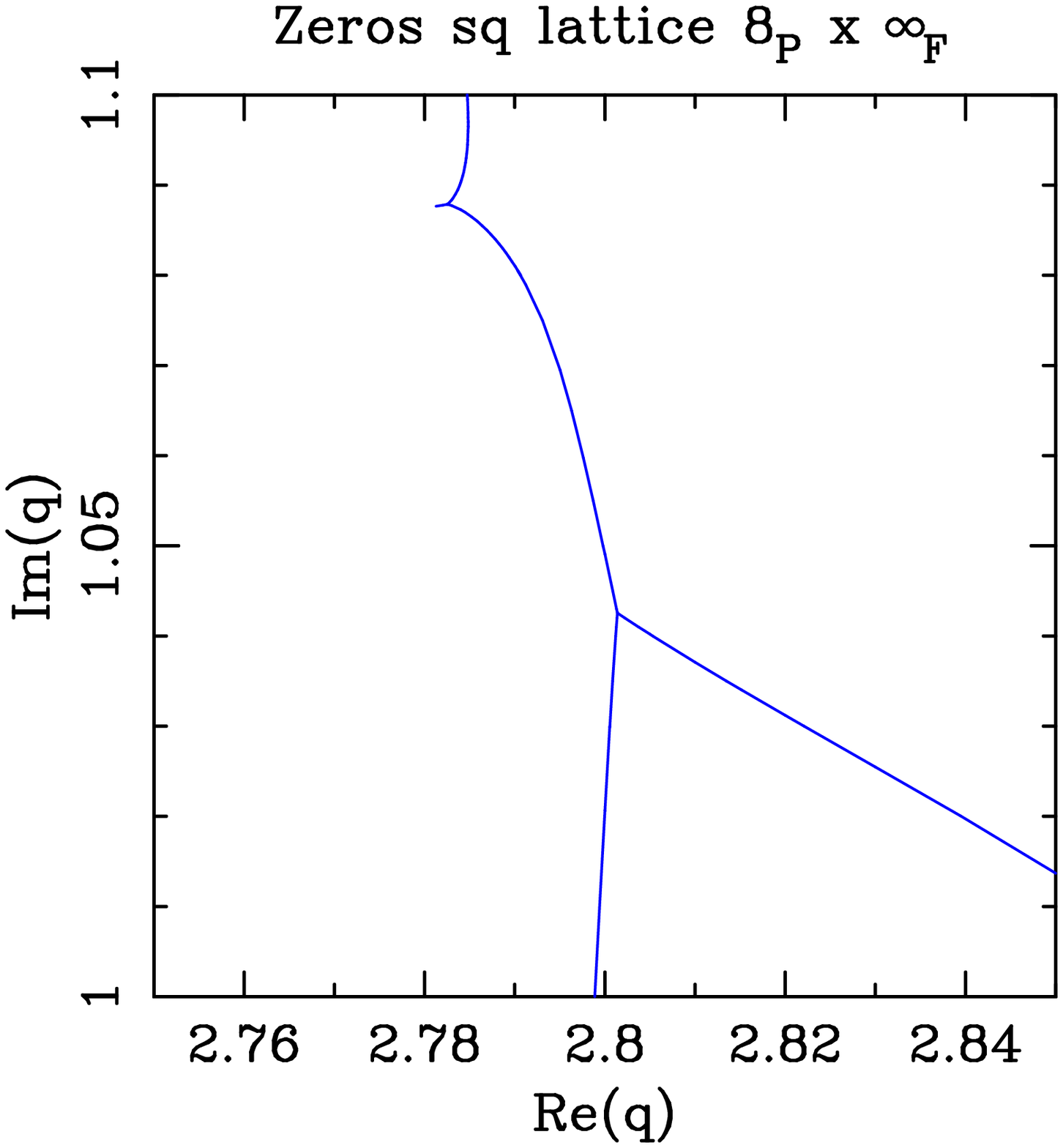} \\
  (a) & (b) \\[4mm]
  \epsfxsize=200pt\epsffile{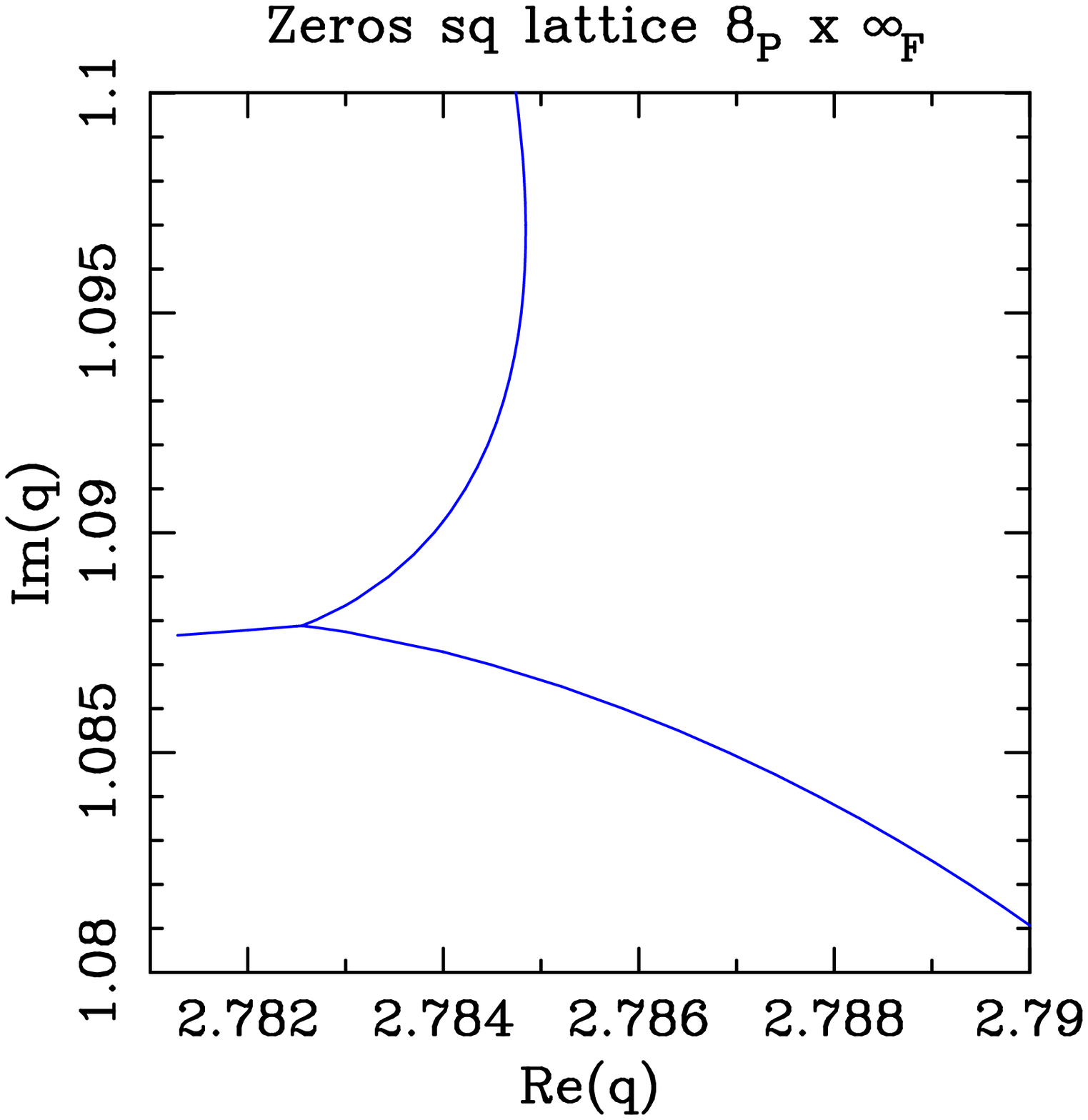} &
  \epsfxsize=200pt\epsffile{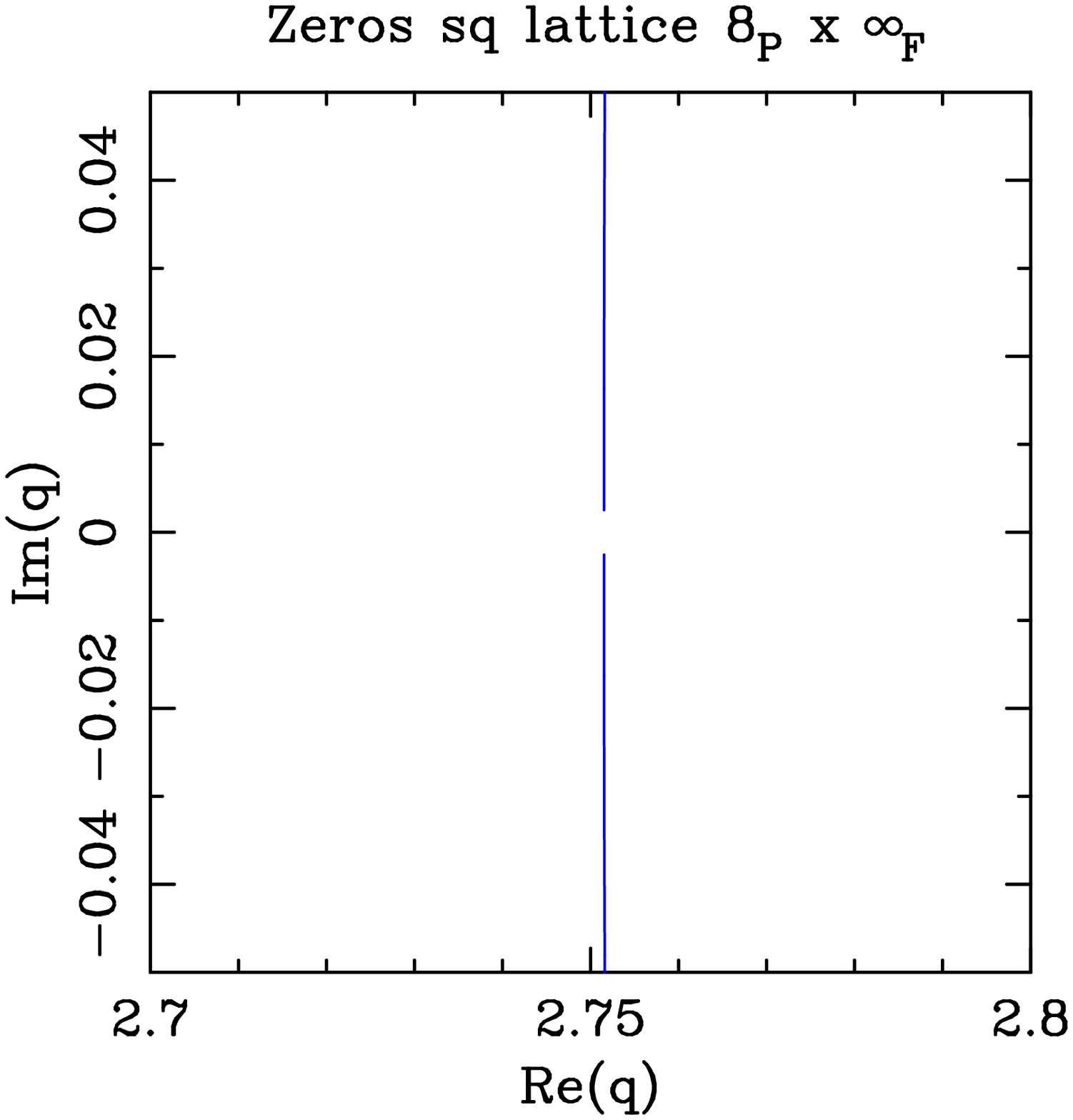} \\
  (c) & (d)
\end{tabular}
  \caption{
  Detail of the limiting curves $\scrb$ for
  the $q$-state Potts antiferromagnet on a square lattice $8_P\times\infty_F$.
  (a) Region around the gap at $q\approx 1.39 + 2.59\,i$.
  (b) Region around the T points $q\approx2.783+1.088\,i$ and
      $q\approx 2.801+1.043\,i$.  At the former T point we have
      $t\approx (1.074,0.0184,1.115)$ and $\theta\approx (1.642,0.037,1.679)$;
      at the latter we have $t\approx (0.119,0.813,1.031)$ and
      $\theta\approx (0.237,1.365,1.601)$.
  (c) Detail of the T point at $q\approx2.783+1.088\,i$.
  (d) Region around the tiny gap at $q\approx 2.7515$.
  }
\label{Figure_sq_8PxInftyF_bis}
\end{figure}

\clearpage
%
%
\begin{figure}
  \epsfxsize=400pt\epsffile{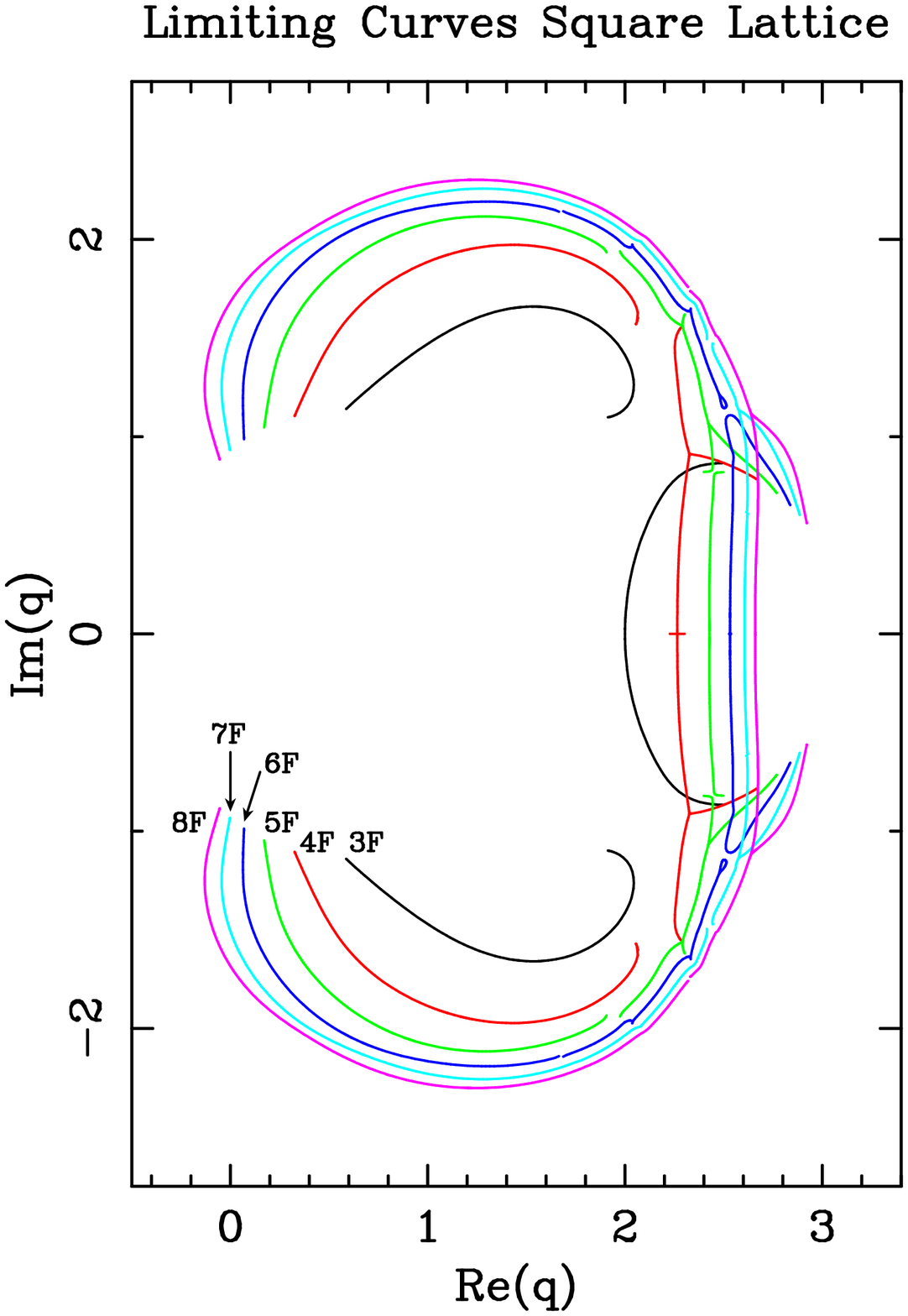}
  \caption{
  Limiting curves for the square-lattice strips
  $L_{\rm F} \times \infty_{\rm F}$ with $3 \leq L \leq 8$.
  }
\label{Figure_limit_F_all}
\end{figure}

\clearpage
%
%
\begin{figure}
  \epsfxsize=400pt\epsffile{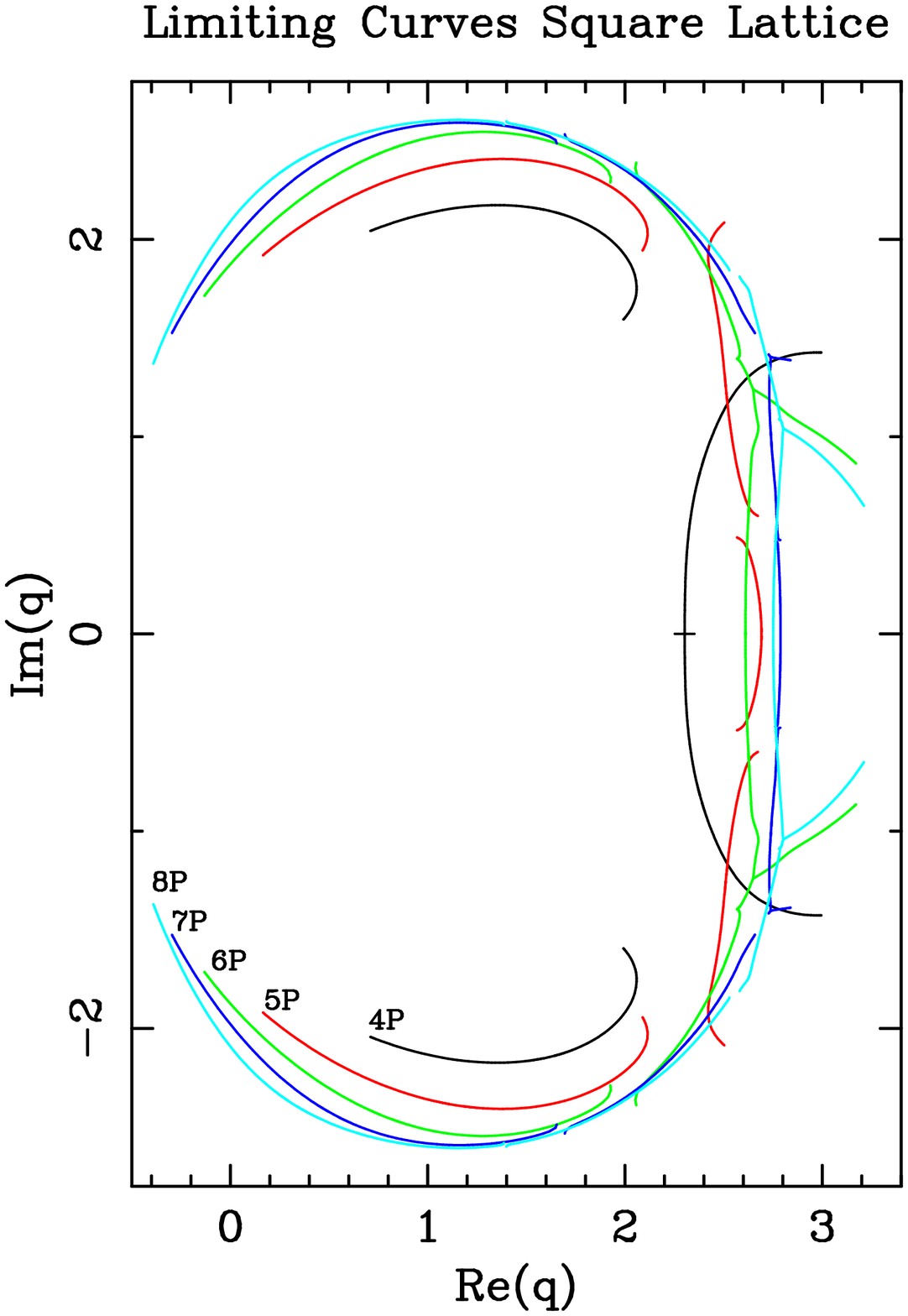}
  \caption{
  Limiting curves for the square-lattice strips
  $L_{\rm P} \times \infty_{\rm F}$ with $4 \leq L \leq 8$.
  }
\label{Figure_limit_P_all}
\end{figure}

\clearpage




\begin{thebibliography}{199}

\bibitem{Potts_52}  R.B. Potts,
   Proc. Cambridge Philos. Soc. {\bf 48}, 106 (1952).

\bibitem{Wu_82}  F.Y. Wu, Rev. Mod. Phys. {\bf 54}, 235 (1982);
   {\bf 55}, 315 (E) (1983).

\bibitem{Wu_84}  F.Y. Wu, J. Appl. Phys. {\bf 55}, 2421 (1984).

\bibitem{Baxter_82}  R.J. Baxter, {\em Exactly Solved Models in Statistical
   Mechanics}\/ (Academic Press, London--New York, 1982).

\bibitem{Itzykson_collection}  C. Itzykson, H. Saleur and J.-B. Zuber, eds.,
  {\em Conformal Invariance and Applications to Statistical Mechanics}\/
  (World Scientific, Singapore, 1988).

\bibitem{DiFrancesco_97}  P. Di Francesco, P. Mathieu and D. S\'en\'echal,
   {\em Conformal Field Theory}\/ (Springer-Verlag, New York, 1997).

\bibitem{Nienhuis_84}  B. Nienhuis, J. Stat. Phys. {\bf 34}, 731 (1984).

\bibitem{Yang-Lee_52}  C.N. Yang and T.D. Lee, Phys. Rev. {\bf 87}, 404 (1952).

\bibitem{Kasteleyn_69}  P.W. Kasteleyn and C.M. Fortuin,
   J. Phys. Soc. Japan {\bf 26} (Suppl.), 11 (1969).

\bibitem{Fortuin_72}  C.M. Fortuin and P.W. Kasteleyn,
   Physica {\bf 57}, 536 (1972).

\bibitem{Shrock_97a}  R. Shrock and S.-H. Tsai,
   Phys. Rev. E {\bf 55}, 5165 (1997), cond-mat/9612249.

\bibitem{Shrock_97b}  R. Shrock and S.-H. Tsai,
   Phys. Rev. E {\bf 56}, 1342 (1997), cond-mat/9703249.

\bibitem{Shrock_97c}  R. Shrock and S.-H. Tsai,
   Phys. Rev. E {\bf 56}, 3935 (1997), cond-mat/9707096.

\bibitem{Shrock_97d}  R. Shrock and S.-H. Tsai,
   Phys. Rev. E {\bf 56}, 4111 (1997), cond-mat/9707306.

\bibitem{Feldmann_98a} H. Feldmann, R. Shrock and S.-H. Tsai, Phys. Rev. E
   {\bf 57}, 1335 (1998), cond-mat/9711058.

\bibitem{Feldmann_98b} H. Feldmann, A.J. Guttmann, I. Jensen, R. Shrock and
   S.-H. Tsai, J. Phys. A {\bf 31}, 2287 (1998), cond-mat/9801305.

\bibitem{Shrock_98a}  M. Ro\v{c}ek, R. Shrock and S.-H. Tsai,
   Physica A {\bf 252}, 505 (1998), cond-mat/9712148.

\bibitem{Shrock_98b}  R. Shrock and S.-H. Tsai,
   Physica A {\bf 259}, 315 (1998), cond-mat/9807105.

\bibitem{Shrock_98c}  M. Ro\v{c}ek, R. Shrock and S.-H. Tsai,
   Physica A {\bf 259}, 367 (1998), cond-mat/9807106.

\bibitem{Tsai_98}  S.-H. Tsai,
   Physica A {\bf 259}, 349 (1998), cond-mat/9807107.

\bibitem{Shrock_98e}  R. Shrock and S.-H. Tsai,
   J. Phys. A {\bf 31}, 9641 (1998),  cond-mat/9810057.

\bibitem{Shrock_99a}  R. Shrock and S.-H. Tsai,
   Physica A {\bf 265}, 186 (1999),  cond-mat/9811410.

\bibitem{Shrock_99b}  R. Shrock and S.-H. Tsai,
   J. Phys. A {\bf 32}, L195 (1999), cond-mat/9903233.

\bibitem{Shrock_99c} R. Shrock and S.-H. Tsai, Phys. Rev. E {\bf 60}, 3512
  (1999), cond-mat/9910377.

\bibitem{Shrock_99d} R. Shrock and S.-H. Tsai, J. Phys. A {\bf 32}, 5053 (1999),
  cond-mat/9905431.

\bibitem{Shrock_99e} N. Biggs and R. Shrock, J. Phys. A {\bf 32}, L489 (1999),
  cond-mat/0001407.

\bibitem{Shrock_99f} R. Shrock, Phys. Lett. A {\bf 261}, 57 (1999),
  cond-mat/9908323.

\bibitem{Shrock_99g} R. Shrock, Chromatic polynomials and their zeros and
   asymptotic limits for families of graphs,
   to appear in the proceedings of the 1999 British Combinatorial Conference
   (University of Kent, Canterbury, July 1999),
   cond-mat/9908307.

\bibitem{Shrock_00a} R. Shrock and S.-H. Tsai,
  Physica A {\bf 275}, 429 (2000), cond-mat/9907403.

\bibitem{Shrock_00b} R. Shrock,
   Physica A {\bf 283}, 388 (2000), cond-mat/0001389.

\bibitem{Shrock_00c} S.-C. Chang and R. Shrock, Ground state entropy of the
   Potts antiferromagnet on triangular lattice strips,
   cond-mat/0004129.

\bibitem{Shrock_00d} S.-C. Chang and R. Shrock, Ground state entropy
   of the Potts antiferromagnet on strips of the square lattice,
   cond-mat/0004161.

\bibitem{Shrock_00e} S.-C. Chang and R. Shrock,
   Physica A {\bf 286}, 189 (2000), cond-mat/0004181.

\bibitem{Biggs_99a}  N. Biggs, A matrix method for chromatic polynomials,
   London School of Economics CDAM Research Report LSE-CDAM-99-03 (1999).

\bibitem{Biggs_99b}  N. Biggs, The chromatic polynomial of the $3 \times n$
   toroidal lattice,
   London School of Economics CDAM Research Report LSE-CDAM-99-05 (1999).

\bibitem{BKW_75}  S. Beraha, J. Kahane and N.J. Weiss,
   Proc. Nat. Acad. Sci. USA {\bf 72}, 4209 (1975).

\bibitem{BKW_78}  S. Beraha, J. Kahane and N.J. Weiss,
   in {\em Studies in Foundations and Combinatorics}\/
   (Advances in Mathematics Supplementary Studies, Vol.~1),
   ed.~G.-C. Rota (Academic Press, New York, 1978).

\bibitem{Beraha_79}  S. Beraha and J. Kahane,
   J. Combin. Theory B {\bf 27}, 1 (1979).

\bibitem{Beraha_80}  S. Beraha, J. Kahane and N.J. Weiss,
   J. Combin. Theory B {\bf 28}, 52 (1980).

\bibitem{Sokal_hierarchical}   A.D. Sokal, Chromatic roots are dense
   in the whole complex plane, in preparation.

\bibitem{Salas_97}  J. Salas and A.D. Sokal,
    J. Stat. Phys. {\bf 86}, 551 (1997), cond-mat/9603068.

\bibitem{Baxter_82b} R.J. Baxter, Proc. Roy. Soc. London A {\bf 383}, 43 (1982).

\bibitem{Onsager_44}  L. Onsager, Phys. Rev. {\bf 65}, 117 (1944).

\bibitem{Lenard_67}  A. Lenard, cited in E.H. Lieb, Phys. Rev. {\bf 162}, 162
   (1967) at pp.~169--170.

\bibitem{Baxter_70}  R.J. Baxter, J. Math. Phys. {\bf 11}, 3116 (1970).

\bibitem{Nijs_82}  M. den Nijs, M.P. Nightingale and M. Schick,
    Phys. Rev.  B {\bf 26}, 2490 (1982).

\bibitem{Burton_Henley_97} J.K. Burton Jr. and C.L. Henley,
         J. Phys. A {\bf 30}, 8385 (1997), cond-mat/9708171.

\bibitem{Salas_98}   J. Salas and A.D. Sokal,
   J. Stat. Phys. {\bf 92}, 729 (1998), cond-mat/9801079.

\bibitem{deQueiroz_99}  S.L.A. de Queiroz, T. Paiva, J.S. de S\'a Martins
   and R.R. dos Santos, Phys. Rev. E {\bf 59}, 2772 (1999), cond-mat/9812341.

\bibitem{Ferreira_99}  S.J. Ferreira and A.D. Sokal,
   J. Stat. Phys. {\bf 96}, 461 (1999), cond-mat/9811345.

%
%

\bibitem{Saleur_90}  H. Saleur,
   Commun. Math. Phys. {\bf 132}, 657 (1990).

\bibitem{Saleur_91}  H. Saleur,
   Nucl. Phys. B {\bf 360}, 219 (1991).

\bibitem{Baxter_78}  R.J. Baxter, H.N.V. Temperley and S.E. Ashley,
   Proc. Roy. Soc. London A {\bf 358}, 535 (1978).

\bibitem{Baxter_86}  R.J. Baxter, J. Phys. A {\bf 19}, 2821 (1986).

\bibitem{Baxter_87}  R.J. Baxter, J. Phys. A {\bf 20}, 5241 (1987).

\bibitem{Nienhuis_82}  B. Nienhuis, Phys. Rev. Lett. {\bf 49}, 1062 (1982).

\bibitem{Stephenson_64}  J. Stephenson, J. Math. Phys. {\bf 5}, 1009 (1964).

\bibitem{Blote_82b} H.W.J. Bl\"ote and H.J. Hilhorst,
   J. Phys. A {\bf 15}, L631 (1982).

\bibitem{Nienhuis_84b}  B. Nienhuis, H.J. Hilhorst and H.W.J. Bl\"ote,
   J. Phys. A {\bf 17}, 3559 (1984).

\bibitem{Baxter_70_TRI}  R.J. Baxter, J. Math. Phys. {\bf 11}, 784 (1970).

\bibitem{Henley_unpublished} C.L. Henley, private communications.

\bibitem{Salas_TRI4state}  J. Salas and A.D. Sokal, unpublished.

\bibitem{vEFS_unpub} A.C.D. van Enter, R. Fern\'andez and A.D. Sokal,
   unpublished (1996).

\bibitem{Adler_95}  J. Adler, A. Brandt, W. Janke and S. Shmulyan,
   J. Phys. A {\bf 28}, 5117 (1995).

\bibitem{transfer5}  J.L. Jacobsen, J. Salas and A.D. Sokal,
   Transfer matrices and partition-function zeros for
   antiferromagnetic Potts models.
   V. Nonzero temperature,
   work in progress.

\bibitem{Blote_82}  H.W.J. Bl\"ote and M.P. Nightingale,
   Physica {\bf 112A}, 405 (1982).

\bibitem{Beraha_unpub}  S. Beraha, unpublished, circa 1974.

\bibitem{Berman_69}  G. Berman and W.T. Tutte,
   J. Combin. Theory {\bf 6}, 301 (1969).

\bibitem{Tutte_70a}  W.T. Tutte,
   J. Combin. Theory {\bf 9}, 289 (1970).

\bibitem{Tutte_70b}  W.T. Tutte, Ann. New York Acad. Sci. {\bf 175}, 391 (1970).

\bibitem{Tutte_74a}  W.T. Tutte, Canad. J. Math. {\bf 26}, 893 (1974).

\bibitem{Tutte_75a}  W.T. Tutte, in {\em Studies in Graph Theory}\/,
Part II, ed.\ D.R. Fulkerson, Studies in Mathematics \#12
(Mathematical Association of America, Washington, 1975), pp.~361--377.

\bibitem{Tutte_82a}  W.T. Tutte, Canad. J. Math. {\bf 34}, 741 (1982).

\bibitem{Tutte_82b}  W.T. Tutte, Canad. J. Math. {\bf 34}, 952 (1982).

\bibitem{Martin_87}  P.P. Martin,
   J. Phys. A {\bf 20}, L399 (1987).

\bibitem{Martin_89}  P.P. Martin,
   J. Phys. A {\bf 22}, 3991 (1989).

\bibitem{Martin_91}  P.P. Martin,
   {\em Potts Models and Related Problems in Statistical Mechanics}\/.
   (World Scientific, Singapore, 1991).

\bibitem{Jaeger_91}  F. Jaeger,
   J. Combin. Theory B {\bf 52}, 259 (1991).

\bibitem{Kauffmann_93}  L.H. Kauffmann and H. Saleur,
   Commun. Math. Phys. {\bf 152}, 565 (1993).

\bibitem{Temperley_93}  H.N.V. Temperley,
   in {\em Low-Dimensional Topology and Quantum Field Theory}\/
   (Cambridge, 1992),
   ed. H.~Osborn,
   NATO Advanced Science Institutes Series B: Physics \#315
   (Plenum, New York, 1993), pp.~203--212.

\bibitem{Tutte_93}  W.T. Tutte, J. Combin. Theory. B {\bf 57}, 269 (1993).

\bibitem{Jackson_94}  D.M. Jackson,
   European J. Combin. {\bf 15}, 245 (1994).

\bibitem{Maillard_94a}  J.-M. Maillard, G. Rollet and F.Y. Wu,
   J. Phys. A {\bf 27}, 3373 (1994).

\bibitem{Maillard_94b}  J.-M. Maillard and G. Rollet,
   J. Phys. A {\bf 27}, 6963 (1994).

\bibitem{Maillard_97}  J.-M. Maillard,
   Math. Comput. Modelling {\bf 26}, 169 (1997).

\bibitem{transfer2}  J.L. Jacobsen and J. Salas,
   Transfer matrices and partition-function zeros for
   antiferromagnetic Potts models.
   II. Extended results for square-lattice chromatic polynomial,
   cond-mat/0011456.

\bibitem{transfer3}  J.L. Jacobsen, J. Salas and A.D. Sokal,
   Transfer matrices and partition-function zeros for
   antiferromagnetic Potts models.
   III. Triangular-lattice chromatic polynomial,
   in preparation.

\bibitem{transfer4}  J.L. Jacobsen, J. Salas and A.D. Sokal,
   Transfer matrices and partition-function zeros for
   antiferromagnetic Potts models.
   IV. Periodic boundary conditions in the longitudinal direction,
   work in progress.

\bibitem{Birkhoff_12}  G.D. Birkhoff, Ann. Math. {\bf 14}, 42 (1912).

\bibitem{Whitney_32a}  H. Whitney, Bull. Amer. Math. Soc. {\bf 38}, 572 (1932).

\bibitem{Tutte_47}  W.T. Tutte, Proc. Cambridge Philos. Soc. {\bf 43},
    26 (1947).

\bibitem{Tutte_54}  W.T. Tutte, Canad. J. Math. {\bf 6}, 80 (1954).

\bibitem{Edwards-Sokal}  R.G. Edwards and A.D. Sokal,
   Phys. Rev. D {\bf 38}, 2009 (1988).

\bibitem{Read_68}  R.C. Read, J. Combin. Theory {\bf 4}, 52 (1968).

\bibitem{Read_88}  R.C. Read and W.T. Tutte,
   in {\em Selected Topics in Graph Theory 3}\/,
   ed.\ L.W. Beineke and R.J. Wilson
   (Academic Press, London, 1988).

\bibitem{Chia_97}  G.L. Chia, Discrete Math. {\bf 172}, 175 (1997).

\bibitem{Stewart_87}  I. Stewart and D. Tall, {\em Algebraic Number Theory}\/,
$2^{nd}$ ed.  (Chapman and Hall, London--New York, 1987).

\bibitem{Cohn_89}  P.M. Cohn, {\em Algebra}\/, $2^{nd}$ ed., vol.~2
   (Wiley, Chichester, 1989).

\bibitem{Stewart_89}  I. Stewart, {\em Galois Theory}\/, $2^{nd}$ ed.
   (Chapman and Hall, London--New York, 1989).

\bibitem{Jackson_93}  B. Jackson, Combin. Probab. Comput. {\bf 2}, 325 (1993).


\bibitem{Birkhoff_46}  G.D. Birkhoff and D.C. Lewis,
   Trans. Amer. Math. Soc. {\bf 60}, 355 (1946).

\bibitem{Woodall_97}  D.R. Woodall, Discrete Math. {\bf 172}, 141 (1997).

\bibitem{Thomassen_97}  C. Thomassen,
    Combin. Probab. Comput. {\bf 6}, 497 (1997).

\bibitem{Woodall_77}  D.R. Woodall,
   in {\em Combinatorial Surveys: Proceedings of the Sixth British
   Combinatorial Conference}\/, ed.\ P.J. Cameron
   (Academic Press, London, 1977).

\bibitem{Woodall_92a}  D.R. Woodall, Discrete Math. {\bf 101}, 327 (1992).

\bibitem{Woodall_92b}  D.R. Woodall, Discrete Math. {\bf 101}, 333 (1992).

\bibitem{Temperley_71} H.N.V. Temperley and E.H. Lieb,
   Proc. Roy. Soc. London A {\bf 322}, 251 (1971).

\bibitem{Lieb_69}  E.H. Lieb and W.A. Beyer, Stud. Appl. Math. {\bf 48},
   77 (1969).

\bibitem{Biggs_72}  N.L. Biggs, R.M. Damerell and D.A. Sands,
   J. Combin. Theory B {\bf 12}, 123 (1972).

\bibitem{Sands_72}  D.A. Sands,  Dichromatic polynomials of linear graphs,
   Ph.D.~thesis, University of London (1972).

\bibitem{Biggs_76}  N.L. Biggs and G.H.J. Meredith,
   J. Combin. Theory B {\bf 20}, 5 (1976).

\bibitem{Derrida_80}  B. Derrida and J. Vannimenus, J. Physique Lett.
   {\bf 41}, L-473 (1980).

\bibitem{Blote_84}  H.W.J. Bl\"ote and M.P. Nightingale,
   Physica {\bf 129A}, 1 (1984).

\bibitem{Blote_89}   H.W.J. Bl\"ote and B. Nienhuis,
   J. Phys. A {\bf 22}, 1415 (1989).

\bibitem{Jacobsen_98}  J.L. Jacobsen and J. Cardy,
   Nucl. Phys. B {\bf 515}[FS], 701 (1998), cond-mat/9711279.

\bibitem{Dotsenko_99}  V. Dotsenko, J.L. Jacobsen, M.-A. Lewis and
   M. Picco, Nucl. Phys. B {\bf 546}[FS], 505 (1999), cond-mat/9812227.

\bibitem{Stanley_86}  R.P. Stanley, {\em Enumerative Combinatorics}\/,
   vol. 1 (Wadsworth \& Brooks/Cole, Monterey, CA, 1986).
   Reprinted by Cambridge University Press, 1999.

\bibitem{Stanley_99}  R.P. Stanley, {\em Enumerative Combinatorics}\/,
   vol. 2 (Cambridge University Press, Cambridge--New York, 1999).

\bibitem{Sloane_on-line}
N.J.A. Sloane, Sloane's On-Line Encyclopedia of Integer Sequences,\hfill\break
   \verb+http://www.research.att.com/~njas/sequences/index.html+

\bibitem{deBruijn_61}  N.G. De Bruijn, {\em Asymptotic Methods in Analysis}\/,
   $2^{nd}$ ed. (North-Holland, Amsterdam, 1961).

\bibitem{Sachkov_96}  V.N. Sachkov, {\em Combinatorial Methods in Discrete
   Mathematics}\/, Encyclopedia of Mathematics and its Applications \#55
   (Cambridge University Press, Cambridge, 1996).

\bibitem{Odlyzko_95}  A.M. Odlyzko, in {\em Handbook of Combinatorics}\/,
   vol.~2, ed.\ R.L. Graham, M. Gr\"otschel and L. Lov\'asz
   (Elsevier/MIT Press, Amsterdam/Cambridge, 1995), pp.~1063--1229.

\bibitem{Corless_96}  R.M. Corless, G.H. Gonnet, D.E.G. Hare, D.J. Jeffrey
   and D.E. Knuth, Adv. Comput. Math. {\bf 5}, 329 (1996).

\bibitem{Simion_91}  R. Simion and D. Ullman, Discrete Math. {\bf 98}, 193
   (1991).

\bibitem{Klazar_98}  M. Klazar, Discrete Appl. Math. {\bf 82}, 263 (1998).

\bibitem{Donaghey_77}  R. Donaghey and L.W. Shapiro, J. Combin. Theory A
   {\bf 23}, 291 (1977).

\bibitem{Riordan_75}  J. Riordan, J. Combin. Theory A {\bf 19}, 214 (1975).

\bibitem{Gouyou_88}  D. Gouyou-Beauchamps and G. Viennot,
   Adv. Appl. Math. {\bf 9}, 334 (1988).

\bibitem{Bernhart_99}  F.R. Bernhart, Discrete Math. {\bf 204}, 73 (1999).

\bibitem{Lang_93}  S. Lang, {\em Algebra}\/, $3^{rd}$ ed.
   (Addison-Wesley, Reading, Mass., 1993).

\bibitem{Cox_98}  D. Cox, J. Little and D. O'Shea, {\em Using Algebraic
   Geometry}\/ (Springer-Verlag, New York, 1998).

\bibitem{Bini_package}  D.A. Bini and G. Fiorentino,
   Numerical computation of polynomial roots: MPSolve -- Version 2.0.
   FRISCO report (1998).
   Available at
   \hfill\break
   {\tt http://www.dm.unipi.it/pages/bini/public\_html/papers/mpsolve.ps.Z}.
   Software package available at
   \hfill\break
   {\tt http://www.dm.unipi.it/pages/bini/public\_html/software/mps2.tar.gz}.

\bibitem{Bini-Fiorentino}  D.A. Bini and G. Fiorentino,
   Numer. Algorithms {\bf 23}, 127 (2000).

\bibitem{Kato_80}  T. Kato, {\em Perturbation Theory for Linear Operators}\/,
   $2^{nd}$ ed., corrected printing (Springer-Verlag, Berlin--New York, 1980).

\bibitem{Gibson_98}  C.G. Gibson, {\em Elementary Geometry of
   Algebraic Curves}\/ (Cambridge University Press, Cambridge, 1998).

\bibitem{Matveev_95} V. Matveev and R. Shrock, J. Phys. A {\bf 28}, L533 (1995).

\bibitem{Farrell_80}  E.J. Farrell, Discrete Math. {\bf 29}, 161 (1980).

\bibitem{Read_91}  R.C. Read and G.F. Royle,
  in {\em Graph Theory, Combinatorics, and Applications}\/
  (Proceedings of the Sixth Quadrennial International Conference on the
  Theory and Applications of Graphs, Western Michigan University, 1988),
  Volume 2, ed.\ Y. Alavi {\em et al.}\/
  (Wiley, New York, 1991).

\bibitem{Bakaev_94}  A.V. Bakaev and V.I. Kabanovich, J. Phys. A {\bf 27},
   6731 (1994).

\bibitem{Bollobas_98}  B. Bollobas, {\em Modern Graph Theory}\/
   (Springer-Verlag, New York--Berlin--Heidelberg, 1998).

\bibitem{Appel_77a}  K. Appel and W. Haken,
   Illinois J. Math. {\bf 21}, 429 (1977).

\bibitem{Appel_77b}  K. Appel, W. Haken and J. Koch,
   Illinois J. Math. {\bf 21}, 491 (1977).

\bibitem{Appel_89}  K. Appel and W. Haken,
   {\em Every Planar Map is Four Colorable}\/,
   Contemporary Mathematics \#98
   (American Mathematical Society, Providence RI, 1989).

\bibitem{Robertson_97}  N.Robertson, D.P. Sanders, P.D. Seymour and
   R.Thomas, J. Combin. Theory B {\bf 70}, 2 (1997).

\bibitem{Thomas_98}  R. Thomas, Notices Amer. Math. Soc. {\bf 45}, 848 (1998).


\end{thebibliography}
\end{document}